\documentclass[12pt]{article}
\pdfoutput=1
\usepackage[a4paper]{geometry}
\usepackage{jheppub, amsmath,amssymb,amsfonts,amsxtra, mathrsfs, makeidx,graphics,graphicx,amsthm,epsfig, ytableau,bm,longtable,float, color,tikz,mathtools,xfrac,footnote,rotating, lscape, makecell, environ,mathtools, empheq, colortbl, xcolor,enumerate}

\usetikzlibrary{positioning}
\usetikzlibrary{chains}
\usetikzlibrary{arrows, arrows.meta ,fit,decorations.pathreplacing}
\tikzstyle{every picture}+=[remember picture]
\tikzstyle{na} = [baseline]

\usetikzlibrary{arrows, decorations.markings, calc, fadings, decorations.pathreplacing, patterns, decorations.pathmorphing, positioning}
\usetikzlibrary{plotmarks}
\tikzset{>={Latex[width=1.5mm,length=1.5mm]}}

\usepackage{mdframed}

\def\node#1#2{\overset{#1}{\underset{#2}{{\color{gray} \bullet}}}}

\def\node#1#2{\overset{#1}{\underset{#2}{\circ}}}

\tikzstyle{every picture}+=[remember picture]
\tikzstyle{na} = [baseline=-.5ex]

\usepackage{bbm}
\usepackage{subfigure}

\newcommand{\eg}{\textit{e.g.}}

\newcommand{\ie}{\textit{i.e.}}

\numberwithin{equation}{section}
\newcommand{\bes}[1]{\begin{equation} \begin{split} #1\end{split} \end{equation}}

\newcommand{\nn}{\nonumber}

\newcommand{\be}{\begin{equation}} \newcommand{\ee}{\end{equation}}
\newcommand{\bea}{\begin{equation} \begin{aligned}} \newcommand{\eea}{\end{aligned} \end{equation}}

\def\tilde{\widetilde}

\def\hat{\widehat}

\def\bar{\overline}

\def\rt2{\sqrt{2}}

\def\Tr{\mathop{\rm Tr}}

\def\CC{{\cal C}}

\def\CH{{\cal H}}

\def\CM{{\cal M}}
\def\CN{{\cal N}}

\def\CT{{\cal T}}

\def\1{{\ds 1}}

\def\repa{\raise4pt\hbox{$\square$}\mkern-14mu\raise-4pt\hbox{$\square$}}
\def\repab{\overline{\raise4pt\hbox{$\square$}\mkern-14mu\raise-4pt\hbox{$\square$}\mkern-1mu}}

\def\smileface{\ensuremath{\hbox{\large$\bigcirc$}\mkern-15mu\raise-1pt\hbox{\scriptsize$\smallsmile$}%
\mkern-10mu\raise4pt\hbox{..}\mkern4mu}}
\def\frownface{\ensuremath{\hbox{\large$\bigcirc$}\mkern-15mu\raise-1pt\hbox{\scriptsize$\smallfrown$}%
\mkern-10mu\raise4pt\hbox{..}\mkern4mu}}

\newcommand{\ba}{\begin{array}}
\newcommand{\ea}{\end{array}}
\newcommand{\bi}{\begin{itemize}}
\newcommand{\ei}{\end{itemize}}
\def\vec#1{\bm{#1}}
\def\bea#1\eea{\allowdisplaybreaks \begin{align}#1\end{align}}
 \newcommand{\ben}{\begin{enumerate}}
\newcommand{\een}{\end{enumerate}}
\newcommand{\bean}{\begin{eqnarray*}}
\newcommand{\eean}{\end{eqnarray*}}
\newcommand{\eref}[1]{(\ref{#1})}

\newcommand{\PE}{\mathop{\rm PE}}
\newcommand{\PL}{\mathop{\rm PL}}

\newcommand{\BC}{\mathbb{C}}

\newcommand{\BZ}{\mathbb{Z}}

\newcommand{\BH}{\mathbb{H}}

\newcommand{\comment}[1]{}
\newcommand{\Sym}{\mathrm{Sym}}

\definecolor{lightgray}{gray}{0.7}
\definecolor{lightblue}{rgb}{0.12, 0.56, 1.0}
\definecolor{lightred}{rgb}{0.99, 0.56, 0.67}

\def\aup#1 {\overset{#1}{\uparrow} \, \overset{\tilde{#1}}{\downarrow}}

\tikzset{snake it/.style={decorate, decoration={snake, amplitude=.4mm, segment length=2mm,
                       post length=0mm,pre length=0mm}}}
                       
\tikzset{->-/.style={decoration={
  markings,
  mark=at position #1 with {\arrow{>}}},postaction={decorate}}}    
\tikzset{-<-/.style={decoration={
  markings,
  mark=at position #1 with {\arrow{<}}},postaction={decorate}}}          
  
\title{3d mirrors of the circle reduction of twisted $A_{2N}$ theories of class $\mathsf{S}$}
\author[a,b]{Emanuele Beratto,} 
\author[c]{Simone Giacomelli,}
\author[b,d]{Noppadol Mekareeya,}
\author[a,b]{\\ and Matteo Sacchi}
\affiliation[a]{Dipartimento di Fisica, Universit\`a di Milano-Bicocca, \\ Piazza della Scienza 3, I-20126 Milano, Italy}
\affiliation[b]{INFN, sezione di Milano-Bicocca, \\Piazza della Scienza 3, I-20126 Milano, Italy}
\affiliation[c]{Mathematical Institute, University of Oxford, \\ Woodstock Road, Oxford, OX2 6GG, United Kingdom}
\affiliation[d]{Department of Physics, Faculty of Science, \\
Chulalongkorn University, Phayathai Road, \\
Pathumwan, Bangkok 10330, Thailand}
\emailAdd{emanuele.beratto@gmail.com}
\emailAdd{simone.giacomelli@maths.ox.ac.uk}
\emailAdd{n.mekareeya@gmail.com}
\emailAdd{m.sacchi13@campus.unimib.it}
\abstract{Mirror symmetry has proven to be a powerful tool to study several properties of higher dimensional superconformal field theories upon compactification to three dimensions. We propose a quiver description for the mirror theories of the circle reduction of twisted $A_{2N}$ theories of class $\mathsf{S}$ in four dimensions.  Although these quivers bear a resemblance to the star-shaped quivers previously studied in the literature, they contain unitary, symplectic and special orthogonal gauge groups, along with hypermultiplets in the fundamental representation. The vacuum moduli spaces of these quiver theories are studied in detail.  The Coulomb branch Hilbert series of the mirror theory can be matched with that of the Higgs branch of the corresponding four dimensional theory, providing a non-trivial check of our proposal.  Moreover various deformations by mass and Fayet-Iliopoulos terms of such quiver theories are investigated.  The fact that several of them flow to expected theories also gives another strong support for the proposal.  Utilising the mirror quiver description, we discover a new supersymmetry enhancement renormalisation group flow.}

\begin{document}
\maketitle

\section{Introduction}

Three dimensional mirror symmetry \cite{Intriligator:1996ex} is one of the most important infrared (IR) duality in supersymmetric quantum field theory.  It relates a pair of theories with different descriptions by exchanging the Higgs and Coulomb branches of the theories in question.  Mirror symmetry is a powerful tool to study a reduction of superconformal field theories (SCFTs) with eight supercharges in four, five and six dimensions on a circle, a two-torus and a three-torus, respectively, to three dimensions.  In several cases, mirror theories of these resulting three dimensional theories admit Lagrangian descriptions (see \eg ~\cite{Benini:2010uu, Nanopoulos:2010bv, Xie:2012hs, DelZotto:2014kka, Buican:2014hfa, Xie:2016uqq, Mekareeya:2017sqh, Mekareeya:2017jgc, Ferlito:2017xdq, Hanany:2018uhm, Hanany:2018vph, Cabrera:2018jxt, Cabrera:2019izd, Cabrera:2019dob, Eckhard:2020jyr, Bourget:2020asf}).  The latter allows for the study of a number of important properties of those SCFTs in higher dimensions, such as the (enhanced) flavour symmetry and the Higgs branch operators. 
For several 4d SCFTs, including theories of class $\mathsf{S}$ and a number of theories of the Argyres-Douglas type \cite{Argyres:1995jj, Argyres:1995xn, Gaiotto:2009hg, Xie:2012hs}, quiver descriptions of the corresponding 3d mirrors theories have been known for a long time \cite{Benini:2010uu, Nanopoulos:2010bv, Cecotti:2010fi, Cecotti:2011rv, Xie:2012hs, Dey:2020hfe}.  In \cite{Buican:2014hfa, Buican:2017fiq}, for example, the precise flavour symmetry of a number of 4d SCFTs was determined using the corresponding 3d mirror theory.  In 5d and 6d, SCFTs may arise at the infinite coupling points of certain gauge theories \cite{Seiberg:1996bd, Morrison:1996xf, Intriligator:1997pq, Ganor:1996mu, Seiberg:1996vs, Aharony:1997bh}.  It is possible to use 3d mirror descriptions of the reduced theory to study extra massless degrees of freedom emerging at infinite coupling, in comparison to those at finite coupling, of the higher dimensional SCFTs \cite{Mekareeya:2017sqh, Mekareeya:2017jgc, Ferlito:2017xdq, Hanany:2018uhm, Hanany:2018vph, Cabrera:2018ann, Cabrera:2018jxt, Cabrera:2019izd, Cabrera:2019dob}.  In many cases, it is possible to realise the quiver description of the 3d mirror theory from a magnetic phase of the brane system.  Such a description of the 3d mirror theory is referred to as a magnetic quiver \cite{Cabrera:2018jxt, Cabrera:2019izd, Cabrera:2019dob, Bourget:2020gzi}.

In this paper we are interested in the $S^1$ reduction of 4d $\CN=2$ theories of class $\mathsf{S}$ involving twisted $A_{\text{even}}$ punctures. This type of 4d theories was proposed in \cite[Section 7.2]{Chacaltana:2012ch} and was further explored in \cite{Chacaltana:2014nya}.  In particular, the twisted $SU(2N+1)$ theory associated with a sphere with one minimal untwisted punture (labelled by $[2N,1]$) and two maximal twisted punctures (each of which is labelled by $[1^{2N}]_t$\footnote{In this paper, we use the subscript $t$ to indicate a twisted puncture, which is labelled by a $C$-partition of $2N$. A $C$-partition of an even number $m$ is an integer partition of $m$ which satisfies the condition that any odd part must appear an even number of times.}), known as the $R_{2,2N}$ theory, was studied extensively in \cite{Chacaltana:2014nya}.  Each maximal twisted puncture gives rise to an $USp(2N)$ global symmetry, whereas the minimal puncture gives rise to a $U(1)$ global symmetry.  The $USp(2N)^2$ symmetry gets enhanced to $USp(4N)$, and the $R_{2, 2N}$ theory has a global symmetry $USp(4N) \times U(1)$. In \cite{Tachikawa:2018rgw}, it was pointed out that the $USp(2N)$ global symmetry carried by each maximal twisted puncture as well as the enhanced $USp(4N)$ flavour symmetry of the $R_{2,2N}$ theory has a global $\BZ_2$ anomaly, introduced by Witten \cite{Witten:1982fp}.  The latter was shown by turning on the mass term associated with the minimal untwisted puncture of the class $\mathsf{S}$ description of the $R_{2,2N}$ theory.  This flows to an IR free theory that is described by the $SO(2N+1)$ gauge theory with $2N$ hypermultiplets in the vector representation, where it is clear that $USp(4N)$ flavour symmetry of this theory has a Witten anomaly.  In this sense, the $R_{2,2N}$ theory can be regarded as the ultraviolet completion of the $SO(2N+1)$ gauge theory with $2N$ flavours. Let us briefly discuss the brane configuration of the latter theory (see also \cite[section 4]{Tachikawa:2018rgw}).  This will turn out to be useful for the construction of the theories studied in this paper. 

The $SO(2N+1)$ gauge theory with $2N$ flavours admits the Type IIA brane realisation \cite{Witten:1997sc} involving an O4 plane, D4 branes and two half NS5 branes, with the following configuration:\footnote{Throughout the paper, unless stated otherwise, we adopt the following notations. Each red node with a label $N$ denotes an $SO(N)$ group, each blue node with an even label $2N$ denotes a $USp(2N)$ group, and each transparent node with a label $m$ denotes a $U(m)$ group. Each circular node denotes a gauge group and each rectangular node denotes a flavour group.} 
\be \label{brane}
\begin{tikzpicture}[baseline]
\draw [thick] (0,-1)--(0,1) node[xshift =0cm, yshift=0.4cm] {\tiny $\frac{1}{2}\mathrm{NS5}$};
\draw [thick] (2,-1)--(2,1) node[xshift =0cm, yshift=0.4cm] {\tiny $\frac{1}{2}\mathrm{NS5}$};
\draw [very thick] (0.05,0)--(1.95,0) node[black,midway, yshift=-0.3cm] {\scriptsize $N$ D4};
\draw [very thick] (2,0)--(4,0) node[black,midway, yshift=-0.3cm] {\scriptsize $N$ D4};
\draw [very thick] (-2,0)--(0,0) node[black,midway, yshift=-0.3cm] {\scriptsize $N$ D4};
\draw [lightblue, dashed, very thick] (2,0.04)--(4,0.04) node[black,xshift=-1cm, yshift=0.3cm] {\scriptsize \color{lightblue}  $\tilde{\mathrm{O4}}^+$};
\draw [lightred, dashed, very thick] (0.05,0.04)--(1.95,0.04) node[black,xshift=-1cm, yshift=0.3cm] {\scriptsize \color{lightred}  $\tilde{\mathrm{O4}}^-$};
\draw [lightblue, dashed, very thick] (-2,0.04)--(0,0.04) node[black,xshift=-1cm, yshift=0.3cm] {\scriptsize \color{lightblue}  $\tilde{\mathrm{O4}}^+$};
\end{tikzpicture} 
\qquad \qquad
\begin{tikzpicture}[baseline]
\tikzstyle{every node}=[font=\footnotesize]
\node[draw, rectangle,  fill=lightblue] (node1) at (-2,0) {};
\node[draw=none] at (-2,-0.4) {\tiny $USp(2N)$};
\node[draw, circle,  fill=lightred] (node2) at (0,0) {};
\node[draw=none] at (0,-0.4) {\tiny $SO(2N+1)$};
\node[draw, rectangle, fill=lightblue] (node3) at (2,0) {};
\node[draw=none] at (2,-0.4) {\tiny $USp(2N)$};
\draw[draw=black,solid]  (node1) to (node2) to (node3);
\end{tikzpicture}
\ee
There are $N$ physical D4 branes stretched between two half NS5 branes on top of the $\tilde{\mathrm{O4}}^-$ plane, and on each half NS5 brane $N$ physical semi-infinite D4 brane on top of the $\tilde{\mathrm{O4}}^+$ plane terminate.  Note that the ${\mathrm{O4}}$ plane changes sign every time it crosses a half-NS5 brane.  The $SO(2N+1)$ gauge group is realised on the D4 brane segment on top of the $\tilde{\mathrm{O4}}^-$ plane.  The $N$ flavours of hypermultiplets arise when two stacks of $N$ physical D4 branes end on a half NS5-brane from opposite sides. Indeed, the worldvolume of {\it each set} of semi-infinite D4 branes on top of the $\tilde{\mathrm{O4}}^+$ plane realises a 5d $USp(2N)$ symmetry with the discrete theta angle $\theta= \pi$ controlled by $\pi_4(USp(2N))=\BZ_2$ \cite{Hori:1998iv}, which also controls the Witten anomaly on the 4d $USp(2N)$ symmetry. Since there are in total $2N$ flavours of hypermultiplets transforming under the vector representation of $SO(2N+1)$, the theory has a $USp(4N)$ flavour symmetry.   

As pointed out in \cite{Tachikawa:2018rgw}, when the two half-NS5 branes are on top of each other, the coupling of the $SO(2N+1)$ gauge group become infinite and this brane system should realise the $R_{2,2N}$ theory.  Indeed, the two half-NS5 branes becomes a full NS5 brane, corresponding to the minimal untwisted puncture, and the two semi-infinite D4 branes on top of $\tilde{\mathrm{O4}}^+$ on each side of the brane system corresponds to each maximal twisted puncture.  This picture provides a nice way of realising the Witten anomaly carried by the maximal twisted puncture.

Our main interest is in the 3d mirror theories of the $S^1$ reductions of the twisted $A_{\text{even}}$ theories.  We propose that they admit a quiver description that can be determined using the method of \cite{Benini:2010uu}, with a simple modification.  Before discussing such a modification in detail, let us first briefly review the method of \cite{Benini:2010uu} for the untwisted $A_{N-1}$ theory associated with a sphere with punctures ${\vec \rho}_1$, ${\vec \rho}_2$ and ${\vec \rho}_3$. The 3d mirror of the $S^1$ reduction of such a theory can be described by a star-shaped quiver with 3 legs, where each leg is determined by the $T_{{\vec \rho}_1}(SU(N))$, $T_{{\vec \rho}_2}(SU(N))$ and $T_{{\vec \rho}_3}(SU(N))$ theories \cite{Gaiotto:2008ak}, whose quiver is depicted in \eref{quivTSUsigrho}, with their $U(N)$ flavour nodes being commonly gauged as a central node.  In this star-shaped quiver, an overall $U(1)$ symmetry needs to be modded out and this can be done at the central node; in which case the central node is taken to be $U(N)/U(1)$.  The theory of our interest is the twisted $A_{2N}$ theory associated with a sphere with untwisted puncture $\vec \rho$ and twisted punctures $\vec \sigma_t$ and $\vec \lambda_t$. Here $\vec \rho$ is a partition of $2N+1$ and $\vec \sigma$ and $\vec \lambda$ are $C$-partitions of $2N$.  We propose that the 3d mirror in question can obtained as follows:
\ben
\item We consider the following theories: 
\be \label{3theories}
T_{\vec \rho} (SU(2N+1))~, \qquad T_{\vec \sigma}(USp'(2N))~, \qquad T_{\vec \lambda}(USp'(2N))~.
\ee
The $T_{\vec \rho} (SU(N))$, as discussed in \cite{Gaiotto:2008ak}, has a brane configuration as a chain of NS5 branes joined by D3 branes such that there are semi-infinite D3 branes terminating on one of the outermost NS5 branes at one end. Its quiver is depicted in \eref{quivTSUsigrho}.   On the other hand, the $T_{\vec \sigma}(USp'(2N))$ is less-known in the literature.  It was introduced in \cite{Cremonesi:2014uva} (see also \cite{Feng:2000eq}) and the corresponding brane configuration is similar to that of $T_{\vec \rho} (SU(N))$, except that an $\mathrm{O3}$ plane is put into the brane system such that the semi-infinite D3 branes are on top of the $\tilde{\mathrm{O3}}^+$ plane.  Note that the $\mathrm{O3}$ plane changes sign every time it crosses a half NS5 brane.  As a result, the quiver of $T_{\vec \sigma}(USp'(2N))$ contains alternating $SO/USp$ gauge groups; it is depicted in \eref{quivTCrhoP}.
\item The $USp(2N)$ symmetry from the flavour symmetry of the theories listed in \eref{3theories} are then gauged.  It plays the role of the central gauge node in the star-shaped quiver as mentioned in \cite{Benini:2010uu}.  Note that in doing this, the $USp(2N)$ flavour node of $T_{\vec \sigma}(USp'(2N))$ and $T_{\vec \lambda}(USp'(2N))$ turns into a gauge node in the star-shaped quiver in a straightforward manner.  However, the flavour node of $T_{\vec \rho} (SU(2N+1))$ is $U(2N+1)$ and we need to decompose the bifundamental hypermultiplet between the $U(2N+1)$ flavour node and the gauge node next to it, say $U(p)$, into 
\bi
\item one hypermultiplet under the $U(p)$ gauge group, and
\item  the bifundamental hypermultiplet between $U(p) \times USp(2N)$.  
\ei
The latter $USp(2N)$ symmetry is then gauged.
\item The resulting 3d mirror quiver is an `almost' star-shaped quiver with the central node being $USp(2N)$ and with one flavour of the fundamental hypermultiplet under the unitary group $U(p)$ located next to the central $USp(2N)$ node.
\een 
We present an example of the 3d mirror theory of the $S^1$ reduction of the $R_{2,2}$ theory, which is also known as the $C_2U_1$ theory, in \eref{mirrC2U1}.  The corresponding mirror theory for $R_{2,2N}$, for a general $N$, is given in \eref{mirrR22N}.  This can be easily generalised to other theories of the same class, as demonstrated throughout the paper.   A feature of such mirror theories is that the quiver description contains unitary, symplectic and orthogonal gauge groups.

Let us briefly comment on the motivation for using the $T_{\vec \sigma}(USp'(2N))$.  We have seen from the aforementioned brane realisation of the $R_{2, 2N}$ theory that the Witten anomaly carried by the maximal twisted puncture can be realised on semi-infinite D4 branes on top of the $\tilde{\mathrm{O4}}^+$ plane.  Upon reduction on $S^1$, we expect that this corresponds to semi-infinite D3 branes on top of the $\tilde{\mathrm{O3}}^+$ plane.  This indeed shows up in the brane configuration of the $T_{\vec \sigma}(USp'(2N))$ as discussed above\footnote{In fact, we remark that the 3d $\CN=4$ $SO(2N+1)$ gauge theory with $2N$ flavours can be written as $T_{[N^2]}(USp'(2N))$.}.  In this paper, we demonstrate this proposal through a number of examples.

Recently there was an interesting proposal by Beem and Peelaers \cite{Beem:2020pry} in using twisted $A_{2N}$ theories of class $\mathsf{S}$, associated with a sphere with only regular punctures, to a number of 4d SCFTs whose Coulomb branch generators have non-integer scaling dimensions, \ie those of the Argyres-Douglas type.  It had been believed that this type of the SCFTs can only be described by theories of class $\mathsf{S}$ with irregular punctures.  The results of \cite{Beem:2020pry} allow us to make progress beyond the $R_{2, 2N}$ theory.  In this paper, we extensively use such results as a testing ground for our proposal for the 3d mirror theories of the $S^1$ reduction of twisted $A_{2N}$ theories.  In particular, we heavily rely on the observation that the Higgs branch of the 4d SCFT should match with the Coulomb branch of the 3d mirror theory of its $S^1$ reduction, and that the rank of the 4d SCFT (\ie~ the complex dimension of the Coulomb branch) should match with the quaternionic Higgs branch dimension of the corresponding mirror theory.  For the former, we match the Coulomb branch Hilbert series of the 3d mirror theory with the Higgs branch Hilbert series of the 4d theory.  We also study the Higgs branch Hilbert series of the mirror theory in detail. In some cases, there are more than one description of the mirror theory for a given 4d SCFT.  The Hilbert series between those mirror theories are matched and we conjecture that they are dual to each other.  In this way, we obtain new dual pairs between 3d $\CN=4$ gauge theories that have not be studied elsewhere in the literature.  Moreover, we study deformations of some of the proposed mirror theories by mass and Fayet-Iliopoulos (FI) terms.  We find that several theories flow to an expected theory and this provides a highly non-trivial check of our proposal.  We also discover a new supersymmetry enhancement renormalisation group flow from the $\tilde{T}_3$ theory \cite{Beem:2020pry} (or the $\mathcal{T}^{(2)}_{A_2,2}$ theory \cite{Giacomelli:2020jel})  to the $SO(4)$ super-Yang-Mills utilising the mirror description of the former.

The paper is organised as follows.  In section \ref{sec:twistedA2trinions}, we study the mirror theories of the circle reduction of the twisted $A_2$ theories associated with a sphere with three punctures.  The derivation of the mirror theories and the Hilbert series calculations are spell out explicitly, especially in the first two subsections.  In section \ref{sec:massandFI}, deformations by mass and FI terms of the mirrors of the circle reduction of some models studied in section \ref{sec:twistedA2trinions}. In section \ref{sec:twistedA2wfour}, we discuss mirror theories associated with the $\CT_{2,\frac{3}{2},\frac{3}{2}}$ and $\CT_{3,2,\frac{3}{2},\frac{3}{2}}$ theories, described by twisted $A_2$ theories with four punctures. In section \ref{sec:twistedA2N}, we discuss the generalisation of the results for the $A_2$ case to the $A_{2N}$ case.  We conclude the paper in section \ref{sec:conclusion}.

\section{Twisted $A_2$ trinions}  \label{sec:twistedA2trinions}
Let us begin by examining the circle reduction of the twisted $A_2$ theories associated with a sphere with three punctures. 

\subsection{Two copies of the $(A_1,D_4)$ theory}
The class $\mathsf{S}$ description of this theory was proposed in \cite{Beem:2020pry} and was referred to as Theory 5 in that reference.  It can be constructed by compactifying 6d (2,0) theory of the type $A_2$ on a sphere with the following punctures:
\bes{
[1^3]~, \qquad [2]_t~, \qquad [2]_t
}
where the subscript $t$ indicates the twisted puncture. Upon compactifying this theory on $S^1$, it is expected that the 3d mirror of the resulting theory can be constructed by adapting the prescription proposed in \cite{Benini:2010uu}. In particular, we conjecture that such a mirror theory admits a `star-shaped' quiver description constructed by `gluing' together the following theories:
\bes{
T_{[1^3]}(SU(3)):  &\quad (U(1))-(U(2))-[U(3)] \\
T_{[2]}(USp'(2)): &\quad  (SO(1))-[USp(2)] \\
T_{[2]}(USp'(2)): &\quad (SO(1))-[USp(2)]
}
where these theories are discussed in Appendix \ref{app:Ttheories}.
By gluing, we mean gauging the common symmetry $USp(2)$ of the above theories, whereby it is the central node of the star-shaped quiver.  Since $U(1)$ is the commutant of $USp(2)$ in $U(3)$, we should split the part $(U(2))-[U(3)]$ of $T_{[1^3]}(SU(3))$ into $[U(1)]-(U(2))-[USp(2)]$.  Gluing together the above theory along $USp(2)$ results in the following mirror theory
\be \label{mirrtwocopiesA1D4}
\begin{tikzpicture}[baseline]
\tikzstyle{every node}=[font=\footnotesize]
\node[draw, circle] (node1) at (-4,1) {$1$};
\node[draw, circle] (node2) at (-2,1) {$2$};
\node[draw, circle, fill=lightblue] (node3) at (0,1) {$2$};
\node[draw, circle, fill=lightred] (node4) at (2,1) {$1$};
\node[draw, rectangle] (sqnode1) at (-2,-0.5) {$1$};
\node[draw, circle, fill=lightred] (sqnode2) at (0,-0.5) {$1$};
\draw[draw=black,solid]  (node1) to (node2) to (node3) to (node4);
\draw[draw=black,solid]  (node2) to (sqnode1);
\draw[draw=black,solid]  (node3) to (sqnode2);
\end{tikzpicture}
\ee
Note that each of the two red circular nodes denotes the $SO(1)$ group, and so the corresponding gauge symmetry is trivial.  We can therefore rewrite this quiver as
\be \label{mirrtwocopiesA1D4a}
\begin{tikzpicture}[baseline]
\tikzstyle{every node}=[font=\footnotesize]
\node[draw, circle] (node1) at (-4,1) {$1$};
\node[draw, circle] (node2) at (-2,1) {$2$};
\node[draw, circle, fill=lightblue] (node3) at (0,1) {$2$};
\node[draw, rectangle, fill=lightred] (node4) at (2,1) {$2$};
\node[draw, rectangle] (sqnode1) at (-2,-0.5) {$1$};
\draw[draw=black,solid]  (node1) to (node2) to (node3) to (node4);
\draw[draw=black,solid]  (node2) to (sqnode1);
\end{tikzpicture}
\ee
where the rightmost red square node denotes the $SO(2)$ flavour symmetry.  

In the following, we discuss about the Coulomb and Higgs branches of the mirror theory \eref{mirrtwocopiesA1D4} or \eref{mirrtwocopiesA1D4a}.  Since upon compactification on $S^1$ the Higgs branch of the 4d theory is expected to be the same as that of the resulting 3d theory, it follows that the Coulomb branch of the mirror theory should match with the Higgs branch of the 4d theory, namely the product of two copies of the closure of the minimal nilpotent orbit $\bar{\mathrm{min}}_{SU(3)}$ of $SU(3)$.  Moreover, since the circle compactification of the $(A_1,D_4)$ theory is identified with 3d $\CN=4$ $U(1)$ gauge theory with 3 flavours (see \eg~ \cite{Xie:2012hs} and \cite{Buican:2015hsa}\footnote{This reference studied carefully dimensional reductions for various Argyres-Douglas theories, including $(A_1, D_4)$, by utilising the reduction of the index in \cite{Buican:2015ina}.}), we expect that the Higgs branch of the mirror theory \eref{mirrtwocopiesA1D4} or \eref{mirrtwocopiesA1D4a} should be $(\BC^2/\BZ_3)^2$.

Let us first comment on the enhanced Coulomb branch symmetry of quiver \eref{mirrtwocopiesA1D4a} along the line of \cite{Gaiotto:2008ak}.  Observe that the $U(1)$ and $U(2)$ gauge nodes in \eref{mirrtwocopiesA1D4a} are balanced.  As a consequence, one expects an $SU(3)$ enhanced symmetry in the IR.  Since the $USp(2)$ gauge node is also balanced, according to \cite[section 5.3]{Gaiotto:2008ak}, this $SU(3)$ symmetry gets doubled and so the symmetry of the Coulomb branch is expected to be $SU(3) \times SU(3)$.  This is in agreement with the symmetry of $(\bar{\mathrm{min}}_{SU(3)})^2$.  Subsequently we confirm such an enhanced symmetry using the Coulomb branch Hilbert series.  

The quaternionic dimension Coulomb branch of \eref{mirrtwocopiesA1D4} or \eref{mirrtwocopiesA1D4a} is
\be
\dim_\BH \CC[\eref{mirrtwocopiesA1D4} \, \text{or}\, \eref{mirrtwocopiesA1D4a}] =  1+2+1 = 4~.
\ee
This agrees with the dimension of the Higgs branch of the 4d $\CN=2$ theory, given by $24(c-a) = 24 \left( \frac{4}{3} - \frac{7}{6} \right) =4$, where $a$ and $c$ are the conformal anomalies given in (3.65) of \cite{Beem:2020pry}.  In particular, this is equal to the dimension of $\left( \bar{\mathrm{min}}_{SU(3)} \right)^2$. On the other hand, the quaternionic dimension of the Higgs branch of \eref{mirrtwocopiesA1D4} or \eref{mirrtwocopiesA1D4a} is
\be
\dim_\BH \CH[\eref{mirrtwocopiesA1D4} \, \text{or}\, \eref{mirrtwocopiesA1D4a}] =  2+2+4+\frac{1}{2}(2 \times2)-(1+4+3) = 2~.
\ee
This is in agreement with the fact that the $S^1$ compactification of two copies of rank-one $(A_1,D_4)$ yields a 3d theory with two quaternionic dimensional Coulomb branch, whose mirror theory has two quaternionic dimensional Higgs branch.  In particular, this is equal to the dimension of $\left( \BC^2/\BZ_3 \right)^2$

Let us now study the Coulomb and Higgs branches of the mirror theory in detail using the Hilbert series.  For the Coulomb branch, we present two methods in computing the Hilbert series, namely the monopole formula \cite{Cremonesi:2013lqa} and the Hall-Littlewood formula \cite{Cremonesi:2014kwa, Cremonesi:2014vla}.  For the Higgs branch, the Hilbert series can be computed using the Molien integral in the usual way \cite{Butti:2007jv} (see also \cite{Benvenuti:2010pq}).

\subsubsection*{The Coulomb branch Hilbert series}
The Coulomb branch Hilbert series computed from the monopole formula \cite{Cremonesi:2013lqa} reads
\bes{ \label{monHCmirrtwocopiesA1D4}
&H^{\text{mon}}_\CC [ \eref{mirrtwocopiesA1D4} \, \text{or}\, \eref{mirrtwocopiesA1D4a}] (t; w_1, w_2) =\\
&\sum_{m \in \BZ}~ \sum_{ n_1 \geq n_2 > -\infty} ~\sum_{a=0}^\infty t^{2 \Delta( m, \vec n, a)} P_{U(1)} (t; m) P_{U(2)} (t; \vec n)  P_{USp(2)} (t; a) \, w_1^m w_2^{n_1+n_2}~, 
}
where we denote by $m$, $\vec n = (n_1, n_2)$ and $a$ the magnetic fluxes associated with the gauge group $U(1)$, $U(2)$ and $USp(2)$ respectively; the function $\Delta( m, \vec n, a)$ is the dimension of the monopole operator with magnetic fluxes $( m, \vec n, a)$
\bes{
\Delta( m, \vec n, a) &= \frac{1}{2} \sum_{i=1}^2 \Big[ |m- n_i| + |n_i|+\left( |n_i + a| +|n_i -a| \right) \Big] \\
&\qquad +\frac{1}{2} \cdot \frac{1}{2} \left( 2 |a|+2|-a| \right) - |n_1-n_2| - |a-(-a)|~;
}
and the dressing factors are given by
\bes{
P_{U(1)}(t;m)  &= (1-t^2)^{-1}\\
P_{U(2)}(t; \vec n) &= \begin{cases} (1-t^2)^{-2} &\quad \text{if $n_1\neq n_2$} \\ (1-t^2)^{-1}(1-t^4)^{-1} & \quad \text{if $n_1= n_2$} \end{cases}   \\
P_{USp(2)}(t; a) &= \begin{cases} (1-t^2)^{-1} &\quad \text{if $a \neq 0$} \\ (1-t^4)^{-1} & \quad \text{if $a=0$}~. \end{cases} 
}
The variables $w_1$ and $w_2$ are the topological fugacities associated with the $U(1)$ and $U(2)$ gauge group, respectively.  Note that we turn off the background magnetic flux for the flavour symmetry in the above expression.   Upon computing the summation, we may rewrite \eref{monHCmirrtwocopiesA1D4} as\footnote{In this notation, the adjoint representation of $SU(3)$ is written as $\chi^{SU(3)}_{[1,1]} (w_1, w_2) = 2+w_1+w_1^{-1} +w_2+w_2^{-1}+ w_1 w_2 +w_1^{-1} w_2^{-1}$.  In the convention where the fundamental representation of $SU(3)$ is written as $\chi^{SU(3)}_{[1,0]} (x_1, x_2) = x_1 + x_2 x_1^{-1} + x_2^{-1}$, this amounts to the change of variables $w_1 = x_1 x_2^{-2}$ and $w_2=x_2 x_1^{-2}$.}
\bes{ \label{twocopiesminorbSU3}
H^{\text{mon}}_\CC [ \eref{mirrtwocopiesA1D4} \, \text{or}\, \eref{mirrtwocopiesA1D4a}] (t; w_1, w_2) = \left[ \sum_{k=0}^\infty  \chi^{SU(3)}_{[k,k]} (w_1, w_2) t^{2k}\right]^2~.
}
Note that the quantity in the square bracket is the Hilbert series of the closure of the minimal nilpotent orbit $\bar{\mathrm{min}}_{SU(3)}$ of $SU(3)$ \cite{Benvenuti:2010pq}.  This result also agrees with the Hall-Littlewood limit $q \rightarrow 0$ of the Macdonald index (3.66) of \cite{Beem:2020pry}.  It can be seen that the topological symmetry $U(1) \times U(1)$, associated with the fugacities $w_1$ and $w_2$, gets enhanced to $SU(3)$.  Note, however, that this $SU(3)$ symmetry can be identified as the diagonal subgroup of $SU(3) \times SU(3)$, which is an isometry of the product $\left( \bar{\mathrm{min}}_{SU(3)} \right)^2$ and is also full flavour symmetry of the 4d $\CN=2$ theory.  Indeed, the mirror theory \eref{mirrtwocopiesA1D4} \text{or} \eref{mirrtwocopiesA1D4a} only allows for the refinement of such a diagonal subgroup in the Coulomb branch Hilbert series \eref{monHCmirrtwocopiesA1D4}, and the rest of the full symmetry is `hidden' in the part of quiver \eref{mirrtwocopiesA1D4a} containing the $USp(2)$ gauge group in the same way as \cite{Kapustin:1998fa}.   A similar observation was made in the context of the punctures of the trinion in the class $\mathsf{S}$ description of the 4d theory; see the discussion below (3.67) in \cite{Beem:2020pry}.  

Let us now discuss the Hall-Littlewood formula for computing the Coulomb branch Hilbert series. It reads
\bes{  \label{HLHCmirrtwocopiesA1D4}
&H^{\text{HL}}_\CC [ \eref{mirrtwocopiesA1D4} \, \text{or}\, \eref{mirrtwocopiesA1D4a}] (t; y_1, y_2,y_3) =\\
&\sum_{a=0}^\infty t^{-2 |a-(-a)|} P_{USp(2)} (t; a) \times H_\CC[T_{[1^3]} (SU(3))](t; y_1, y_2, y_3; a,0,-a) \\
& \qquad   H_\CC[T_{[2]}(USp'(2))] (t;  a) H_\CC[T_{[2]}(USp'(2))]  (t;  a) ~,
}
where the expression for each of the above Coulomb branch Hilbert series is given in Appendix \ref{app:Ttheories}.
We find that
\be
H^{\text{HL}}_\CC [ \eref{mirrtwocopiesA1D4} \, \text{or}\, \eref{mirrtwocopiesA1D4a}] (t; w_1, w_2^{-1},1) = H^{\text{mon}}_\CC [ \eref{mirrtwocopiesA1D4} \, \text{or}\, \eref{mirrtwocopiesA1D4a}] (t; w_1,w_2) = \eref{twocopiesminorbSU3}~. 
\ee
One of the advantages of the Hall-Littlewood formula \eref{HLHCmirrtwocopiesA1D4} is that one only needs the information about the partitions, corresponding to the punctures of the 4d theory of class $\mathsf{S}$, and not the detailed information about the quiver of the 3d mirror theory.  Moreover, this formula takes the same form as the TQFT's structure constant of the Macdonald index \cite{Gadde:2011uv, Gadde:2009kb, Gadde:2011ik, Gaiotto:2012xa, Lemos:2012ph, Mekareeya:2012tn} of the 4d theory; see (2.9) of \cite{Beem:2020pry}.

\subsubsection*{The Higgs branch Hilbert series}
The Higgs branch Hilbert series reads
\bes{
&H_\CH [ \eref{mirrtwocopiesA1D4a}] (t; x,y) \\
&= \oint_{|u|=1} \frac{du}{2\pi i u }  \oint_{|q|=1} \frac{dq}{2\pi i q } \oint_{|z|=1} \frac{dz}{2\pi i z } (1-z^2) \oint_{|v|=1} \frac{dv}{2\pi i v } (1-v^2) \times \\
& \qquad H_\CH [[1]_u-[2]_{q,z}] (t; u, q, z) ~H_\CH [[1]_x-[2]_{q,z}] (t; x, q, z) \times \\
& \qquad  H_\CH [[2]_{q,z}-[USp(2)]_v] (t; q, z, v) \times \\
& \qquad H_\CH [[USp(2)]_v-[SO(2)]_y] (t; v, y) \times \\
& \qquad \PE \left[ -2t^2 - (z^2+1+z^{-2}) t^2- (v^2+1+v^{-2}) t^2\right]~,
}
where where $\PE$ denotes the {plethystic exponential}%
\footnote{The \textit{plethystic exponential} of a multivariate function $f(t_1, . . . , t_n)$ that vanishes at the origin is defined as ${\rm PE} \left[ f(t_1, t_2, \ldots, t_n) \right] = \exp \left( \sum_{k=1}^\infty \frac{1}{k} f(t_1^k, \cdots, t_n^k) \right)$. For instance $ \PE[n t^m]=(1-t^m)^{-n}$.}; $x$ and $y$ are fugacities for the $U(1)$ and the $SO(2)$ flavour symmetries respectively; and
\bes{ \label{HCH1}
H_\CH [[1]_u-[2]_{q,z}](t, u, q, z) &= \PE \left[ t (u q^{-1}+ u^{-1} q) (z+z^{-1}) \right]   \\
H_\CH [[2]_{q,z}-[USp(2)]_v](t, q, z, v) &= \PE \left[ t  (q^{-1}+ q) (z+z^{-1})(v+v^{-1}) \right]  \\
H_\CH [[USp(2)]_v-[SO(2)]_y](t, v, y) &= \PE \left[ t  (v+v^{-1})(y+y^{-1}) \right]  ~.
}
Evaluating the integrals, we obtain the Hilbert series of $(\BC^2/\BZ_3)^2$ as expected:
\bes{
H_\CH [ \eref{mirrtwocopiesA1D4a}] (t; x,y) &= H[\BC^2/\BZ_3] (t; x y) H[\BC^2/\BZ_3] (t; x y^{-1})
}
where $H[\BC^2/\BZ_3] (t; w)$ is the Hilbert series of $\BC^2/\BZ_3$ given by
\be \label{HSC2Z3}
H[\BC^2/\BZ_3] (t; w) = \PE \left[  t^2 + t^3 (w+w^{-1}) - t^6\right]~.
\ee

We emphasise that the $SO(2)$ symmetry in quiver \eref{mirrtwocopiesA1D4a} arises due to the proposal that each red circular node in quiver \eref{mirrtwocopiesA1D4} is in fact $SO(1)$, and not $O(1)$.  This proposal is justified by the above Higgs branch Hilbert series, since it reproduces the Hilbert series of $(\BC^2/\BZ_3)^2$ correctly.  Note that if each red circular node in quiver \eref{mirrtwocopiesA1D4} were taken to be $O(1)$, the quantities that carry fugacity $t^3 (x y^{-1}+ x^{-1} y)$, for example, are not invariant under the $O(1)$ gauge symmetry\footnote{In this case, we would have to replace the factor $H_\CH [[USp(2)]-[SO(2)]](t, z, y)$ by the square of the Higgs branch Hilbert series of $[USp(2)]-(O(1))$.  The latter is the Hilbert series of $\BC^2/\BZ_2$; see (3.32) of \cite{Benvenuti:2010pq}:
\be
H_\CH [[USp(2)]-(O(1))] (t; z) = \PE[ t^2(z^2+1+z^{-2}) -t^4]~.
\ee
The result is no longer the Hilbert series of $(\BC^2/\BZ_3)^2$.  In particular, there is no generator of the Higgs branch at order $t^3$.}.  This is also a justification to take the red circular nodes in quiver \eref{quivTCrhoP} to be of the special orthogonal type.

\subsection{The $(A_1,D_4)$ theory with a free hypermultiplet} \label{sec:A1D4wfree}
The class $\mathsf{S}$ description of this theory was proposed in \cite{Beem:2020pry} and was referred to as Theory 4 in that reference.  It can be constructed by compactifying 6d (2,0) theory of the type $A_2$ on a sphere with the following punctures:
\be
[2,1] ~, \qquad [1^2]_t~, \qquad [2]_t~.
\ee
where the subscript $t$ denotes the twisted puncture.  The mirror of the 3d theory arising from compactifying such a 4d theory on a circle admits a `star-shaped' quiver description constructed by gauging the common $USp(2)$ symmetry of the following theories \cite{Benini:2010uu}:
\bes{
T_{[2,1]}(SU(3)):  &\quad (U(1))-[U(3)] \\
T_{[1^2]}(USp'(2)): &\quad  (SO(3))-[USp(2)] \\
T_{[2]}(USp'(2)): &\quad (SO(1))-[USp(2)]
}
where $USp(2)$ plays the role of the central node of the star-shaped quiver.  Since $U(1)$ is the commutant of $USp(2)$ in $U(3)$, we need to first rewrite the quiver for $T_{[2,1]}(SU(3))$ as $[U(1)]-(U(1))-[USp(2)]$ and then gauge the $USp(2)$ group.  The 3d mirror theory in question is then
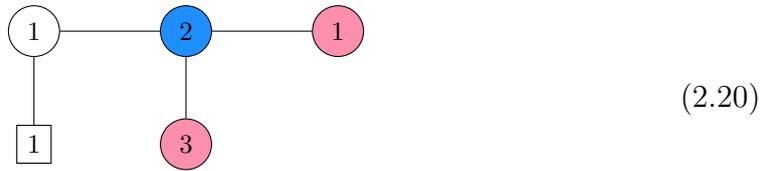
\be \label{mirrA1D4free}
\begin{tikzpicture}[baseline]
\tikzstyle{every node}=[font=\footnotesize]
\node[draw, circle] (node2) at (-2,1) {$1$};
\node[draw, circle, fill=lightblue] (node3) at (0,1) {$2$};
\node[draw, circle, fill=lightred] (node4) at (2,1) {$1$};
\node[draw, rectangle] (sqnode1) at (-2,-0.5) {$1$};
\node[draw, circle, fill=lightred] (sqnode2) at (0,-0.5) {$3$};
\draw[draw=black,solid]  (node2) to (node3) to (node4);
\draw[draw=black,solid]  (node2) to (sqnode1);
\draw[draw=black,solid]  (node3) to (sqnode2);
\end{tikzpicture}
\ee

As we have proposed and justified in the previous subsection, the rightmost red circular node with the label 1 denotes the $SO(1)$ group, and the corresponding gauge symmetry is trivial.  The line connecting it with the blue node thus denotes a half-hypermultiplet in the fundamental representation of the $USp(2)$ gauge group.  In the following we study the Coulomb and Higgs branches of \eref{mirrA1D4free}.  Since the Higgs branch of the 4d theory is $\BC^2 \times \bar{\mathrm{min}}_{SU(3)}$, we expect that the Coulomb branch of the 3d mirror theory \eref{mirrA1D4free} is isomorphic to this space also.  Moreover, similarly to the previous subsection, we also expect that the Higgs branch of \eref{mirrA1D4free} is isomorphic to $\BC^2/\BZ_3$.  Due to these properties of the moduli space, we also conjecture that theory \eref{mirrA1D4free} is {\bf dual} to the following quiver \cite{deBoer:1996mp, Porrati:1996xi, Cremonesi:2014xha}:
\bes{
\scalebox{0.8}
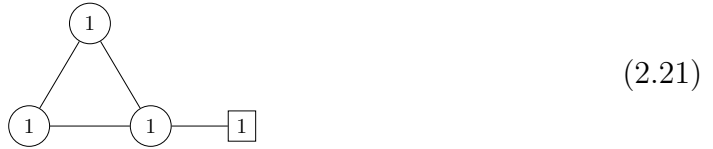
{
\begin{tikzpicture}[baseline]
\tikzstyle{every node}=[font=\footnotesize]
\node[draw, circle] (node1) at (-1,0) {$1$};
\node[draw, circle] (node2) at (1,0) {$1$};
\node[draw, circle] (node3) at (0,1.7) {$1$};
\node[draw, rectangle] (sqnode2) at (2.5,0) {$1$};
\draw[draw=black,solid]  (node1) to (node2) to (node3) to (node1);
\draw[draw=black,solid]  (sqnode2) to (node2);
\end{tikzpicture}}
}
Note that the mirror of this quiver is the well-known ADHM gauge theory for one $SU(3)$ instanton on $\BC^2$, namely the $U(1)$ gauge theory with one adjoint and three fundamental hypermultiplets \cite{deBoer:1996mp, Porrati:1996xi, Cremonesi:2014xha}:
\bes
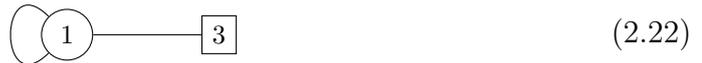
{
\begin{tikzpicture}[baseline]
\tikzstyle{every node}=[font=\footnotesize]
\node[draw, circle] (node1) at (0,0) {$1$};
\node[draw, rectangle] (node2) at (2,0) {$3$};
\draw[draw=black,solid]  (node1) to (node2);
\draw[black,solid] (node1) edge [out=135,in=-135,loop,looseness=5]  (node1);
\end{tikzpicture}
}

The Coulomb branch of \eref{mirrA1D4free} is $1+1+1=3$ quaternionic dimensional; this is in agreement with that of $\BC^2 \times \bar{\mathrm{min}}_{SU(3)}$. On the other hand, the computation of the Higgs branch dimension of \eref{mirrA1D4free} is more subtle than the previous subsection, since the $SO(3)$ gauge group is not completely broken at a generic point on the hypermultiplet moduli space.  In fact, it was argued in Footnote 7 of \cite{Cremonesi:2014uva} that the Higgs branch of the theory $(SO(3))-[USp(2)]$ is the equal to that of $(O(1))-[USp(2)]$; the latter turns out to be $\BC^2/\BZ_2$, which is one quaternionic dimensional.  The quaternionic Higgs branch dimension of \eref{mirrA1D4free} is therefore $(1\times 1)+(1\times 2)+\frac{1}{2}(2 \times 1)+1 - (1+3)= 1$, which is equal to that of $\BC^2/\BZ_3$.  In the following we study both branches of the moduli space in more detail using the Hilbert series.

\subsubsection*{The Coulomb branch Hilbert series}
Since the $SO(3)$ gauge group in \eref{mirrA1D4free} has only one flavour of the hypermultiplet transforming under the vector representation, this renders quiver \eref{mirrA1D4free} a bad theory in the sense of \cite{Gaiotto:2008ak}. In this case, the monopole formula diverges due to the presence of the monopole operators whose dimension is zero.  Nevertheless, it is possible to compute the Coulomb branch Hilbert series using the Hall-Littlewood formula.  This reads
\bes{  \label{HLmirrA1D4free}
&H_\CC [ \eref{mirrA1D4free}] (t; x_1, x_2, y) =\\
&\sum_{a=0}^\infty t^{-2 |a-(-a)|} P_{USp(2)} (t; a) \times H_\CC[T_{[2,1]} (SU(3))](t; x_1, x_2; a,0,-a) \\
& \qquad   H_\CC[T_{[1^2]}(USp'(2))] (t; y ;  a) H_\CC[T_{[2]}(USp'(2))]  (t;  a) ~,
}
where the expression for each of the above Coulomb branch Hilbert series is given in Appendix \ref{app:Ttheories}, {and the fugacities $x_1, x_2$ have to satisfy the constraint \eqref{fugconstr}:
\be
x_1^2x_2=1~.
\ee}
Evaluating the summation, we obtain
\bes{ \label{HLmirrA1D4freea}
H_\CC [ \eref{mirrA1D4free}] (t; x_1, x_2, y) = \PE \left[ (y+y^{-1}) t \right] \times \left[ \sum_{k=0}^\infty \chi^{SU(3)}_{[k,k]} (\vec u) t^{2k} \right]
}
where in this notation the character of the adjoint representation $[1,1]$ of $SU(3)$ is written as
\be
 \chi^{SU(3)}_{[1,1]} (\vec u) = u_1 u_2+ \frac{u_1^2}{u_2}+\frac{u_1}{u_2^2}+\frac{1}{u_1 u_2} +\frac{u_2}{u_1^2}+\frac{u_2^2}{u_1}+2~,
\ee
with
\be \label{u1u2}
u_1 = (x_1 x_2^{-1})^{\frac{1}{3}} y~, \qquad u_2 = (x_1^{-1} x_2)^{\frac{1}{3}} y~.
\ee
The Hilbert series \eref{HLmirrA1D4freea} is indeed that of $\BC^2 \times \bar{\mathrm{min}}_{SU(3)}$.   Note that the free hypermultiplet arises from the $(SO(3))-[USp(2)]$ part of the quiver.  The can be seen from the the fact that the fugacity $y$ associated with the $SU(2)$ symmetry of $\BC^2$, parametrised by the expectation values of the free hypermultiplet, comes from the factor $H_\CC[T_{[1^2]}(USp'(2))]$ in the Hall-Littlewood formula. It is worth pointing out that this $SU(2)$ is not manifest in the description $T_{[1^2]}(USp'(2)): (SO(3))-[USp(2)]$ but is enhanced in the IR; the reason for this is that the theory is self-mirror and that its flavour symmetry is $SU(2)$. Similarly, the $SU(3)$ symmetry of the space $\bar{\mathrm{min}}_{SU(3)}$ is also not manifest in quiver \eqref{mirrA1D4free} and is enhanced in the IR.  As can be seen from \eref{u1u2}, the generators of the Cartan subalgebra of this $SU(3)$ symmetry is a linear combination of the generator of the Cartan subalgebra of $SU(2)$, which is the symmetry of $\BC^2$, and a generator of the $U(1)$ topological symmetry in \eref{mirrA1D4free}.

\subsubsection*{The Higgs branch Hilbert series}
The Higgs branch Hilbert series can be computed as follows:
\bes{
H_\CH [ \eref{mirrA1D4free}] (t; w) &= \oint_{|u|=1} \frac{du}{2\pi i u }  \oint_{|v|=1} \frac{dv}{2\pi i v } (1-v^2) \times \\
& \qquad H_\CH [[1]_u-[1]_w] (t; u, w) \, H_\CH [[1]_u-[USp(2)]_v] (t; u, v) \times \\
& \qquad H_\CH [[USp(2)]_v-(SO(3))] (t; v) \times \\
& \qquad H_\CH [[USp(2)]_v-(SO(1))] (t; v) \times \\
& \qquad \PE \left[ -t^2- (v^2+1+v^{-2}) t^2\right]~,
}
where
\bes{ \label{HCH2}
H_\CH [[1]_u-[1]_w] (t; u, w)  &= \PE \left[ (u w^{-1}+ u^{-1} w) t \right] \\
H_\CH [[1]_u-[USp(2)]_v] (t; u, v) &= \PE \left[ (u+ u^{-1}) (v+v^{-1}) t\right] \\
H_\CH [[USp(2)]_v-(SO(3))] (t;  v) &=H [\BC^2/\BZ_2] (t;  v) = \PE[t^2(v^2+1+v^{-2}) -t^4] \\
H_\CH [[USp(2)]_v-(SO(1))] (t;  v) &=  \PE[(v+v^{-1})t] \\
}
Note that, in the third line, we have used the fact, which has been discussed earlier, that Higgs branch of the theory $(SO(3))-[USp(2)]$ is isomorphic to $\BC^2/\BZ_2$.
Evaluating the integrals, we obtain the Hilbert series of $\BC^2/\BZ_3$ as expected:
\bes{
H_\CH [ \eref{mirrA1D4free}] (t; w) &= H[\BC^2/\BZ_3] (t; w) = \PE \left[  t^2 + t^3 (w+w^{-1}) - t^6\right]~.
}

\subsection{The rank-two $SU(3)$ instanton SCFT} \label{sec:ranktwoSU3inst}
This 4d SCFT was studied extensively in \cite{Buican:2017fiq}, where it was dubbed $\CT_X$\footnote{In fact, in \cite{Buican:2014hfa, Buican:2017fiq}, the authors studied the $\CT_{3,\frac{3}{2}}$ theory, which flows to a free hypermultiplet and the interacting SCFT called $\CT_X$. The Higgs branch of the $\CT_{3,\frac{3}{2}}$ theory is the full moduli space of two $SU(3)$ instantons on $\BC^2$, which also includes the $\BC^2$ factor due to the centre of the instantons.  Upon decoupling the free hypermultiplet, the Higgs branch of the $\CT_X$ theory is identified with the reduced instanton moduli space.} (see also \cite{Beem:2019snk}).  The class $\mathsf{S}$ description of this theory was recently proposed in \cite{Beem:2020pry} and was referred to as Theory 3 or $\CT^{(2)}_{SU(3)}$ in that reference.  It can be constructed by compactifying 6d (2,0) theory of the type $A_2$ on a sphere with the following punctures:
\be
[1^3] ~, \qquad [1^2]_t~, \qquad [2]_t~.
\ee
where the subscript $t$ denotes the twisted puncture.  The mirror of the 3d theory arising from compactifying such a 4d theory on a circle can constructed by gauging the common $USp(2)$ symmetry of the following theories:
\bes{
T_{[1^3]}(SU(3)):  &\quad (U(1))-(U(2)-[U(3)] \\
T_{[1^2]}(USp'(2)): &\quad  (SO(3))-[USp(2)] \\
T_{[2]}(USp'(2)): &\quad (SO(1))-[USp(2)]
}
where $USp(2)$ plays the role of the central node of the star-shaped quiver.  Similarly to the preceding subsections, the 3d mirror theory in question is then
\be \label{mirrrank2SU3}
\begin{tikzpicture}[baseline]
\tikzstyle{every node}=[font=\footnotesize]
\node[draw, circle] (node1) at (-4,1) {$1$};
\node[draw, circle] (node2) at (-2,1) {$2$};
\node[draw, circle, fill=lightblue] (node3) at (0,1) {$2$};
\node[draw, circle, fill=lightred] (node4) at (2,1) {$1$};
\node[draw, rectangle] (sqnode1) at (-2,-0.5) {$1$};
\node[draw, circle, fill=lightred] (sqnode2) at (0,-0.5) {$3$};
\draw[draw=black,solid]  (node1) to (node2) to (node3) to (node4);
\draw[draw=black,solid]  (node2) to (sqnode1);
\draw[draw=black,solid]  (node3) to (sqnode2);
\end{tikzpicture}
\ee
In the following we study the Coulomb and Higgs branches of \eref{mirrrank2SU3}.  The ADHM gauge theory of the moduli space of two $SU(3)$ instantons on $\BC^2$ is the $U(2)$ gauge theory with one adjoint and three fundamental hypermultiplets:
\be \label{ADHM2SU3inst}
\begin{tikzpicture}[baseline]
\tikzstyle{every node}=[font=\footnotesize]
\node[draw, circle] (node1) at (0,0) {$2$};
\node[draw, rectangle] (node2) at (2,0) {$3$};
\draw[draw=black,solid]  (node1) to (node2);
\draw[black,solid] (node1) edge [out=135,in=-135,loop,looseness=5]  (node1);
\end{tikzpicture}
\ee
The Higgs branch of \eref{ADHM2SU3inst} is $\BC^2 \times \tilde{\CM}_{2,SU(3)}$, where $\tilde{\CM}_{2,SU(3)}$ is the reduced (or centred) moduli space of two $SU(3)$ instantons on $\BC^2$, and the Coulomb branch of \eref{ADHM2SU3inst} is the second symmetric power of $\BC^2/\BZ_3$ \cite{deBoer:1996mp, Hanany:2012dm}, denoted by $\Sym^2( \BC^2/\BZ_3)$.  We thus expect that the Coulomb branch of theory \eref{mirrrank2SU3} is isomorphic to $\tilde{\CM}_{2,SU(3)}$ and that the Higgs branch of \eref{mirrrank2SU3} is isomorphic to $\Sym^2( \BC^2/\BZ_3)$.  Below we show that these are indeed the case.

The Coulomb branch of \eref{mirrrank2SU3} is $1+2+1+1=5$ quaternionic dimensional; this is in agreement with that of $\tilde{\CM}_{2,SU(3)}$. On the other hand, the computation of the Higgs branch of \eref{mirrrank2SU3} can be performed similarly to the previous subsection, \ie~ by noting that the Higgs branch of the theory $(SO(3))-[USp(2)]$ is the equal to $\BC^2/\BZ_2$ \cite[Footnote 7]{Cremonesi:2014uva}, which is one quaternionic dimensional.  The quaternionic Higgs branch dimension of \eref{mirrrank2SU3} is therefore $(1\times 2)+(2\times 1)+(2 \times 2)+\frac{1}{2}(2 \times 1)+1 - (1+4+3)= 2$, which is equal to that of $\Sym^2( \BC^2/\BZ_3)$.  In the following we study both branches of the moduli space in more detail using the Hilbert series. 

We now discuss the Coulomb branch. Since the $SO(3)$ gauge group has one flavour transforming under its vector representation, the theory is `bad'. As a result, the monopole formula diverges.  However, as in the previous subsection, we can use the Hall-Littlewood formula to compute the Coulomb branch Hilbert series
\bes{  \label{HLmirrrank2SU3}
&H_\CC [ \eref{mirrrank2SU3}] (t; x_1, x_2,x_3, y) =\\
&\sum_{a=0}^\infty t^{-2 |a-(-a)|} P_{USp(2)} (t; a) \times H_\CC[T_{[1^3]} (SU(3))](t; x_1, x_2,x_3; a,0,-a) \\
& \qquad   H_\CC[T_{[1^2]}(USp'(2))] (t; y ;  a) H_\CC[T_{[2]}(USp'(2))]  (t;  a) ~,
}
where the expression for each of the above Coulomb branch Hilbert series is given in Appendix \ref{app:Ttheories}, and the fugacities $x_1, x_2, x_3$ have to satisfy the constraint \eqref{fugconstr}:
\be
x_1x_2x_3=1~.
\ee
Evaluating the summation, we obtain the Hilbert series of $\tilde{\CM}_{2,SU(3)}$ (see \cite[(3.23)]{Hanany:2012dm}):
\bes{
&H_\CC [ \eref{mirrrank2SU3}] (t; x_1, x_2,x_3, y) \\
&= \PE \left[ \left( \chi^{SU(3)}_{[1,1]}(\vec x) +\chi^{SU(2)}_{[2]}(\vec y) \right)  t^2 +  \left( \chi^{SU(3)}_{[1,1]}(\vec x) \chi^{SU(2)}_{[1]}(\vec y) \right)  t^3 - t^4 + \ldots \right]~.
}

Let us now turn to the Higgs branch. The Higgs branch Hilbert series is given by
\bes{
&H_\CH [ \eref{mirrrank2SU3}] (t; x) \\
&= \oint_{|u|=1} \frac{du}{2\pi i u }  \oint_{|q|=1} \frac{dq}{2\pi i q } \oint_{|z|=1} \frac{dz}{2\pi i z } (1-z^2) \oint_{|v|=1} \frac{dv}{2\pi i v } (1-v^2) \times \\
& \qquad H_\CH [[1]_u-[2]_{q,z}] (t; u, q, z) ~H_\CH [[1]_x-[2]_{q,z}] (t; x, q, z) \\ 
& \qquad  H_\CH [[2]_{q,z}-[USp(2)]_v] (t; q, z, v) \times \\
& \qquad H_\CH [[USp(2)]_v-(SO(3))] (t; v) \times \\
& \qquad H_\CH [[USp(2)]_v-(SO(1))] (t; v) \times \\
& \qquad \PE \left[ -2t^2 - (z^2+1+z^{-2}) t^2- (v^2+1+v^{-2}) t^2\right]~,
}
where the notations are as described in \eref{HCH1} and \eref{HCH2}.  Evaluating the integrals, we find that
\bes{
H_\CH [ \eref{mirrrank2SU3}] (t; x)  = \frac{1}{2} \left[  \left( H[\BC^2/\BZ_3] (t; x) \right)^2 +H[\BC^2/\BZ_3] (t^2; x^2) \right]
}
where the Hilbert series of $\BC^2/\BZ_3$ is given by \eref{HSC2Z3}.  This is indeed the Hilbert series of $\Sym^2(\BC^2/\BZ_3)$.

\subsection{The $R_{2,2N}$ theory}
The class $\mathsf{S}$ description of the 4d $R_{2,2N}$ SCFT was proposed in \cite{Chacaltana:2014nya}.  This is a twisted $A_{2N}$ theory associated with a sphere with punctures:
\be
[2N,1]~, \quad [1^{2N}]_t~, \quad [1^{2N}]_t~.
\ee

Let us first focus on the case of $N=1$.  This theory is also referred to as the $C_2 U_1$ theory in the literature and it corresponds to Theory 2 in \cite{Beem:2020pry}.   Following the procedures described in the previous subsections, we obtain the following 3d mirror theory upon reducing this theory on $S^1$:
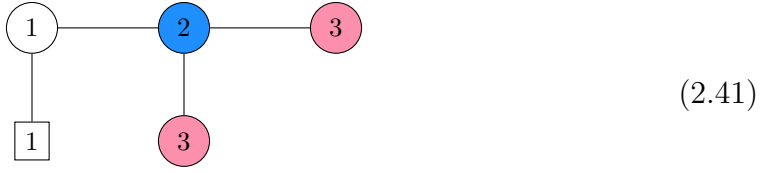
\be \label{mirrC2U1}
\begin{tikzpicture}[baseline]
\tikzstyle{every node}=[font=\footnotesize]
\node[draw, circle] (node2) at (-2,1) {$1$};
\node[draw, circle, fill=lightblue] (node3) at (0,1) {$2$};
\node[draw, circle, fill=lightred] (node4) at (2,1) {$3$};
\node[draw, rectangle] (sqnode1) at (-2,-0.5) {$1$};
\node[draw, circle, fill=lightred] (sqnode2) at (0,-0.5) {$3$};
\draw[draw=black,solid]  (node2) to (node3) to (node4);
\draw[draw=black,solid]  (node2) to (sqnode1);
\draw[draw=black,solid]  (node3) to (sqnode2);
\end{tikzpicture}
\ee
The Coulomb branch of \eref{mirrC2U1} is $1+1+1+1=4$ quaternionic dimensional, in agreement with the Higgs branch dimension of the 4d theory which is equal to $24(c-a) = 24\left( \frac{19}{12}- \frac{17}{12} \right) =4$, where $a=\frac{17}{12}$ and $c=\frac{19}{12}$ are the conformal anomalies of the 4d theory \cite{Chacaltana:2014nya}.  The quaternionic dimension of the Higgs branch of \eref{mirrC2U1} is $(1 \times 1) + (1\times 2) + 1 +1 - (1+3) = 1$, which is in agreement with the fact that the $C_2U_1$ theory is a rank-one 4d theory.  In the following, we use the Hilbert series to show that this Higgs branch is in fact isomorphic to $\BC^2/\BZ_6$.

We remark that the $S^1$ reduction of the $C_2U_1$ theory has recently been investigated in \cite{Bourget:2020asf} using a different approach. In that reference, the theory in question was studied using the magnetic quiver with a non-simply laced edge depicted in \cite[Table 2]{Bourget:2020asf}. We will see that the Coulomb branch Hilbert series computed in that reference is in agreement with ours.

\subsubsection*{The Coulomb branch Hilbert series}
The Coulomb branch Hilbert series is given by the following Hall-Littlewood formula:
\bes{  \label{HLmirrC2U1}
&H_\CC [ \eref{mirrC2U1}] (t; x_1, x_2, y, z) =\\
&\sum_{a=0}^\infty t^{-2 |a-(-a)|} P_{USp(2)} (t; a) \times H_\CC[T_{[2,1]} (SU(3))](t; x_1, x_2; a,0,-a) \\
& \qquad   H_\CC[T_{[1^2]}(USp'(2))] (t; y ;  a) H_\CC[T_{[1^2]}(USp'(2))]  (t; z;  a) ~,
}
where the expression for each of the above Coulomb branch Hilbert series is given in Appendix \ref{app:Ttheories} and the fugacities $x_1, x_2$ satisfy the relation \eref{fugconstr}:
\bes{ 
x_1^2 x_2=1~.
}
This Hilbert series can be written concisely in a closed form in terms of the highest weight generating function (HWG) \cite{Hanany:2014dia} as
\bes{ \label{HWGC2U1}
\mathrm{HWG} \left[ H_\CC [ \eref{mirrC2U1}] \right] = \PE \left[ t^2 (1+ \mu_1^2) + t^3 (w+w^{-1}) \mu_2 + t^4 \mu_2^2 - t^6 \mu_2^2\right]~.
}
where, upon computing the power series of this expression in $t$, $\mu_1^{p_1} \mu_2^{p_2}$ denotes the representation $[p_1, p_2]$, whose character written in terms of $y$ and $z$, of $USp(4)$. Here $w$ is the fugacity for the $U(1)$ symmetry which can be written in terms of $x_1, \, x_2$ as
\be
w= x_2 x_1^{-1}~.
\ee
The highest weight generating function \eref{HWGC2U1} is indeed in agreement with that presented in \cite[Table 11, row 3 with $n=2$]{Bourget:2020asf}.

As can be seen from the coefficient of the orer $t^2$, the symmetry of the Coulomb branch is indeed $USp(4) \times U(1)$. 
Note that, in this notation, the adjoint representation $[2,0]$ of $USp(4)$ can be written as
\bes{
 \chi^{USp(4)}_{[2,0]}( \vec u) = \frac{u_1^2}{u_2}+\frac{u_1^2}{u_2^2}+u_1^2+\frac{u_2^2}{u_1^2}+\frac{u_2}{u_1^2}+u_2+\frac{1}{u_1^2}+\frac{1}{u_2}+2
}
with $u_1= y$ and $u_2= y z$. Recalling that the $T_{[1^2]}(USp'(2))$ theory is self-mirror, we expect the Coulomb branch symmetry of the two copies of it appearing in the quiver \eqref{mirrC2U1} to get enhanced in the IR to $SU(2)\times SU(2)$, corresponding to the fugacities $y$ and $z$. From the above computation we see that this $SU(2)\times SU(2)$ symmetry is, in fact, further enhanced to $USp(4)$. Setting $w=1, y=1, z=1$, we obtain the unrefined Hilbert series, as presented below Eq. (3) in \cite{Chacaltana:2014nya} with $\tau =t^2$ and \cite[Table 3, row 3]{Bourget:2020asf}.  The plethystic logarithm of the Hilbert series \eref{HLmirrC2U1} can be obtain from the argument inside the $\PE$ in Eq. (3.30) of \cite{Beem:2020pry}\footnote{Note that the notation in \cite{Beem:2020pry} can be mapped to ours as follows: $t_{\text{there}} = t_{\text{ours}}^2$ and $a^3 = w$.} by taking the limit $q \rightarrow 0$ of that expression.  The generators of the moduli space and their relations were analysed in that reference.

\subsubsection*{The Higgs branch Hilbert series}
The Higgs branch Hilbert series can be computed as follows:
\bes{
H_\CH [ \eref{mirrC2U1}] (t; w) &= \oint_{|u|=1} \frac{du}{2\pi i u }  \oint_{|v|=1} \frac{dv}{2\pi i v } (1-v^2) \times \\
& \qquad H_\CH [[1]_u-[1]_w] (t; u, w) \, H_\CH [[1]_u-[USp(2)]_v] (t; u, v) \times \\
& \qquad H_\CH [[USp(2)]_v-(SO(3))] (t; v) \times \\
& \qquad H_\CH [[USp(2)]_v-(SO(3))] (t; v) \times \\
& \qquad \PE \left[ -t^2- (v^2+1+v^{-2}) t^2\right]~,
}
where the notations are as in \eref{HCH2}.  Evaluating the integrals, we obtain the Hilbert series of $\BC^2/\BZ_6$:
\bes{
H_\CH [ \eref{mirrC2U1}] (t; w) = \PE \left[ t^2 + (w+ w^{-1}) t^6 - t^{12} \right] = H[\BC^2/\BZ_6](t;w)~.
}

The appearance of $\BC^2/\BZ_6$ can be understood by considering the $\mathcal{S}$-fold realization of the $C_2U_1$ theory \cite{Apruzzi:2020pmv}: In the F-theory context this model arises by probing with a $D3$ brane a background which is obtained by combining a 7-brane of type $H_2$ with a $\BZ_2$ $\mathcal{S}$-fold action whose effect is to act as a sign flip on the Coulomb branch of the $H_2$ (or $(A_1,D_4)$) theory resulting in a $IV^*$ geometry. Upon reduction to three dimensions the Coulomb branch of the $(A_1,D_4)$ theory becomes the hyperkahler singularity $\BC^2/\BZ_3$ as we have seen before. We should then expect the $\BZ_2$ $\mathcal{S}$-fold to act on this geometry, resulting therefore in a $\BC^2/\BZ_6$ singularity.

\subsubsection*{Mirror of the $S^1$ reduction of the $R_{2, 2N}$ theory}
We propose that the 3d mirror theory in question is
\be \label{mirrR22N}
\scalebox{0.65}{
\begin{tikzpicture}[baseline]
\tikzstyle{every node}=[font=\footnotesize]
\node[draw, circle, fill=lightred] (nodem1) at (-2,1) {\tiny $2N+1$};
\node[draw, circle, fill=lightblue] (nodem2) at (-4,1) {\tiny $2N-2$};
\node[draw=none] (nodem3) at (-6,1) {\Large $\cdots$};
\node[draw, circle, fill=lightblue] (nodem4) at (-8,1) {2};
\node[draw, circle, fill=lightred] (nodem5) at (-10,1) {3};
\node[draw, circle, fill=lightblue] (node0) at (0,1) {$2N$};
\node[draw, circle, fill=lightred] (node1) at (2,1) {\tiny $2N+1$};
\node[draw, circle, fill=lightblue] (node2) at (4,1) {\tiny $2N-2$};
\node[draw=none] (node3) at (6,1) {\Large $\cdots$};
\node[draw, circle, fill=lightblue] (node4) at (8,1) {2};
\node[draw, circle, fill=lightred] (node5) at (10,1) {3};
\node[draw, circle] (U1) at (0,-0.5) {$1$};
\node[draw, rectangle] (sqnode) at (0,-2) {$1$};
\draw[draw=black,solid]  (nodem5) to (nodem4) to (nodem3) to (nodem2) to (nodem1) to (node0) to (node1) to (node2) to (node3) to (node4) to (node5);
\draw[draw=black,solid]  (node0) to (U1);
\draw[draw=black,solid]  (U1) to (sqnode);
\end{tikzpicture}}
\ee
Note that the Coulomb branch of this quiver is $2N^2+N+1$ quaternionic dimensional, where we have used the fact that the Coulomb branch of $T_{USp'(2N)}$ is $N^2$ dimensional.  This is in agreement with the Higgs branch dimension of the 4d theory which can be computed from $24(c-a)=2N^2+N+1$, where the conformal anomalies are $a= \frac{14N^2+19N+1}{24}$ and $c=\frac{8N^2+10N+1}{12}$ \cite{Chacaltana:2014nya}.  On the other hand, the Higgs branch of quiver \eref{mirrR22N} is  $2N^2+2N+1-\frac{1}{2}(2N)(2N+1)-1=N$, where we have used the fact that the Higgs branch of $T_{USp'(2N)}$ is also $N^2$ dimensional.  This result is in agreement with the fact that the $R_{2,2N}$ theory has rank $N$.

Again, we remark that there is an alternative description of the mirror theory in terms of a non-simply-laced quiver.  This, together with the corresponding highest weight generating function, were given in \cite[Table 11, row 3]{Bourget:2020asf}, with $n= 2N$.

\subsection{The $\tilde{T}_3$ or $\mathcal{T}^{(2)}_{A_2,2}$ theory}
\label{tildeT3sec}
This theory was proposed and studied in \cite{Beem:2020pry}. It also recently appeared in \cite{Giacomelli:2020jel} where it was called $\mathcal{T}^{(2)}_{A_2,2}$. It has the class $\mathsf{S}$ description as a twisted $A_2$ theory associated with the sphere with punctures
\bes{
[1^3]~, \qquad [1^2]_t~, \qquad [1^2]_t~.
}
Following the procedure described in the previous subsections, we obtain the following quiver description of the 3d mirror theory of the compactification of $\tilde{T}_3$ on $S^1$:
\be \label{mirrT3tilde}
\begin{tikzpicture}[baseline]
\tikzstyle{every node}=[font=\footnotesize]
\node[draw, circle] (node1) at (-4,1) {$1$};
\node[draw, circle] (node2) at (-2,1) {$2$};
\node[draw, circle, fill=lightblue] (node3) at (0,1) {$2$};
\node[draw, circle, fill=lightred] (node4) at (2,1) {$3$};
\node[draw, rectangle] (sqnode1) at (-2,-0.5) {$1$};
\node[draw, circle, fill=lightred] (sqnode2) at (0,-0.5) {$3$};
\draw[draw=black,solid]  (node1) to (node2) to (node3) to (node4);
\draw[draw=black,solid]  (node2) to (sqnode1);
\draw[draw=black,solid]  (node3) to (sqnode2);
\end{tikzpicture}
\ee
The Coulomb branch of \eref{mirrT3tilde} is $1+2+1+1+1 = 6$ quaternionic dimensional, in agreement with the Higgs branch dimension of the 4d theory which is equal to $24(c-a) = 24\left( 3- \frac{11}{4} \right) =6$, where $a=\frac{11}{4}$ and $c=3$ are the conformal anomalies of the 4d theory, as given in (3.1) of \cite{Beem:2020pry}.  On the other hand, the Higgs branch of \eref{mirrC2U1} is $(1 \times 2) + (2\times 1)  + (2\times 2) + 1 +1 - (1+4+3) = 2$, which is in agreement with the claim in \cite{Beem:2020pry} that $\tilde{T}_3$ is a rank-two theory.  Again, in this computation, we have used the fact that the $SO(3)$ gauge theory with one flavour has the Higgs branch isomorphic to $\BC^2/\BZ_2$, which is one quaternionic dimensional \cite{Cremonesi:2014uva}. In the following, we investigate both branches in more detail using the Hilbert series.

As in the previous subsection, the Coulomb branch Hilbert series can be computed using the Hall-Littlewood formula:
\bes{  \label{mirrT3tilde}
&H_\CC [ \eref{mirrT3tilde}] (t; x_1, x_2,x_3, y, z) =\\
&\sum_{a=0}^\infty t^{-2 |a-(-a)|} P_{USp(2)} (t; a) \times H_\CC[T_{[1^3]} (SU(3))](t; x_1, x_2,x_3; a,0,-a) \\
& \qquad   H_\CC[T_{[1^2]}(USp'(2))] (t; y ;  a) H_\CC[T_{[1^2]}(USp'(2))]  (t; z;  a) ~,
}
where the expression for each of the above Coulomb branch Hilbert series is given in Appendix \ref{app:Ttheories} and the fugacities $x_1, x_2, x_3$ have to satisfy the constraint \eqref{fugconstr}:
\be
x_1x_2x_3=1~.
\ee
The highest weight generating function of the Coulomb branch Hilbert series up to $t^{12}$ is
\bes{
&\PE \Big[ t^2 \left(\mu _1 \mu _2+\nu ^2+\sigma ^2\right) + t^4 \left(\mu _1 \mu _2 \nu  \sigma +\mu _1 \mu _2+1\right) \\
& \qquad + t^6 \left(\mu _1^3 \nu  \sigma + \mu _1 \mu _2 \nu  \sigma +\mu _2^3 \nu  \sigma +\mu _1^3+\mu _2^3\right) \\
& \qquad + t^8 \left(\mu _1^3 \nu  \sigma +\mu _2^3 \nu  \sigma \right)- t^{10} \left( \mu _1^4  \mu _2 \nu  \sigma +\mu _1  \mu _2^4  \nu  \sigma \right) \\
& \qquad - t^{12} (\mu _1^4 \mu _2  \nu ^2 \sigma ^2+ \mu _1^3  \mu _2^3 \nu ^2 \sigma ^2+ \mu _1^2  \mu _2^2 \nu ^2 \sigma ^2+ \mu _1  \mu _2^4 \nu ^2 \sigma ^2+ \mu _1^4 \mu _2 \nu  \sigma \\
& \qquad \qquad +2 \mu _1^3  \mu _2^3 \nu  \sigma + \mu _1 \mu _2^4 \nu  \sigma +\mu _2^3 \mu _1^3 )+\ldots \Big]
}
where, upon computing the power series of this expression in $t$, $\mu_1^{p_1} \mu_2^{p_2} \nu^r \sigma^s$ denotes the representation $[p_1, p_2; r; s]$, whose character can be written as $\chi^{SU(3)}_{[p_1,p_2]} (\vec x) \chi^{SU(2)}_{[r]}(y)\chi^{SU(2)}_{[s]}(z)$, of $SU(3) \times SU(2) \times SU(2)$.  This is indeed the symmetry of the Coulomb branch of the theory.  The plethystic logarithm of the Hilbert series \eref{mirrT3tilde} can be obtain from the argument inside the $\PE$ in Eq. (3.3) of \cite{Beem:2020pry} by taking the limit $q \rightarrow 0$ of that expression.  The generators of the moduli space and their relations were analysed in that reference.

Now let us examine the Higgs branch.  The Hilbert series can be computed in a similar way to the previous subsection; it is given by
\bes{
H_\CH [ \eref{mirrT3tilde}] (t; x) &=\oint_{|u|=1} \frac{du}{2\pi i u }  \oint_{|q|=1} \frac{dq}{2\pi i q } \times \\
& \qquad \oint_{|z|=1} \frac{dz}{2\pi i z } (1-z^2)  \oint_{|v|=1} \frac{dv}{2\pi i v } (1-v^2) \times \\
& \qquad H_\CH [[1]_u-[2]_{q,z}] (t; u, q, z) ~H_\CH [[1]_x-[2]_{q,z}] (t; x, q, z) \times \\
& \qquad  H_\CH [[2]_{q,z}-[USp(2)]_v] (t; q, z, v) \times \\
& \qquad H_\CH [[USp(2)]_v-(SO(3))] (t; v) \times \\
& \qquad H_\CH [[USp(2)]_v-(SO(3))] (t; v) \times \\
& \qquad\PE \left[ -2t^2 - (z^2+1+z^{-2}) t^2- (v^2+1+v^{-2}) t^2\right]~,
}
where the notations are as described in \eref{HCH1} and \eref{HCH2}.  Here $x$ is the fugacity of the $U(1)$ flavour symmetry. This can be evaluated and has the following closed form:
\bes{ \label{HLmirrT3tilde}
&H_\CH [ \eref{mirrT3tilde}] (t; x) \\
&= \frac{1}{(1-t^3 x)^2(1-t^3 x^{-1})^2(1+t^3 x)^2(1+t^3 x^{-1})^2 } \times\\
&\qquad  \Big[ 1 + t^2 + 2 t^4 + 3 t^6 +(5+x^2+x^{-2}) t^8  +(6+x^2+x^{-2}) t^{10} \\
& \qquad +(5+x^2+x^{-2}) t^{12}  + \ldots\, \text{(palindrome)} \, \ldots + t^{20} \Big]~.
}
Setting $x=1$, we obtain the following unrefined Hilbert series:
\bes{
H_\CH [ \eref{mirrT3tilde}] (t; x=1)  = \frac{1 - t^2 + t^4 + 2 t^6 + t^8-t^{10}+t^{12}}{(1-t)^4 (1+t)^4 \left(1-t+t^2\right)^2 \left(1+t+t^2\right)^2}~,
}
where the order of the pole at $t=1$ confirms that the Higgs branch is 4 complex dimensional, or equivalently 2 quaternionic dimensional as expected.  The plethystic logarithm\footnote{The plethystic logarithm of a multi-variate function $f(x_1, \ldots, x_n)$ such that $f(0, \ldots, 0)=1$ is 
	\bea
	\PL[f(x_1, \ldots, x_n)] =  \sum_{k=1}^\infty \frac{1}{k} \mu(k) \log f(x_1^k, \ldots, x_n^k) ~. \nn
	\eea} of the Hilbert series \eref{HLmirrT3tilde} is
\bes{
\PL\left[H_\CH [ \eref{mirrT3tilde}] (t; x) \right] &= t^2+ t^4+t^6 \left(2 x^2+\frac{2}{x^2}+1\right)+ t^8 \left(x^2+\frac{1}{x^2}+1\right) \\
& \quad-t^{12} \left(x^2+\frac{1}{x^2}+4\right) -t^{14} \left(2 x^2+\frac{2}{x^2}+4\right) \\
& \quad - t^{16} \left(x^4+\frac{1}{x^4}+x^2+\frac{1}{x^2}+2\right) +\ldots ~.
}

\section{Mass deformations and Fayet-Iliopoulos terms} \label{sec:massandFI}
In this section, we study the deformations by mass and Fayet-Iliopoulous (FI) terms of certain theories discussed in the previous section and investigate the end point of the RG flow. It will be seen that this provides a highly non-trivial test of the quiver descriptions of the 3d mirror theories in this paper. As a byproduct we will predict that the $\tilde{T}_3$ theory can be mass-deformed to $\mathcal{N}=4$ SYM with gauge group $SO(4)$.

\subsection{Mass deformation of the $C_2U_1$ theory} 

A further test one can consider for the proposed 3d mirror \eref{mirrC2U1} of the $C_2U_1$ theory is given by the analysis of mass deformations. We can activate a mass deformation for the $U(1)$ symmetry by turning on a FI parameter at the abelian node in the 3d mirror. The equations of motion are satisfied by activating a vev for the $U(1)_{gauge}\times U(1)_{flavour}$ bifundamental. This has the effect of breaking  $U(1)_{gauge}\times U(1)_{flavour}$ to a diagonal $U(1)$ global symmetry and quiver \eref{mirrC2U1} therefore becomes 

\be \label{mirrC2U1def}
\begin{tikzpicture}[baseline]
\tikzstyle{every node}=[font=\footnotesize]
\node[draw, circle, fill=lightblue] (node3) at (0,1) {$2$};
\node[draw, circle, fill=lightred] (node4) at (2,1) {$3$};
\node[draw, rectangle] (node2) at (-2,1) {$1$};
\node[draw, circle, fill=lightred] (sqnode2) at (0,-0.5) {$3$};
\draw[draw=black,solid]  (node2) to (node3) to (node4);
\draw[draw=black,solid]  (node3) to (sqnode2);
\end{tikzpicture}
\ee  

Let us examine this theory in detail. We start by noticing that the $USp(2)$ flavour can be interpreted as two half-hypermultiplets in the fundamental representation. We can therefore interpret the theory as a $USp(2)$ vector multiplet coupled to two identical sectors; each one describing a $USp(2)-SO(3)$ tail plus a single hypermultiplet charged under $USp(2)$ and neutral under $SO(3)$. Since $SO(3)\simeq SU(2)$, we can interpret the two above-mentioned sectors as $SU(2)$ gauge theories with a triplet (or adjoint) and a singlet and this theory is just a copy of the dimensional reduction of the $T_2$ theory \cite{Gaiotto:2009we} with the diagonal combination of two $SU(2)$ gauged (see also \cite[(2.3) and section 5.1]{Hanany:2010qu}). Said differently, this is the dimensional reduction of the untwisted $A_1$ theory of class $\mathsf{S}$ described by a one-punctured torus. 

We therefore conclude that 
\vspace{-0.8cm}
\be \label{mirrC2U1defnew}
\vspace{-1cm}
\begin{tikzpicture}[baseline]
\tikzstyle{every node}=[font=\footnotesize]
\node[draw, circle, fill=lightblue] (node3) at (0,1) {$2$};
\node[draw, circle, fill=lightred] (node4) at (2,1) {$3$};
\node[draw, rectangle] (node2) at (-2,1) {$1$};
\node[draw, circle, fill=lightred] (sqnode2) at (0,-0.5) {$3$};
\draw[draw=black,solid]  (node2) to (node3) to (node4);
\draw[draw=black,solid]  (node3) to (sqnode2);
\end{tikzpicture}
\qquad = \qquad
\begin{tikzpicture}[baseline]
\node[draw, circle, fill=lightgray] (L1) at (6,0) {\scriptsize $T_2$};
\path[every node/.style={font=\sffamily\small,
  		fill=white,inner sep=1pt}]
(L1) edge [loop, out=120, in=240, looseness=8] (L1);
\node[draw, circle, fill=lightred] (node6) at (4.8,0) {\begin{tiny} $SU(2)$\end{tiny}};
\node[draw, circle, fill=lightblue] (node7) at (7.5,0) {$2$};
\node[draw, circle, fill=lightgray] (L2) at (9,0) {\scriptsize $T_2$};

\path[every node/.style={font=\sffamily\small,
  		fill=white,inner sep=1pt}]
(L2) edge [loop, out=60, in=300, looseness=8] (L2);
\node[draw, circle, fill=lightred] (node6) at (10.2,0) {\begin{tiny} $SU(2)$\end{tiny}};
\draw[draw=black,solid]  (L2) to (node7) to (L1);
\end{tikzpicture}
\ee  
where we have put $SU(2)$ in the red nodes on the right diagram in order to remind the reader that these come from the $SO(3) \simeq SU(2)$ red nodes of left diagram.  Here we denote the circle reduction of the $T_2$ theory \cite{Gaiotto:2009we} in grey.  The theory on the right is the dimensional reduction of the genus 2 class $\mathsf{S}$ theory without punctures, whose mirror dual is known to be a $U(2)/U(1) \cong SU(2)/\BZ_2 \cong SO(3)$ gauge theory with two hypermultiplets in the adjoint representation \cite{Benini:2010uu, Cremonesi:2014vla}.
Putting everything together, we find that the dimensional reduction of the mass deformed $C_2U_1$ theory is the $SU(2)$ gauge theory with two adjoint hypermuiltiplets, whose global symmetry is manifestly $USp(4)$. This confirms that the FI deformation we have considered corresponds to a mass deformation for the $U(1)$ factor. Finally, by turning on a mass for one of the adjoint hypermultiplets, we end up with $\mathcal{N}=8$ $SU(2)/\BZ_2 \cong SO(3)$ super-Yang-Mills (SYM), in agreement with the four-dimensional expectation that upon mass deformation the $C_2U_1$ model flows to a theory with sixteen supercharges; see the third arrow in \cite[(5.4)]{Argyres:2015gha}. This represents a highly nontrivial consistency check of our claim and in particular confirms that the tail $T_{[1^2]}(USp'(2))$ describing the twisted puncture $[1^2]_t$ should involve a $SO(3)$, rather than $O(3)$, gauge group.  

This sequence of mass deformations is analogous to the flow from the $E_6$ Minahan-Nemeschansky theory to $SU(2)$ SQCD with 4 flavours. Also in this case the rank of the global symmetry is decreased by two. This flow can actually be divided into two steps: The $E_6$ theory first flows to $SU(2)$ SQCD with 5 flavours\footnote{This can be also be viewed as the $S^1$ reduction of the mass deformation of the following 5d SCFTs: $E_6 \rightarrow E_5$ \cite{Seiberg:1996bd}.}, whose global symmetry is $SO(10)$, and then to SQCD with four flavours.  Note that the intermediate theory, namely the $SU(2)$ gauge theory with 5 flavours, is not a SCFT in four dimensions but rather an IR free gauge theory, The same happens for the $C_2U_1$ theory, which first flows to the infrared free $SO(3)$ gauge theory with two adjoint hypermultiplets and then to the $\mathcal{N}=4$ $SO(3)$ SYM.

\subsection{Mass deformation for the $(A_1,D_4)$ theory} 
We can perform a similar consistency check for other twisted $A_2$ trinions. Let us consider for example the three punctured sphere describing $(A_1,D_4)$ Argyres-Douglas theory plus a decoupled hypermultiplet. Using again the observation that $SO(3)-[USp(2)]$ plus a free hypermultiplet is equivalent to the class $\mathsf{S}$ one punctured torus, we find 
\vspace{-1.5cm}
\be \label{mirrA1D4}
\begin{tikzpicture}[baseline]
\tikzstyle{every node}=[font=\footnotesize]
\node[draw, circle, fill=lightblue] (node3) at (0,1) {$2$};
\node[draw, circle, fill=lightred] (node4) at (2,1) {$1$};
\node[draw, circle] (node2) at (-2,1) {$1$};
\node[draw, rectangle] (node22) at (-2,-0.5) {$1$};
\node[draw, circle, fill=lightred] (sqnode2) at (0,-0.5) {$3$};
\draw[draw=black,solid]  (node2) to (node3) to (node4);
\draw[draw=black,solid]  (node3) to (sqnode2);
\draw[draw=black,solid]  (node2) to (node22);
\end{tikzpicture}
\qquad = \qquad
\begin{tikzpicture}[baseline]
\node[draw, circle] (node8) at (5,1) {$1$};
\node[draw, rectangle] (node88) at (5,-0.5) {$1$};
\node[draw, circle, fill=lightblue] (node7) at (7,1) {$2$};
\node[draw, circle, fill=lightgray] (L2) at (8.5,1) {\scriptsize$T_2$};
\draw[draw=black,solid]  (node8) to (node88);
\path[every node/.style={font=\sffamily\small,
  		fill=white,inner sep=1pt}]
(L2) edge [loop, out=60, in=300, looseness=8] (L2);
\node[draw, circle, fill=lightred] (node6) at (9.7,1) {\begin{tiny} $SU(2)$\end{tiny}};
\draw[draw=black,solid]  (L2) to (node7) to (node8);
\end{tikzpicture}
\vspace{-0.6cm}
\ee  
We can again study the effect of a mass deformation for the four-dimensional theory by activating a FI parameter at the $U(1)$ node in (\ref{mirrA1D4}). The effect is again to higgs the abelian group and consequently the quiver on the right of (\ref{mirrA1D4}) becomes 
\vspace{-1.5cm}
\be \label{mirrA1D4def}
\begin{tikzpicture}[baseline]
\tikzstyle{every node}=[font=\footnotesize]
\node[draw, rectangle] (node8) at (5,1) {$1$};
\node[draw, circle, fill=lightblue] (node7) at (7,1) {$2$};
\node[draw, circle, fill=lightgray] (L2) at (8.5,1) {\scriptsize$T_2$};
\path[every node/.style={font=\sffamily\small,
  		fill=white,inner sep=1pt}]
(L2) edge [loop, out=60, in=300, looseness=8] (L2);
\node[draw, circle, fill=lightred] (node6) at (9.7,1) {\begin{tiny} $SU(2)$\end{tiny}};
\draw[draw=black,solid]  (L2) to (node7) to (node8);
\end{tikzpicture}
\vspace{-1.2cm}
\ee  

In order to interpret this quiver, it is convenient to look at the mirror dual. This is more easily done by interpreting (\ref{mirrA1D4def}) as the mass deformation of 
\vspace{-1.5cm}
\be \label{mirrA1D4def}
\begin{tikzpicture}[baseline]
\tikzstyle{every node}=[font=\footnotesize]
\node[draw, rectangle] (node8) at (5,1) {$2$};
\node[draw, circle, fill=lightblue] (node7) at (7,1) {$2$};
\node[draw, circle, fill=lightgray] (L2) at (8.5,1) {\scriptsize$T_2$};
\path[every node/.style={font=\sffamily\small,
  		fill=white,inner sep=1pt}]
(L2) edge [loop, out=60, in=300, looseness=8] (L2);
\node[draw, circle, fill=lightred] (node6) at (9.7,1) {\begin{tiny} $SU(2)$\end{tiny}};
\draw[draw=black,solid]  (L2) to (node7) to (node8);
\end{tikzpicture}
\vspace{-1.4cm}
\ee
which is equivalent to the untwisted $A_1$ theory class $\mathsf{S}$ two-punctured torus; this was in fact discussed in section 5.2 of \cite{Hanany:2010qu}. Its mirror dual is therefore 
\be \label{mirrA1D4def2}
\begin{tikzpicture}[baseline]
\tikzstyle{every node}=[font=\footnotesize]

\node[draw, circle] (node8) at (2,1) {$1$};
\node[draw, circle] (node7) at (2,-1) {$1$};
\node[draw, circle] (L2) at (0,0) {$2$}; 
\path[every node/.style={font=\sffamily\small,
  		fill=white,inner sep=1pt}]
(L2) edge [loop, out=140, in=220, looseness=6] (L2);
\draw[draw=black,solid]  (L2) to (node7);
\draw[draw=black,solid]  (L2) to (node8);
\end{tikzpicture}
\ee  
In order to extract the mirror dual of (\ref{mirrA1D4def}) we should remove from (\ref{mirrA1D4def2}) the $SU(2)$ fundamental we have added by hand. This can be done directly in the quiver (\ref{mirrA1D4def2}) by turning on {\bf complex} FI parameters $\lambda_{1,2}$ at the abelian nodes. Notice that they need to satisfy the constraint $\lambda_1=-\lambda_2$. This deformation forces us to turn on a nonzero expectation value for both $U(1)\times U(2)$ bifundamentals, which in turn break the $U(1)^2\times U(2)$ gauge group to $U(1)^2$. At the quiver level we find   
\be \label{mirrA1D4final}
\begin{tikzpicture}[baseline]
\tikzstyle{every node}=[font=\footnotesize]
\node[draw, circle] (node8) at (2,1) {$1$};
\node[draw, circle] (node7) at (2,-1) {$1$};
\node[draw, circle] (L2) at (0,0) {$2$}; 
\path[every node/.style={font=\sffamily\small,
  		fill=white,inner sep=1pt}]
(L2) edge [loop, out=140, in=220, looseness=6] (L2);
\draw[draw=black,solid]  (L2) to (node7);
\draw[draw=black,solid]  (L2) to (node8);
\node[] at (3.5,0) {$\longrightarrow$};
\node[draw, circle] (node9) at (5,0) {$1$};
\node[draw, circle] (node99) at (7,0) {$1$};
\node[] at (9.3,0) {$+\;\; \text{Free Hypermultiplet}$};
\path[every node/.style={font=\sffamily\small,
  		fill=white,inner sep=1pt}]
(node9) edge [loop, out=45, in=135, looseness=.9] (node99);
\path[every node/.style={font=\sffamily\small,
  		fill=white,inner sep=1pt}]
(node9) edge [loop, out=315, in=225, looseness=.9] (node99);
\end{tikzpicture}
\ee  
where the $U(1)\times U(1)$ bifundamentals arise from the off-diagonal components of the adjoint hypermultiplet, whereas the free hypermultiplet arises from the Cartan component of the adjoint representation of $U(2)$. This result is perfectly consistent with the presence of a decoupled hypermultiplet in the trinion. The interacting part of the theory is the self-mirror $U(1)$ theory with two flavors, which is known to be the dimensional reduction of $(A_1,A_3)$ AD theory. We therefore recover the well-known statement that $(A_1,D_4)$ can be mass deformed to $(A_1,A_3)$ \cite{Argyres:1995xn} (see also the fifth arrow in \cite[(5.3)]{Argyres:2015gha}), with the decoupled hypermultiplet being a spectator\footnote{We recall that the original theory has a decoupled hypermultiplet.}.

\subsection{Mass deformation from the $C_3C_1$ theory to the $C_2U_1$ theory} 
We can analogously study the mass deformation from the $C_3C_1$ theory to $C_2U_1$ at the level of 3d mirrors.  The $C_3C_1$ theory is described by a $A_3$ twisted trinion.  We remind the reader that twisted $A_3$ punctures are labelled by $B$-partitions\footnote{A $B$-partition of an odd integer $m$ is an integer partition of $m$ which satisfies the condition that any even part must appear an even number of times.} of 5 and in the case of the $C_3C_1$ theory the punctures are
\be\label{c3c1} [2,1^2] ~, \qquad [2^2,1]_t~, \qquad [2^2,1]_t~,\ee
where this was discussed in row 2 of the bottom of Page 62 of \cite{Chacaltana:2012ch}.

More precisely, the trinion (\ref{c3c1}) is a mixed fixture describing the $C_3C_1$ theory plus a decoupled hypermultiplet charged under the $SU(2)$ symmetry carried by the untwisted puncture. Since we are going to turn on a mass term for that $SU(2)$, the hypermultiplet will become massive and therefore can be ignored. The quiver tails associated with $A_{2n-1}$ twisted punctures are the models $T_{\vec \rho}(USp(2n))$ \cite{Gaiotto:2008ak} where $\vec{\rho}$ is the $B$-partition labelling the puncture. The 3d mirror of the $S^1$ reduction of (\ref{c3c1}) is then  
\be \label{mirrC3C1}
\begin{tikzpicture}[baseline]
\tikzstyle{every node}=[font=\footnotesize]
\node[draw, circle, fill=lightblue] (node3) at (0,1) {$4$};
\node[] at (-3,1.2) {$\tilde{q},q$};
\node[] at (-1,1.2) {$\widetilde{Q},Q$};
\node[] at (1,1.2) {$B_1$};
\node[] at (0.3,0.25) {$B_2$};
\node[draw, circle, fill=lightred] (node4) at (2,1) {$4$};
\node[draw, circle] (node1) at (-4,1) {$1$};
\node[draw, circle] (node2) at (-2,1) {$2$};
\node[draw, circle, fill=lightred] (sqnode2) at (0,-0.5) {$4$};
\draw[draw=black,solid]  (node1) to (node2) to (node3) to (node4);
\draw[draw=black,solid]  (node3) to (sqnode2);
\end{tikzpicture}
\ee  
where the red node labelled by $4$ denotes the $SO(4)$ gauge symmetry.   Next to each line in the quiver, we label the chiral mulitplets in the hypermultiplet or half-hypermultiplet that we shall use in the subsequent analysis.

Notice that the dimension of the Coulomb branch is $1+2+2+2+2=1+8$, in agreement with the Higgs branch of the 4d theory which is $8+1$ dimensional, where the 1 comes from the free hypermultiplet. The dimension of the Higgs branch of (\ref{mirrC3C1}) is $2+8+4\times4-1-4-10-5-5=1$, in agreement with the expectation that the 4d theory has rank 1.  Note that the $SO(4)$ gauge nodes contribute $-5$ because the group can be Higgsed at most to $U(1)$ by activating a vev for the bifundamentals.

We now turn on the {\bf complex} FI parameters $\lambda_{1,2}$ at the $U(1)$ and $U(2)$ nodes respectively, imposing the constraint $\lambda_1+2\lambda_2=0$. As was mentioned before, this corresponds to turning on a mass term for the $SU(2)$ global symmetry carried by the untwisted puncture. Let us now discuss the effect of the deformation. The equations of motion can be solved by setting 
\be \langle \tilde{q}\rangle=\sqrt{\lambda_2}\left(\begin{array}{c}1 \\ 1 \\ \end{array}\right);\quad \langle q\rangle=\sqrt{\lambda_2}(1,1),\ee 
and consequently the vev of $\widetilde{Q}$ and $Q$ satisfy the relation 
\be \langle Q\widetilde{Q}\rangle=\lambda_2 \left(\begin{array}{cc} 0 &1 \\ 1 & 0 \\\end{array}\right),\ee
where the $USp(4)$ indices are contracted in this expression. 
The Lagrangian includes the coupling \be\label{lagrangian}\Tr(\Phi_{USp(4)}\widetilde{Q}Q),\ee
where $\Tr$ denotes the trace over the fundamental representation of $USp(4)$ and $\Phi_{USp(4)}$ is a $4\times 4$ matrix in the adjoint representation of $USp(4)$ which can be written as
\be\label{USpadj} \Phi_{USp(4)}=\left(\begin{array}{cccc}
a & b & \alpha & \gamma \\
c & d & \gamma & \beta \\
\delta & \epsilon & -a & -c \\
\epsilon & \mu & -b & -d \\
\end{array}\right).\ee
Setting the vev of $Q$ and $\widetilde{Q}$ to 
\be\label{vevchoice} \langle Q\rangle= \sqrt{\lambda_2}  \left(\begin{array}{cccc}
1 & 0 & 0 & 0 \\
0 & 0 & 1 & 0 \\
\end{array}\right);\quad \langle\widetilde{Q}\rangle=  \sqrt{\lambda_2} \left(\begin{array}{cc}
 0 & 1 \\
 0 & 0 \\
 1 & 0 \\
 0 & 0 \\
\end{array}\right),\ee 
we can easily see that a $USp(2)$ subgroup of $USp(4)$ remains unbroken. Let us now look at the mass term in detail. With the following notation for the fluctuation of $Q$ and $\widetilde{Q}$ around the vev 
\be\label{fluct}  Q=   \sqrt{\lambda_2} \left(\begin{array}{cccc}
Q_{11} & Q_{12} & Q_{13} & Q_{14} \\
Q_{21} & Q_{22} & Q_{23} & Q_{24} \\
\end{array}\right);\quad \widetilde{Q}=   \sqrt{\lambda_2} \left(\begin{array}{cc}
 \widetilde{Q}_{11} &  \widetilde{Q}_{12} \\
  \widetilde{Q}_{21} &  \widetilde{Q}_{22} \\
  \widetilde{Q}_{31} &  \widetilde{Q}_{32} \\
  \widetilde{Q}_{41} &  \widetilde{Q}_{42} \\
\end{array}\right),\ee 
we see that expanding (\ref{lagrangian}) around the vev (\ref{vevchoice}) we find (among others) the following mass terms 
\be\label{masses}\mathcal{L}\supset (Q_{12}+ \widetilde{Q}_{41})\gamma+(Q_{22}- \widetilde{Q}_{42})c+ (\widetilde{Q}_{21}-Q_{14})b+(Q_{24}+\widetilde{Q}_{22})\epsilon+\dots~,\ee
where we have absorbed $\lambda_2$ in the coefficients $\gamma$, $c$, $b$, $\epsilon$, etc., in this expression.

If all the components in (\ref{USpadj}) were independent, or equivalently the blue gauge node were $U(4)$, all the $Q$ and $\widetilde{Q}$ components appearing in (\ref{masses}) would acquire a mass. In the case at hand instead only half of them receive a mass and can be integrated out. The other four form a hypermultiplet doublet under the unbroken $USp(2)$. They are also charged under the further $U(1)$ subgroup\footnote{More explicitly, this is a combination of the $U(1)$ gauge group, a $U(1)$ subgroup of the $U(2)$ gauge group, and the part of the $USp(4)$ gauge group corresponding to $d, \, \beta, \, \mu, \, -d$, left unbroken by the vev.} of $U(1)\times U(2)\times USp(4)$ left unbroken by the vev. Upon considering the full expression of \eref{masses}, we also find an extra massless hypermultiplet charged under $U(1)$, which becomes a flavour symmetry of \eref{mirrC2U1}.  Finally, from (\ref{vevchoice}) we see that the $F$-terms with respect to the $USp(4)$ gauge group provide the constraint 
\be \langle B_1B_1\rangle + \langle B_2B_2\rangle= \lambda_2 \left(\begin{array}{cccc}
0 & 0 & 1 & 0 \\
0 & 0 & 0 & 0 \\
1 & 0 & 0& 0 \\
0 & 0 & 0 & 0 \\
\end{array}\right),\ee and therefore $ \langle B_1B_1\rangle$ and $\langle B_2B_2\rangle$ (with the $SO$ indices contracted) can both be chosen nilpotent, resulting in the breaking pattern $SO(4)\rightarrow SO(3)$\footnote{Each $USp(4)-SO(4)$ leg of \eref{mirrC3C1} can be viewed as the $A_1$ theory of class $\mathsf{S}$ associated with a torus with two punctures, whose global symmetry is $USp(4)$. Upon giving the aforementioned nilpotent vev, a puncture is closed and so we are left with a torus with one puncture. Recall that this can be viewed as $USp(2)-SO(3)$, which is a leg of \eref{mirrC2U1}. Indeed, $SO(4)$ is broken to $SO(3)$ due to such a nilpotent vev.}. Overall, we end up with the 3d mirror \eref{mirrC2U1} of the $C_2U_1$ theory, in agreement with the four dimensional expectation.

\subsection{Mass deformation of the $\tilde{T}_3$ theory}

With similar techniques we can also study the mass deformation of the $\tilde{T}_3$ theory. As we will see, this theory can be deformed to $SO(4)$ $\mathcal{N}=4$ SYM. Starting from the 3d mirror
\be \label{mirrT3t}
\begin{tikzpicture}[baseline]
\tikzstyle{every node}=[font=\footnotesize]
\node[draw, circle, fill=lightblue] (node3) at (0,1) {$2$};
\node[] at (-3,1.2) {$\tilde{q},q$};
\node[] at (-1.7,0.3) {$\tilde{p},p$};
\node[draw, circle, fill=lightred] (node4) at (2,1) {$3$};
\node[draw, circle] (node1) at (-4,1) {$1$};
\node[draw, circle] (node2) at (-2,1) {$2$};
\node[draw, rectangle] (node11) at (-2,-0.5) {$1$};
\node[draw, circle, fill=lightred] (sqnode2) at (0,-0.5) {$3$};
\draw[draw=black,solid]  (node1) to (node2) to (node3) to (node4);
\draw[draw=black,solid]  (node3) to (sqnode2);
\draw[draw=black,solid]  (node2) to (node11);
\end{tikzpicture}
\ee  
we turn on a complex FI parameter $\lambda$ at the $U(1)$ node on the left. The equations of motion can be solved by giving vev to the $U(1)\times U(2)$ bifundamentals $q,\tilde{q}$ and $p,\tilde{p}$: 
\be \langle \tilde{q}\rangle= \langle \tilde{p}\rangle=\sqrt{\lambda}\left(\begin{array}{c}1 \\ 0 \\ \end{array}\right);\quad \langle q\rangle=\langle p\rangle=\sqrt{\lambda}(1,0).\ee 
This vev breaks the $U(1)\times U(2)$ gauge group to $U(1)$ and reduces the quiver to the following form: 
\vspace{-1.2cm}
\be \label{mirrT3tfinal}
\begin{tikzpicture}[baseline]
\tikzstyle{every node}=[font=\footnotesize]
\node[draw, circle, fill=lightblue] (node3) at (0,1) {$2$};
\node[draw, circle, fill=lightred] (node4) at (2,1) {$3$};
\node[draw, circle] (node2) at (-2,2) {$1$};
\node[draw, rectangle] (node11) at (-2,0) {$1$};
\node[draw, circle, fill=lightred] (sqnode2) at (0,-0.5) {$3$};
\draw[draw=black,solid]  (node2) to (node3) to (node4);
\draw[draw=black,solid]  (node3) to (sqnode2);
\draw[draw=black,solid]  (node3) to (node11);

\node[] (node5) at (3.2,1) {$=$}; 
\node[draw, circle, fill=lightgray] (L1) at (6,1) {\scriptsize$T_2$};

\path[every node/.style={font=\sffamily\small,
  		fill=white,inner sep=1pt}]
(L1) edge [loop, out=120, in=240, looseness=8] (L1);
\node[draw, circle, fill=lightred] (node6) at (4.8,1) {\begin{tiny} $SU(2)$\end{tiny}};
\node[draw, circle, fill=lightblue] (node7) at (7.5,1) {$2$};
\node[draw, circle] (node21) at (7.5,-0.5) {$1$};
\node[draw, circle, fill=lightgray] (L2) at (9,1) {\scriptsize$T_2$};

\path[every node/.style={font=\sffamily\small,
  		fill=white,inner sep=1pt}]
(L2) edge [loop, out=60, in=300, looseness=8] (L2);
\node[draw, circle, fill=lightred] (node6) at (10.2,1) {\begin{tiny} $SU(2)$\end{tiny}};
\draw[draw=black,solid]  (L2) to (node7) to (L1);
\draw[draw=black,solid]  (node21) to (node7);

\end{tikzpicture}
\ee   
where we have used the relation to the class $\mathsf{S}$ one-punctured torus. We claim that the quiver on the right is mirror dual to 
\be \label{finalmirr}
\left[
\begin{tikzpicture}[baseline]
\tikzstyle{every node}=[font=\footnotesize]
\node[draw, circle, fill=lightblue] (node7) at (2,0) {$2$};
\node[draw, circle, fill=lightblue] (L2) at (0,0) {$2$}; 
\path[every node/.style={font=\sffamily\small,
  		fill=white,inner sep=1pt}]
(L2) edge [loop, out=140, in=220, looseness=6] (L2);
\draw[draw=black,solid]  (L2) to (node7);
\path[every node/.style={font=\sffamily\small,
  		fill=white,inner sep=1pt}]
(node7) edge [loop, out=40, in=320, looseness=6] (node7);
\end{tikzpicture}
\right]/\BZ_2
\qquad  = \qquad 
\begin{tikzpicture}[baseline]
\node[draw, circle, fill=lightred] (L1) at (5,0) {$4$};
\path[every node/.style={font=\sffamily\small,
  		fill=white,inner sep=1pt}]
(L1) edge [loop, out=140, in=220, looseness=6] (L1);
\node[draw, rectangle] (node1) at (7,0) {$1$};
\draw[draw=black,solid]  (L1) to (node1);
\end{tikzpicture}
\ee  
where the quiver on the left represents an $(SU(2)\times SU(2))/\mathbb{Z}_2$ gauge theory with one adjoint hypermultiplet for each $SU(2)$ node and a bifundamental hypermultiplet, which is equivalent to the $SO(4)$ gauge theory with one adjoint hypermultiplet and a hypermultiplet in the vector representation on the right. 

In the following, we compute the Coulomb and Higgs branch Hilbert series of \eqref{mirrT3tfinal} and verify that they coincide with the Higgs and Coulomb branch Hilbert series of \eqref{finalmirr} respectively.  Since the Coulomb branch Hilbert series is sensitive to the $\BZ_2$ discrete action in \eqref{finalmirr}, this provide a highly non-trivial test to the claim.  Below we also sketch the argument for the mirror pair \eqref{mirrT3tfinal} and \eqref{finalmirr}.

From the $SO(4)$ gauge theory with one adjoint and one hypermultiplet in the vector representation of \eqref{finalmirr}, we can turn on a further mass deformation for the vector, flowing to 3d $\mathcal{N}=8$ $SO(4)$ SYM.  This leads us to conclude that the 4d $\CN=2$ $\tilde{T}_3$ theory can be mass-deformed to the 4d $\mathcal{N}=4$ $SO(4)$ SYM. This is a new prediction of our construction.

Let us make some comments regarding the above prediction.  First of all, a similar technique was used by the authors of \cite{Buican:2018ddk, Buican:2019kba} to study various supersymmetry enhancement RG flows from theories with 16 supercharges to those with 32 supercharges.  It was shown in \cite{Buican:2019kba} that this type of RG flows is generic in a certain class (known as type III in the nomenclature of \cite{Xie:2012hs}) of the 4d $\CN=2$ theories.  Secondly, as pointed out in \cite{Buican:2018ddk}, one could be more cautious about the above conclusion whether the 4d theory in question needs to be $\CN=4$ SYM\footnote{We thank Matthew Buican for pointing this out to us after the first version of this article has appeared on the arXiv.}.  To be accurate, there could be a more `exotic' 4d $\CN=4$ SCFT, which is not a SYM and is not known to date, that reduces on a circle to the same aforementioned 3d $N=8$ SYM. It would be nice to confirm the above prediction in the future.

\subsubsection*{Sketch of argument for the mirror pair \eqref{mirrT3tfinal} and \eqref{finalmirr}}
Let us first consider the right hand side of \eqref{mirrT3tfinal} without $(USp(2))-(U(1))$.  As is well-known from \cite{Benini:2010uu}, the mirror of the reduction of the $A_1$ theory of class $\mathsf{S}$ associated with a Riemann surface with genus $2$ without a puncture is a $U(2)/U(1) \cong SU(2)/\BZ_2$ gauge theory with $2$ adjoint hypermultiplets.  

Let us now add one flavour of the fundamental hypermultiplet to the blue $USp(2)$ node of the right hand side of \eqref{mirrT3tfinal} (\ie~ ungauging the $U(1)$ for now).  According to the prescription in \cite[Section 6]{Nanopoulos:2010bv}, at the level of the mirror theory, we should split the gauge node in the aforementioned theory into two gauge group, where each gauge group has an adjoint hypermultiplet charge under it and the two gauge groups are connected by a bifundamental hypermultiplet.  Now let us gauge the $U(1)$ symmetry in the original description, \ie~ theory \eqref{mirrT3tfinal}. This corresponds to ungauging a $U(1)$ symmetry from the aforementioned mirror theory.  It is expected that the gauge algebra for the end result is $su(2) \times su(2)$, with a bifundamental hyermultiplet and an adjoint hypermultiplet transforming under each factor.  This is depicted in the square brackets in the left hand side of \eqref{finalmirr}. The prescription of \cite{Nanopoulos:2010bv}, however, does not allow us to fix the global structure of the gauge group.  We determine the latter using the Coulomb branch Hilbert series (see below).  It turns out that there is a $\BZ_2$ action on the $SU(2) \times SU(2)$ gauge symmetry.  Since $SO(4) \cong (SU(2) \times SU(2))/\BZ_2$, the left diagram in \eqref{finalmirr} can be rewritten as the right one.

\subsubsection*{The Hilbert series of \eqref{mirrT3tfinal} and \eqref{finalmirr}}

The Coulomb branch Hilbert series of \eqref{mirrT3tfinal} can be computed from the quiver description on the left side similarly to what we did in subsection \ref{tildeT3sec} for the original $\tilde{T}_3$ theory
\bes{  \label{CBmirrT3tfinal}
&H_\CC [ \eref{mirrT3tfinal}_{\text{LHS}}] (t; x, y, z) =\\
&\sum_{m=-\infty}^\infty\sum_{a=0}^\infty t^{-2 |a-(-a)|+\frac{1}{2}(|a|+|-a|)+\frac{1}{2}(|m+a|+|m-a|)} x^{m} P_{U(1)} (t; m)P_{USp(2)} (t; a) \times \\
& \qquad   H_\CC[T_{[1^2]}(USp'(2))] (t; y ;  a) H_\CC[T_{[1^2]}(USp'(2))]  (t; z;  a) ~.
}
This should be compared with the Higgs branch Hilbert series of \eqref{finalmirr} which, using the description in terms of the quiver on the left side, can be written as
\bes{
H_\CH [ \eref{finalmirr}_{\text{LHS}}] (t; x,y,z) &=\oint_{|u|=1} \frac{du}{2\pi i u } (1-u^2)  \oint_{|v|=1} \frac{dv}{2\pi i v } (1-v^2) \times \\
& \quad H_\CH [[2]_u-[2]_{v}] (t; u, v, x) H_\CH [[3]_{v}] (t; v,y) ~H_\CH [[3]_{u}] (t; u,z) \times \\
& \quad\PE \left[ - (u^2+1+u^{-2}) t^2- (v^2+1+v^{-2}) t^2\right]~,
}
where 
\bes{ 
H_\CH [[2]_u-[2]_{v}] (t; u, v, x)&= \PE \left[ t (z+z^{-1})(v+v^{-1}) (x+x^{-1}) \right]   \\
H_\CH [[3]_{v}] (t; v,y)&= \PE \left[ t (v^2+1+v^{-2}) (y+y^{-1}) \right]   \\
H_\CH [[3]_{u}] (t; u,z)&= \PE \left[ t (u^2+1+u^{-2}) (z+z^{-1}) \right]    ~.
}
Evaluating these two Hilbert series, we find
\bes{ 
&\PL\left[H_\CC [ \eref{mirrT3tfinal}] (t; x^2, y, z)\right]=  \PL\left[H_\CH [ \eref{finalmirr}] (t; x,y,z)\right] \\
&= ([2]_x+[2]_y+[2]_z)t^2 +(2+[2]_x[1]_y[1]_z)t^4 \\
& \qquad - (2\times[2]_x[1]_y[1]_z+[1]_y[1]_z)t^6+ (3\times[2]_x[1]_y[1]_z+\\
&\qquad2\times[1]_y[1]_z-[4]_x-[2]_x[2]_y -[2]_y[2]_z-[2]_x[2]_z)t^8+\ldots ~.
}

The Higgs branch Hilbert series of \eqref{mirrT3tfinal} can be computed either from the quiver description on the left side using the standard Molien integral or from the description involving two class $\mathsf{S}$ theories on a one-punctured torus. The Higgs branch Hilbert series of the latter theory has been computed in \cite[(5.8)]{Hanany:2010qu} and we shall denote it by
\bes{
H_\CH^{g=1,n=1} (t; w)=(1-t^4)\PE\left[[1]_wt+[2]_wt^2\right]~,
}
where $g=1$ is the genus and $n=1$ is the number of punctures of the Riemann surface, while $w$ is the fugacity for the $SU(2)$ global symmetry of the puncture. The Higgs branch Hilbert series of \eqref{mirrT3tfinal} can then be written as
\bes{
H_\CH [ \eref{mirrT3tfinal}_{\text{RHS}}] (t) &=\oint_{|w|=1} \frac{dw}{2\pi i w } (1-w^2)  \oint_{|q|=1} \frac{dq}{2\pi i q } \times \\
& \quad H_\CH^{g=1,n=1} (t; w) ~H_\CH [[1]_q-[2]_{w}] (t; q,w) \times \\
& \quad \PE \left[ - (1+w^2+1+w^{-2}) t^2\right]~.
}
This should be compared with the Coulomb branch Hilbert series of \eqref{finalmirr} which can be computed using the monopole formula \cite{Cremonesi:2013lqa}.  For the quiver on the right hand side of \eqref{finalmirr}, we have
\bes{  
&H_\CC [ \eref{finalmirr}_{\text{RHS}}] (t) =\sum_{m_2=-\infty}^\infty\sum_{m_1=|m_2|}^\infty t^{2 |m_1|+2|m_2|} P_{SO(4)} (t; m_1,m_2) ~,
}
where
\bes{  
P_{SO(4)} (t; m_1,m_2)=  \begin{cases} (1-t^2)^{-2} &\quad \text{if $m_1 \neq m_2$} \\ (1-t^4)^{-2} & \quad \text{if $m_1=m_2$}~. \end{cases} 
}
For the quiver on the left hand side of \eqref{finalmirr}, we have 
\bes{
H_\CC [ \eref{finalmirr}_{\text{LHS}}] (t) =\sum_{a \, \in\,  \frac{1}{2} \BZ_{\geq 0}} ~ \sum_{b\,  \in \, \BZ_{\geq 0}}  t^{2|a+b|+2|a-b|} P_{USp(2)}(a) P_{USp(2)}~(b)~, 
}
where it should be noted that $a$ is summed over $\frac{1}{2} \BZ_{\geq 0}$, not  $\BZ_{\geq 0}$, because of the $\BZ_2$ action.  Evaluating these  series, we find
\bes{ 
\PL\left[H_\CH [ \eref{mirrT3tfinal}]  (t)\right]=  \PL\left[H_\CC [ \eref{finalmirr}_{\text{LHS} ,\text{RHS}} ] (t)\right]= t^2 +6t^4- 3t^6-6t^8+\ldots ~.
}

\section{Twisted $A_2$ theories with four punctures} \label{sec:twistedA2wfour}
In this section, we discuss the mirror theories associated with the $\CT_{2,\frac{3}{2},\frac{3}{2}}$ and $\CT_{3,2,\frac{3}{2},\frac{3}{2}}$ theories, described by twisted $A_2$ theories with four punctures.

\subsection{The $\CT_{2,\frac{3}{2},\frac{3}{2}}$ theory: a sphere with punctures $[2,1]$, $[2,1]$, $[2]_t$, $[2]_t$} \label{sec:T23h3h}
The 4d $\CN=2$ $\CT_{2,\frac{3}{2},\frac{3}{2}}$ SCFT was studied in \cite{Buican:2014hfa} as an $SU(2)$ gauge theory coupled to a doublet of hypermultiplets and two copies of the $(A_1, D_4)$ theory, where an $SU(2)$ subgroup of the $SU(3)$ global symmetry of each copy is gauged.  In that reference, it was proposed that this theory is dual to another 4d $\CN=2$ SCFT known as  the $I_{4,4}$ or $(A_3,A_3)$ theory \cite{Xie:2012hs}. Upon compactifying the latter on $S^1$, the 3d mirror theory was proposed in \cite[Figure 8]{Xie:2013jc} (see also \cite{Benvenuti:2017bpg} for a derivation) to be
\be
\begin{tikzpicture}[baseline]
\tikzstyle{every node}=[font=\footnotesize]
\node[draw, circle] (node1) at (-1,1) {$1$};
\node[draw, circle] (node2) at (1,1) {$1$};
\node[draw, circle] (node3) at (1,-1) {$1$};
\node[draw, circle] (node4) at (-1,-1) {$1$};
\draw[draw=black,solid]  (node1) to (node2) to (node3) to (node4) to (node1);
\draw[draw=black,solid]  (node2) to (node4);
\draw[draw=black,solid]  (node1) to (node3);
\end{tikzpicture}
\ee
where an overall $U(1)$ needs to be decoupled from this quiver.  Upon doing so, one obtains the following equivalent description of the above mirror theory \cite[(3.3)]{Buican:2014hfa}:
\be \label{mirrT33232a}
\begin{tikzpicture}[baseline]
\tikzstyle{every node}=[font=\footnotesize]
\node[draw, circle] (node1) at (-1,0) {$1$};
\node[draw, circle] (node2) at (1,0) {$1$};
\node[draw, circle] (node3) at (0,1.7) {$1$};
\node[draw, rectangle] (sqnode1) at (-2.2,0) {$1$};
\node[draw, rectangle] (sqnode2) at (2.2,0) {$1$};
\node[draw, rectangle] (sqnode3) at (0,2.7) {$1$};
\draw[draw=black,solid]  (node1) to (node2) to (node3) to (node1);
\draw[draw=black,solid]  (sqnode1) to (node1);
\draw[draw=black,solid]  (sqnode2) to (node2);
\draw[draw=black,solid]  (sqnode3) to (node3);
\end{tikzpicture}
\ee
This quiver has two interesting properties:
\ben
\item It is self-mirror.
\item Both Higgs and Coulomb branches are isomorphic to the moduli space of one $SU(3)$ instanton\footnote{Strictly speaking, this should be called a $PU(3) \cong PSU(3)$ instanton.} on $\BC^2/\BZ_3$ with the holonomy at infinity such that $SU(3)$ is broken to $U(1)^3/U(1) \cong U(1)^2$.
\een
The first property can be understood from the Type IIB Hanany-Witten brane construction \cite{Hanany:1996ie} involving one complete D3 brane wrapping a circle and stretching between three NS5 brane, with one D5 brane within each NS5 brane interval.  The mirror symmetry can be realised by an action that involves interchanging the NS5 and D5 branes, and this leaves the brane system invariant.  We thus conclude that \eref{mirrT33232a} is self-mirror.  The second property follows from \cite{kronheimer1990yang, Douglas:1996sw, Bianchi:1996zj, deBoer:1996mp, Porrati:1996xi, Fucito:2004ry, Cherkis:2008ip, Witten:2009xu, Cherkis:2009jm, Dey:2013fea, Nakajima:2015txa, Mekareeya:2015bla}.

On the other hand, the class $\mathsf{S}$ description, without an irregular puncture, of the $\CT_{2,\frac{3}{2},\frac{3}{2}}$ theory has recently been proposed in \cite[(5.3)]{Beem:2020pry}.  It is a twisted $A_2$ theory associated with the sphere with punctures
\bes{
[2,1]~, \quad [2,1]~, \quad [2]_t~, \quad [2]_t~.
}
Following the procedure described in the preceding section, we obtain the 3d mirror of this theory compactified on $S^1$ as
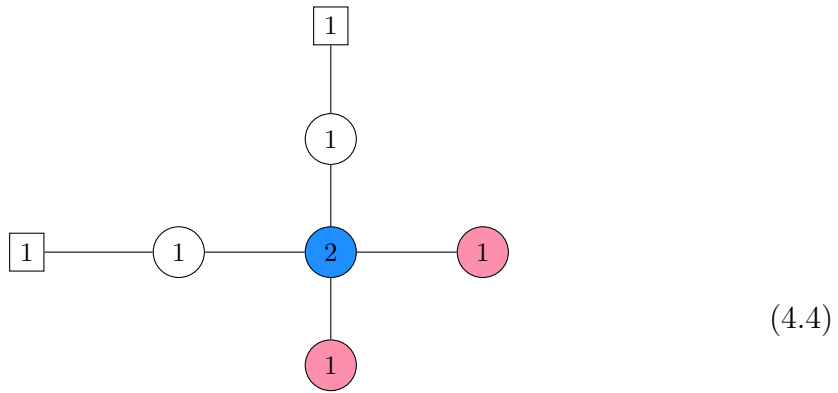
\be  \label{mirrST33232b}
\begin{tikzpicture}[baseline]
\tikzstyle{every node}=[font=\footnotesize]
\node[draw, circle] (node2) at (-2,1) {$1$};
\node[draw, circle, fill=lightblue] (node3) at (0,1) {$2$};
\node[draw, circle, fill=lightred] (node4) at (2,1) {$1$};
\node[draw, circle, fill=lightred] (node5) at (0,-0.5) {$1$};
\node[draw, rectangle] (sqnode1) at (-4,1) {$1$};
\node[draw, rectangle] (sqnode2) at (0,4) {$1$};
\node[draw, circle] (node6) at (0,2.5) {$1$};
\draw[draw=black,solid]  (node2) to (node3) to (node4);
\draw[draw=black,solid]  (node2) to (sqnode1);
\draw[draw=black,solid]  (node3) to (node5);
\draw[draw=black,solid]  (node3) to (node6);
\draw[draw=black,solid]  (node6) to (sqnode2);
\end{tikzpicture}
\ee
Since the red circular node with the label $1$ denotes $SO(1)$, this quiver can be rewritten as
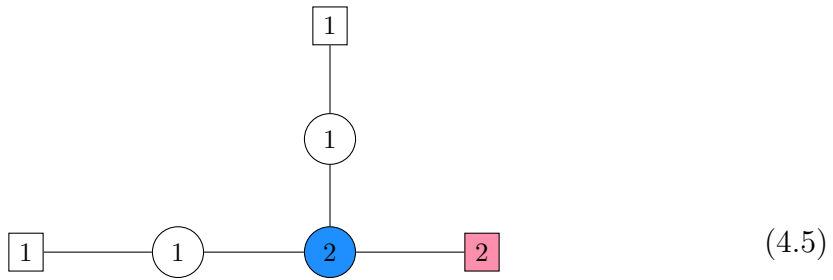
\be \label{mirrST33232a}
\begin{tikzpicture}[baseline]
\tikzstyle{every node}=[font=\footnotesize]
\node[draw, circle] (node2) at (-2,0) {$1$};
\node[draw, circle, fill=lightblue] (node3) at (0,0) {$2$};
\node[draw, rectangle, fill=lightred] (node4) at (2,0) {$2$};
\node[draw, rectangle] (sqnode1) at (-4,0) {$1$};
\node[draw, rectangle] (sqnode2) at (0,3) {$1$};
\node[draw, circle] (node6) at (0,1.5) {$1$};
\draw[draw=black,solid]  (node2) to (node3) to (node4);
\draw[draw=black,solid]  (node2) to (sqnode1);
\draw[draw=black,solid]  (node3) to (node6);
\draw[draw=black,solid]  (node6) to (sqnode2);
\end{tikzpicture}
\ee
Indeed, we conjecture that theory \eref{mirrT33232a} is {\it dual} to theory \eref{mirrST33232a}.  It is thus expected that the two properties discussed above also hold for theory \eref{mirrST33232a}.  In the following we provide some non-trivial checks for these statements.

The Coulomb branch of \eref{mirrST33232a} is $1+1+1=3$ quaternionic dimensional. The Higgs branch of \eref{mirrST33232a} is also $1+2+\frac{1}{2}(2 \times 2)+2+1-(1+3+1) = 3$ quatenionic dimensional. These are also equal to the corresponding quantities of \eref{mirrT33232a}.  The equality of the Higgs and Coulomb branch dimensions is as expected from the property that the theory is self-mirror.  We now study both branches in more detail using the Hilbert series

We first consider the Higgs branch Hilbert series of \eref{mirrST33232a}.  This is given by
\bes{
&H_\CH [ \eref{mirrST33232a}] (t; x,y,q) = \\
&\oint_{|u|=1} \frac{du}{2\pi i u }\oint_{|w|=1} \frac{dw}{2\pi i w }  \oint_{|v|=1} \frac{dv}{2\pi i v } (1-v^2) \times \\
& \qquad H_\CH [[1]_u-[1]_x] (t; u, x) \, H_\CH [[1]_u-[USp(2)]_v] (t; u, v) \times \\
& \qquad H_\CH [[1]_w-[1]_y] (t; u, x) \, H_\CH [[1]_w-[USp(2)]_v] (t; w, v) \times \\
& \qquad H_\CH [[USp(2)]_v-[SO(2)]_q] (t; v, q) \times \\
& \qquad \PE \left[ -2t^2- (v^2+1+v^{-2}) t^2\right]~,
}
where the notations are as in \eref{HCH1} and \eref{HCH2}.  Here, $x$, $y$, $q$ are the fugacities for each of the $U(1)$ in the $U(1)^3$ flavour symmetry of the theory.  Evaluating the integrals, this can be written as
\bes{ \label{HiggsmirrST33232b}
&H_\CH [ \eref{mirrST33232a}] (t; x,y,q)  \\
&=  \PE \Big[ 3t^2 + t^3 \left(q x+\frac{q}{x}+\frac{1}{q x}+\frac{x}{q}+q y+\frac{q}{y}+\frac{1}{q y}+\frac{y}{q}\right) \\
& \qquad +t^4 \left(q^2+\frac{1}{q^2}+x y+\frac{y}{x}+\frac{x}{y}+\frac{1}{x y}\right) \\
& \qquad -2t^6 \left(q^2+\frac{1}{q^2}+ x y+\frac{y}{x}+\frac{x}{y}+\frac{1}{x y}+2\right) \\ 
& \qquad - 3 t^7\left(q x+\frac{q}{x}+\frac{1}{q x}+\frac{x}{q}+q y+\frac{q}{y}+\frac{1}{q y}+\frac{y}{q}\right) + \ldots \Big]~.
}
The closed form for the unrefined Higgs branch Hilbert series, whereby $x=y=q=1$, is
\bes{
&H_\CH [ \eref{mirrST33232a}] (t; x=1,y=1,q=1) \\
& = \frac{1 - 2 t + 3 t^2 + 2 t^3 - 2 t^4 + 2 t^5 + 3 t^6 - 2 t^7 + t^8}{(1-t)^6 (1+t)^2 \left(1+t^2\right) \left(1+t+t^2\right)^2}~.
}
The order of the pole at $t=1$, which is $6$, is indeed the complex dimension of the Higgs branch, and the numerator is palindromic as it should be for a Calabi-Yau variety. Indeed it can be checked using the method described in \cite{Dey:2013fea, Mekareeya:2015bla} that this is indeed the Hilbert series of the moduli space of instanton mentioned below \eref{mirrT33232a}. 
 
We now focus on the Coulomb branch. Since theory \eref{mirrST33232a} is `good' in the sense of \cite{Gaiotto:2008ak}, the Coulomb branch Hilbert series can be computed using either the monopole formula or the Hall-Littlewood formula.  Here we present the latter:
\bes{ \label{HLmirrST33232b}
&H_\CC [ \eref{mirrST33232b}\, \text{or}\, \eref{mirrST33232a} ] (t; x_1, x_2; y_1, y_2) =\\
&\sum_{a=0}^\infty t^{-2 |a-(-a)|} P_{USp(2)} (t; a) \times \\
& \qquad H_\CC[T_{[2,1]} (SU(3))](t; x_1, x_2; a,0,-a) \,H_\CC[T_{[2,1]} (SU(3))](t; y_1, y_2; a,0,-a) \times \\
& \qquad   H_\CC[T_{[2]}(USp'(2))] (t;  a) \, H_\CC[T_{[2]}(USp'(2))]  (t;  a) ~.
}
where the fugacities $x_1, x_2$ and $y_1, y_2$ are subject to the conditions:
\bes{
x_1^2 x_2 = 1~, \qquad y_1^2 y_2 = 1~.
}
After imposing these conditions, we see only two $U(1)$ fugacities appear in formula \eref{HLmirrST33232b}. They are associated with the two $U(1)$ topological symmetries associated with each $U(1)$ gauge node in quiver \eref{mirrST33232a}.  From description \eref{mirrT33232a} and the Higgs branch computation we expect, however, that there should be three $U(1)$ global symmetries. The other $U(1)$ symmetry is indeed `hidden' in the above Coulomb branch computation, in a similar way as described in \cite{Kapustin:1998fa}. In order to match \eref{HiggsmirrST33232b} with \eref{HLmirrST33232b}, we need to unrefine one fugacity in the former:
\bes{
H_\CH [ \eref{mirrST33232a}] (t; x_1^3,y_1^3 ,q = 1) = H_\CC [ \eref{mirrST33232b}\, \text{or}\, \eref{mirrST33232a} ] (t; x_1, x_1^{-2}; y_1, y_1^{-2})~. 
}

It is also interesting to compare these results with the Higgs branch Hilbert series of the 4d $\CT_{2,\frac{3}{2},\frac{3}{2}}$ theory.  Recall that the Higgs branch of the $(A_1, D_4)$ theory is the closure of the minimal nilpotent orbit of $SU(3)$, whose Hilbert series is \cite{Benvenuti:2010pq}
\bes{
H[ \bar{\mathrm{min}}_{SU(3)}] (t; \vec u) =\sum_{p=0}^\infty \chi^{SU(3)}_{[p,p]} (\vec u)t^{2p}~,
}
where $\vec{u} = (u_1, u_2)$ are the fugacities of $SU(3)$ such that the character of the fundamental representation $[1,0]$ is written as $u_1 + u_2 u_1^{-1}+ u_2^{-1}$.  We now take two copies of the $(A_1, D_4)$ theory, gauge a common $SU(2)$ symmetry and then couple it to one flavour of the fundamental hypermultiplets.  For each copy, we need to decompose representations of $SU(3)$ into those of the $SU(2) \times U(1)$ subgroup.  This amounts to using the following fugacity map:
\bes{
u_1 =  x^{1/3} z ~, \qquad u_2 = x^{-1/3} z~,
}
where $z$ is the $SU(2)$ fugacity, $x$ is the $U(1)$ fugacity, and the power $1/3$ is the normalisation of the $U(1)$ charge such that we have the following decomposition: $[1,0] \rightarrow [1]_{\frac{1}{3}} + [0]_{\frac{2}{3}}$.
The Higgs branch Hilbert series of the $\CT_{2,\frac{3}{2},\frac{3}{2}}$ theory is then
\bes{
& H_\CH [\CT_{2,\frac{3}{2},\frac{3}{2}}] (t; x,y,q)  \\
&=\oint_{|z|=1} \frac{dz}{2 \pi i z} (1-z^2)  \,  \PE \left[ -t^2 \chi^{SU(2)}_{[2]} (z) \right] \, \PE\left[ (z+z^{-1} )(q+q^{-1}) t \right]  \times \\
& \qquad H[ \bar{\mathrm{min}}_{SU(3)}] (t; x^{1/3} z, x^{-1/3} z  ) H[ \bar{\mathrm{min}}_{SU(3)}] (t; y^{1/3} z, y^{-1/3} z )
}
where $z$ is the $SU(2)$ gauge fugacity, and each of $x$, $y$ and $q$ is the $U(1)$ fugacity.  Evaluating the integral, we find that
\bes{
H_\CH [\CT_{2,\frac{3}{2},\frac{3}{2}}] (t; x,y,q) = H_\CH [ \eref{mirrST33232a}] (t; x,y,q)~,
}
which is given by \eref{HiggsmirrST33232b}.

\subsection{The $\CT_{3,2,\frac{3}{2},\frac{3}{2}}$ theory: a sphere with punctures $[1^3]$, $[2,1]$, $[2]_t$, $[2]_t$}
The 4d $\CN=2$ $\CT_{3,2,\frac{3}{2},\frac{3}{2}}$ SCFT was studied in \cite{Buican:2014hfa} (see also \cite{Buican:2017fiq}).  It admits two known descriptions: (1) an $SU(3)$ gauge theory coupled to two  $(A_1,D_4)$ theories with three flavours of fundamental hypermultiplets, and (2) an $SU(2)$ gauge theory coupled to the $(A_1,D_4)$ theory and the $\CT_{3, \frac{3}{2}}$ theory\footnote{Equivalently, this is an $SU(2)$ gauge theory coupled to the $(A_1,D_4)$ theory and the $\CT^{(2)}_{SU(3)}$ theory (see section \ref{sec:ranktwoSU3inst}), with a half-hypermultiplet in the doublet of the $SU(2)$ gauge group.}, where the Higgs branch of the latter is the full moduli space of two $SU(3)$ instantons on $\BC^2$.  These two descriptions are related by the Argyres-Seiberg duality \cite{Argyres:2007cn}.  In \cite{Buican:2014hfa}, it was proposed that the $\CT_{3,2,\frac{3}{2},\frac{3}{2}}$ theory is dual to another 4d $\CN=2$ SCFT known as  the $III_{6,6}^{3 \times [2,2,1,1]}$ theory. Upon compactifying the latter on $S^1$, the 3d mirror theory can be obtain using the method described in \cite{Xie:2012hs} and the result was presented in \cite[(4.3)]{Buican:2014hfa}:
\be
\begin{tikzpicture}[baseline]
\tikzstyle{every node}=[font=\footnotesize]
\node[draw, circle] (node1) at (-1,1) {$1$};
\node[draw, circle] (node2) at (1,1) {$1$};
\node[draw, circle] (node3) at (1,-1) {$2$};
\node[draw, circle] (node4) at (-1,-1) {$2$};
\draw[draw=black,solid]  (node1) to (node2) to (node3) to (node4) to (node1);
\draw[draw=black,solid]  (node2) to (node4);
\draw[draw=black,solid]  (node1) to (node3);
\end{tikzpicture}
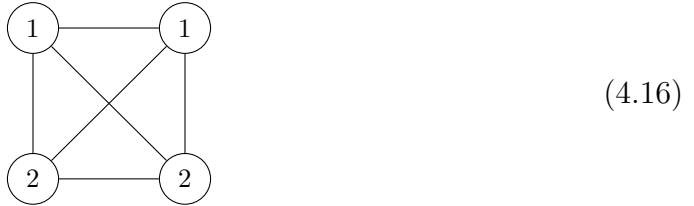
\ee
where an overall $U(1)$ needs to be decoupled from this quiver.  Doing so from one of the $U(1)$ gauge node, we obtain the following equivalent quiver:
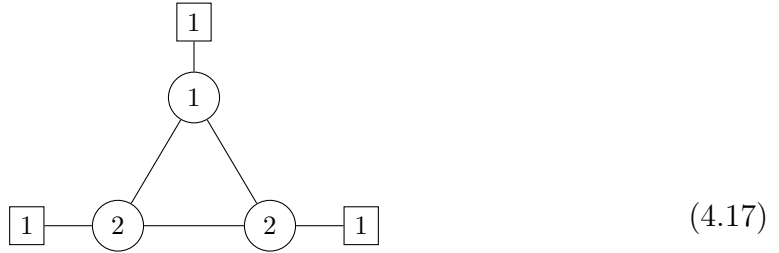
\be \label{mirrT323232a}
\begin{tikzpicture}[baseline]
\tikzstyle{every node}=[font=\footnotesize]
\node[draw, circle] (node1) at (-1,0) {$2$};
\node[draw, circle] (node2) at (1,0) {$2$};
\node[draw, circle] (node3) at (0,1.7) {$1$};
\node[draw, rectangle] (sqnode1) at (-2.2,0) {$1$};
\node[draw, rectangle] (sqnode2) at (2.2,0) {$1$};
\node[draw, rectangle] (sqnode3) at (0,2.7) {$1$};
\draw[draw=black,solid]  (node1) to (node2) to (node3) to (node1);
\draw[draw=black,solid]  (sqnode1) to (node1);
\draw[draw=black,solid]  (sqnode2) to (node2);
\draw[draw=black,solid]  (sqnode3) to (node3);
\end{tikzpicture}
\ee

The class $\mathsf{S}$ description of the $\CT_{3,2,\frac{3}{2},\frac{3}{2}}$ theory has recently been proposed in \cite[(5.1), (5.2)]{Beem:2020pry}.  It is a twisted $A_2$ theory associated with the sphere with punctures
\bes{
[1^3]~, \quad [2,1]~, \quad [2]_t~, \quad [2]_t~.
}
Following the procedure described in the preceding sections, we obtain the 3d mirror of this theory compactified on $S^1$ as
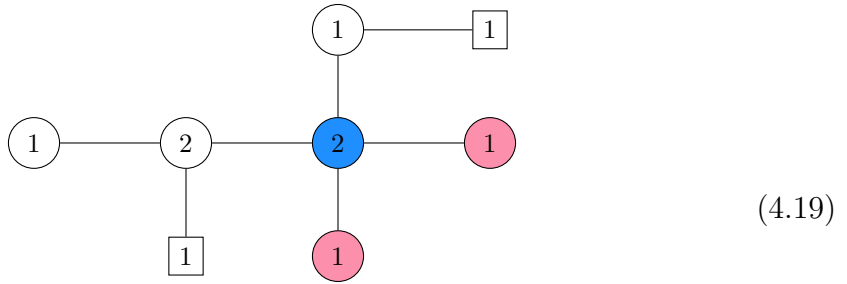
\be
\begin{tikzpicture}[baseline]
\tikzstyle{every node}=[font=\footnotesize]
\node[draw, circle] (node1) at (-4,1) {$1$};
\node[draw, circle] (node2) at (-2,1) {$2$};
\node[draw, circle, fill=lightblue] (node3) at (0,1) {$2$};
\node[draw, circle, fill=lightred] (node4) at (2,1) {$1$};
\node[draw, circle, fill=lightred] (node5) at (0,-0.5) {$1$};
\node[draw, rectangle] (sqnode1) at (-2,-0.5) {$1$};
\node[draw, rectangle] (sqnode2) at (2,2.5) {$1$};
\node[draw, circle] (node6) at (0,2.5) {$1$};
\draw[draw=black,solid]  (node1) to (node2) to (node3) to (node4);
\draw[draw=black,solid]  (node2) to (sqnode1);
\draw[draw=black,solid]  (node3) to (node5);
\draw[draw=black,solid]  (node3) to (node6);
\draw[draw=black,solid]  (node6) to (sqnode2);
\end{tikzpicture}
\ee
Since the red circular node denotes $SO(1)$, this quiver can be rewritten as
\be \label{mirrT323h3h}
\begin{tikzpicture}[baseline]
\tikzstyle{every node}=[font=\footnotesize]
\node[draw, circle] (node1) at (-4,0) {$1$};
\node[draw, circle] (node2) at (-2,0) {$2$};
\node[draw, circle, fill=lightblue] (node3) at (0,0) {$2$};
\node[draw, rectangle, fill=lightred] (node4) at (2,0) {$2$};
\node[draw, rectangle] (sqnode1) at (-2,1.5) {$1$};
\node[draw, rectangle] (sqnode2) at (2,1.5) {$1$};
\node[draw, circle] (node6) at (0,1.5) {$1$};
\draw[draw=black,solid]  (node1) to (node2) to (node3) to (node4);
\draw[draw=black,solid]  (node2) to (sqnode1);
\draw[draw=black,solid]  (node3) to (node6);
\draw[draw=black,solid]  (node6) to (sqnode2);
\end{tikzpicture}
\ee
We conjecture that theories \eref{mirrT323232a} and \eref{mirrT323h3h} are dual to each other.  In the following we provide number of non-trivial checks.

The Coulomb branch of \eref{mirrT323h3h} is $1+2+1+1=5$ quaternionic dimensional. The Higgs branch of \eref{mirrT323h3h} is also $2+2+4+\frac{1}{2}(2 \times 2)+2+1-(1+4+3+1) = 4$ quatenionic dimensional, in agreement with the fact that the 4d $\CT_{3,2,\frac{3}{2},\frac{3}{2}}$ theory is a rank-four theory. Note that these are also equal to the corresponding quantities of \eref{mirrT323232a}. We now study both branches in more detail using the Hilbert series

As in the previous subsection, the theory is `good' and so we can compute the Coulomb branch Hilbert series of \eref{mirrT323h3h} using either the monopole formula or the Hall-Littlewood formula.  The latter reads
\bes{ \label{HLmirrT323h3h}
&H_\CC [ \eref{mirrT323h3h}] (t; x_1, x_2,x_3; y_1, y_2) =\\
&\sum_{a=0}^\infty t^{-2 |a-(-a)|} P_{USp(2)} (t; a) \, H_\CC[T_{[1^3]} (SU(3))](t; x_1, x_2,x_3; a,0,-a)  \\
& \qquad \,H_\CC[T_{[2,1]} (SU(3))](t; y_1, y_2; a,0,-a) \times \\
& \qquad   H_\CC[T_{[2]}(USp'(2))] (t;  a) \, H_\CC[T_{[2]}(USp'(2))]  (t;  a) ~.
}
with the following conditions on the fugacities due to \eref{fugconstr}:
\bes{
x_1 x_2 x_3=1~, \qquad y_1^2 y_2 =1~.
}
Upon evaluating the summation, the result of \eref{HLmirrT323h3h} can be summarised as the highest weight generating function up to order $t^8$ as follows:
\bes{
&\mathrm{HWG} \left[ H_\CC [ \eref{mirrT323h3h}]\right] \\
&= \PE \Big[ \left(\mu _1 \mu _2+1\right) t^2+\left(b+\frac{1}{b}\right) t^3+\left(2 \mu _1 \mu _2+1\right) t^4 \\
& \qquad +\left(b+\frac{1}{b}\right) \mu _1 \mu _2 t^5+\left(2 \mu _1^3+\mu _2 \mu _1+2 \mu _2^3-1\right) t^6 \\
& \qquad +\left(b+\frac{1}{b}\right) \left(\mu _1^3+\mu _2^3\right) t^7 +\left(\mu _1^3-2 \mu _2 \mu _1+\mu _2^3\right) t^8 + \ldots \Big]~,
}
where, upon computing the power series of this expression in $t$, $\mu_1^{p_1} \mu_2^{p_2}$ denotes the representation $[p_1, p_2]$, whose character written in terms of $x_1, x_2, x_3$, of $SU(3)$. Here $b$ is the fugacity for the $U(1)$ symmetry which can be written in terms of $y_1, \, y_2$ as
\be
b = y_2 y_1^{-1}~.
\ee
As can be seen from the order $t^2$, the Coulomb branch symmetry of this theory is $U(3)$.  This is in agreement with that of theory \eref{mirrT323h3h} and the flavour symmetry of the 4d theory.

The Higgs branch Hilbert series is
\bes{
&H_\CH [ \eref{mirrT323h3h}] (t; x,y,b) = \\
&\oint_{|u|=1} \frac{du}{2\pi i u }\oint_{|w|=1} \frac{dw}{2\pi i w }  \oint_{|q|=1} \frac{dq}{2\pi i q } \times \\
& \qquad \oint_{|z|=1} \frac{dz}{2\pi i z } (1-z^2)  \oint_{|v|=1} \frac{dv}{2\pi i v } (1-v^2) \times \\
& \qquad H_\CH [[1]_u-[2]_{q,z}] (t; u, q, z) ~H_\CH [[1]_x-[2]_{q,z}] (t; x, q, z) \times \\
& \qquad H_\CH [[2]_{q,z}-[USp(2)]_v] (t; q, z, v) \, H_\CH [[1]_w-[USp(2)]_v] (t; w, v) \times \\
& \qquad H_\CH [[1]_w-[1]_y] (t; w, y)  \, H_\CH [[USp(2)]_v-[SO(2)]_b] (t; v, b) \times \\
& \qquad \PE \left[ -3t^2- (v^2+1+v^{-2}) t^2- (z^2+1+z^{-2}) t^2\right]~,
}
where the notations are as in \eref{HCH1} and \eref{HCH2}. Evaluating the integrals, we obtain
\bes{
&H_\CH [ \eref{mirrT323h3h}] (t; x,y,b) \\
&= \PE \Big[ 3 t^2 + t^3 \left(b x+\frac{b}{x}+\frac{1}{b x}+\frac{x}{b}+b y+\frac{b}{y}+\frac{1}{b y}+\frac{y}{b}\right)\\
& \qquad + t^4 \left(b^2+\frac{1}{b^2}+x y+\frac{y}{x}+\frac{x}{y}+\frac{1}{x y}+1\right) \\
& \qquad + t^5 \left(b x+\frac{b}{x}+\frac{1}{b x}+\frac{x}{b}\right) - \ldots \Big]~.
}
The order $t^2$ indicates that the Higgs branch symmetry is $U(1)^3$.  Setting $x=y=b=1$, we obtain the closed form of the unrefined Hilbert series as 
\bes{
&H_\CH [ \eref{mirrT323h3h}] (t; x=1,y=1,b=1) \\
&\frac{1}{(1 - t)^8 (1 + t)^4 (1 + t^2)^2 (1 - t + t^2) (1 + t + t^2)^3} \Big[ 1 - 2 t + 3 t^2 + 2 t^3 \\
& \qquad - 2 t^4 + 6 t^5 + 3 t^6 - 2 t^7 + 12 t^8 -  2 t^9 + \ldots \,\text{(palindrome)} \,\ldots + t^{16} \Big]~.
}
This Higgs branch Hilbert series is in agreement with that for \eref{mirrT323232a}. 

\section{Twisted $A_{2N}$ theories with $N\geq1$} \label{sec:twistedA2N}
In this section, we discuss the generalisation of our results for the twisted $A_2$ theories to the case of $A_{2N}$ with $N\geq 1$.

\subsection{The $D_2[SU(2N+1)]$ theory with $N$ free hypermultiplets}
The class $\mathsf{S}$ description (without an irregular puncture) was proposed in \cite[(6.3)]{Beem:2020pry}.  It is a twisted $A_{2N}$ theory associated with a sphere with punctures
\bes{
[N+1,N]~, \qquad [1^{2N}]_t~, \qquad [2N]_t~.
}
For $N=1$, this was discussed in section \ref{sec:A1D4wfree}, where the low energy theory is the $(A_1,D_4)$ SCFT with a free hypermultiplet..  For a general $N$, the 3d mirror theory of the reduction of the 4d theory in question on a circle admits the following quiver description:
\be \label{mirrD2wNfree}
\begin{tikzpicture}[baseline]
\tikzstyle{every node}=[font=\footnotesize]
\node[draw, circle] (node2) at (-2,1) {$N$};
\node[draw, circle, fill=lightblue] (node3) at (0,1) {$2N$};
\node[draw, circle, fill=lightred] (node4) at (2,1) {\tiny $2N+1$};
\node[draw, circle, fill=lightblue] (node5) at (4,1) {\tiny $2N-2$};
\node[draw=none] (node6) at (6,1) {\Large $\cdots$};
\node[draw, circle, fill=lightblue] (node7) at (8,1) {2};
\node[draw, circle, fill=lightred] (node8) at (10,1) {3};
\node[draw, rectangle] (sqnode1) at (-2,-0.5) {$1$};
\node[draw, circle, fill=lightred] (sqnode2) at (0,-0.5) {$1$};
\draw[draw=black,solid]  (node2) to (node3) to (node4) to (node5) to (node6) to (node7) to (node8);
\draw[draw=black,solid]  (node2) to (sqnode1);
\draw[draw=black,solid]  (node3) to (sqnode2);
\end{tikzpicture}
\ee
The Coulomb branch of \eref{mirrD2wNfree} is $N+\sum_{j=1}^N 2j = N^2+2N$ quaternionic dimension.  This is in agreement with the Higgs branch of the 4d theory: the Higgs branch of $D_2[SU(2N+1)]$ is $N(N+1)$ quaternionic dimensional (see Appendix \ref{sec:D2SUodd}), and the Higgs branch of the theory of $N$ free hypermultiplets is $N$ quaternionic dimensional; in total we have $N^2+2N$ quaternionic dimensions.  The Higgs branch of \eref{mirrD2wNfree} is
$N+ 2N^2+N + \dim_\BH \CH[T(USp'(2N))] -N^2-\frac{1}{2}(2N)(2N+1) = N$, where we have used the fact that $ \dim_\BH \CH[T(USp'(2N))] =N^2$.  This is in agreement with the fact that $D_2[SU(2N+1)]$ is a rank $N$ theory.

The Coulomb branch Hilbert series of \eref{mirrD2wNfree} can be computed using the Hall-Littlewood formula as follows:
\be
\scalebox{0.9}{$
\begin{split}
&H_\CC [ \eref{mirrD2wNfree}] (t; x_1, x_2, y_1, y_2, \ldots, y_N) =\\
&\sum_{n_1 \geq n_2 \geq n_N \geq 0} t^{-2 \left[ \sum_{j=1}^N |2n_j| + \sum_{1 \leq i < j \leq N} (|n_i-n_j|+|n_i+n_j|) \right]} P_{USp(2N)} (t; n_1,\ldots, n_N) \times \\
&~ H_\CC[T_{[N+1,N]} (SU(2N+1))](t; x_1, x_2; n_1, n_2, \ldots, n_N,0,-n_N, -n_{N-1}, \ldots, -n_1) \times \\
& ~  H_\CC[T_{[1^{2N}]}(USp'(2N))] (t; y_1, y_2,\ldots, y_N  ;  n_1, n_2, \ldots, n_N) \times \\
& ~ H_\CC[T_{[2N]}(USp'(2N))]  (t;  n_1, \ldots, n_N) ~,
\end{split}$}
\ee
where 
\be
x_1^{N+1} x_2^N =1~.
\ee
Evaluating the summations, we obtain
\bes{
&H_\CC [ \eref{mirrD2wNfree}] (t; x_1, x_2, y_1, y_2, \ldots, y_N) \\
&=  \PE \left[ \sum_{j=1}^N( y_j +y_j^{-1})\right] \times H[D_2[SU(2N+1)]] (x_1, x_2, y_1, \ldots, y_N) ~,
}
where the first factor is the Hilbert series of $\BH^N \cong \BC^{2N}$ and the second factor is as described in Appendix \ref{sec:D2SUodd}.  We have tested this expression for $N=1,2,3$.  This confirms that the moduli space of the Coulomb branch is a product of $\BH^N$ and that of the Higgs branch fo the $D_2[SU(2N+1)]$ theory, as expected from the 4d theory.

The Higgs branch Hilbert series, on the other hand, can be computed as follows:
\bes{
& H_\CH [ \eref{mirrD2wNfree}] (t; x) = \int d \mu_{U(N)} (\vec z)\, \int d \mu_{USp(2N)} (\vec v)  \times \\
& \quad H_\CH [[1]_x-[N]_{\vec z}]\, H_\CH [[N]_{z}-[USp(2N)]_{\vec v}] \times \\
& \quad H_\CC[T_{[1^{2N}]}(USp'(2N))] (t; \vec v  ;  0, 0, \ldots, 0)  \, H_{\CH}[ [SO(1)]-(USp(2N))_{\vec v}] \times \\
& \quad \PE \left[ -\left( \sum_{i,j=1}^N z_i z_j^{-1} \right) t^2 - \chi^{USp(2N)}_{[2,0,\ldots,0]} (\vec v) t^2\right]~,
}
where we have used the fact that $T_{[1^{2N}]}(USp'(2N))$ is self-mirror and so the Higgs branch Hilbert series of such a theory can be computed from the Coulomb branch one.  Here
\be
\scalebox{0.8}{$
\begin{split}
H_\CH [[1]_x-[N]_{\vec z}] &= \PE \left[ \left( x^{-1} \sum_{j=1}^N z_j + x \sum_{j=1}^N z^{-1}_j  \right)t \right] \\
H_\CH [[N]_z-[USp(2N)]_{\vec v}] &= \PE \left[\left( \sum_{j=1}^N z_j +  \sum_{j=1}^N z^{-1}_j  \right) \left( \sum_{k=1}^N v_k + \sum_{k=1}^N v^{-1}_k \right)t \right] \\
\chi^{USp(2N)}_{[2,0,\ldots,0]} (\vec v) &= N+\sum_{j=1}^N (v_j^2+ v_j^{-2}) + \sum_{1 \leq i < j \leq N} (v_i v_j + v^{-1}_i v^{-1}_j+ v_i v^{-1}_j+  v^{-1}_i v_j) \\
d \mu_{USp(2N)} (\vec v) &= \prod_{j=1}^N v_j^{-1} (1-v_j^2) \prod_{1 \leq i < j \leq N} (1-v_i v_j )(1- v_i v^{-1}_j ) \\
d \mu_{U(N)} (\vec z) &= \prod_{j=1}^N z_j^{-1} \prod_{1 \leq i < j \leq N} (1-z_i z^{-1}_j )~.
\end{split} $}
\ee
In the case of $N=2$, for example, we have
\bes{ \label{HiggsHmirrD2wNfree}
 &H_\CH [ \eref{mirrD2wNfree}_{N=2} ] (t; x) \\
 &= \PE\left[ t^2 + (x+x^{-1}) t^3 + t^4 + (x+x^{-1})  t^5 - t^8 -t^{10} +\ldots \right]~.
}
It can be checked that this is in agreement with the Higgs branch Hilbert series of \eref{mirrD2SU2Np1}, with $N=2$.  This is indeed the Coulomb branch Hilbert series of the $S^1$ reduction of the $D_2[SU(2N+1)]$ theory.

\subsection{Two copies of the $D_2[SU(2N+1)]$ theory}
The class $\mathsf{S}$ description (without an irregular puncture) was proposed in \cite[(6.4)]{Beem:2020pry}.  It is a twisted $A_{2N}$ theory associated with a sphere with punctures
\bes{
[1^{2N+1}]~, \qquad [2N]_t~, \qquad [2N]_t~.
}
The 3d mirror of the reduction of this 4d theory on $S^1$ can be described by the following quiver:
\be \label{mirrtwocopiesD2}
\begin{tikzpicture}[baseline]
\tikzstyle{every node}=[font=\footnotesize]
\node[draw, circle] (nodem2) at (-10,1) {1};
\node[draw, circle] (nodem1) at (-8,1) {2};
\node[draw=none] (node0) at (-6,1) {\Large $\cdots$};
\node[draw, circle] (node1) at (-4,1) {\tiny $2N-1$};
\node[draw, circle] (node2) at (-2,1) {$2N$};
\node[draw, circle, fill=lightblue] (node3) at (0,1) {$2N$};
\node[draw, circle, fill=lightred] (node4) at (2,1) {$1$};
\node[draw, rectangle] (sqnode1) at (-2,-0.5) {$1$};
\node[draw, circle, fill=lightred] (sqnode2) at (0,-0.5) {$1$};
\draw[draw=black,solid]  (nodem2) to (nodem1) to (node0) to (node1) to (node2) to (node3) to (node4);
\draw[draw=black,solid]  (node2) to (sqnode1);
\draw[draw=black,solid]  (node3) to (sqnode2);
\end{tikzpicture}
\ee
Since the red circular node is $SO(1)$, this quiver can be rewritten as 
\be \label{mirrtwocopiesD2a}
\begin{tikzpicture}[baseline]
\tikzstyle{every node}=[font=\footnotesize]
\node[draw, circle] (nodem2) at (-10,1) {1};
\node[draw, circle] (nodem1) at (-8,1) {2};
\node[draw=none] (node0) at (-6,1) {\Large $\cdots$};
\node[draw, circle] (node1) at (-4,1) {\tiny $2N-1$};
\node[draw, circle] (node2) at (-2,1) {$2N$};
\node[draw, circle, fill=lightblue] (node3) at (0,1) {$2N$};
\node[draw, rectangle, fill=lightred] (node4) at (2,1) {$2$};
\node[draw, rectangle] (sqnode1) at (-2,-0.5) {$1$};
\draw[draw=black,solid]  (nodem2) to (nodem1) to (node0) to (node1) to (node2) to (node3) to (node4);
\draw[draw=black,solid]  (node2) to (sqnode1);
\end{tikzpicture}
\ee
For $N=1$, we recover quiver \eref{mirrtwocopiesA1D4a}.  The Coulomb branch of \eref{mirrtwocopiesD2a} is $\sum_{j=1}^{2N} j + N = 2N(N+1)$, in agreement with the Higgs branch dimension of a product of two $D_2[SU(2N+1)]$.  The Higgs branch of \eref{mirrtwocopiesD2a} is $\sum_{j=1}^{2N-1} j(j+1) +2N+ 4N^2+2N-\sum_{j=1}^{2N} j^2-N(2N+1) = 2N$, in agreement with the fact that $D_2[SU(2N+1)]$ is a rank $N$ theory.

Similarly to the case of quiver \eref{mirrtwocopiesA1D4a}, we can see the enhanced Coulomb branch symmetry of quiver \eref{mirrtwocopiesD2a} using the observation of \cite{Gaiotto:2008ak}.  Since all of the $U(s)$ gauge nodes, with $s=1, \ldots, 2N$, in \eref{mirrtwocopiesD2a} are all balanced, one expects an $SU(2N+1)$ enhanced symmetry in the IR.  Moreover, since the $USp(2N)$ gauge node is also balanced, according to \cite[section 5.3]{Gaiotto:2008ak}, this $SU(2N+1)$ symmetry gets doubled and so the symmetry of the Coulomb branch is expected to be $SU(2N+1) \times SU(2N+1)$.  This is in agreement with the Higgs branch symmetry of a product of two $D_2[SU(2N+1)]$.  Shortly we confirm this using the Coulomb branch Hilbert series.

The Coulomb branch Hilbert series of \eref{mirrtwocopiesD2a} can be computed using either the monopole formula or the the Hall-Littlewood formula.  The latter reads
\be
\scalebox{0.9}{$
\begin{split}
&H_\CC [ \eref{mirrtwocopiesD2a}] (t; x_1, \ldots, x_{2N+1}) =\\
&\sum_{n_1 \geq n_2 \geq n_N \geq 0} t^{-2 \left[ \sum_{j=1}^N |2n_j| + \sum_{1 \leq i < j \leq N} (|n_i-n_j|+|n_i+n_j|) \right]} P_{USp(2N)} (t; n_1,\ldots, n_N) \times \\
&~ H_\CC[T_{[1^{2N+1}]} (SU(2N+1))](t; x_1, \ldots, x_{2N+1}; n_1, n_2, \ldots, n_N,0,-n_N, -n_{N-1}, \ldots, -n_1) \times \\
& ~  H_\CC[T_{[2N]}(USp'(2N))]  (t;  n_1, \ldots, n_N)  \, H_\CC[T_{[2N]}(USp'(2N))]  (t;  n_1, \ldots, n_N) ~,
\end{split}$}
\ee
where 
\be
x_1 x_2 \cdots x_{2N+1}=1~.
\ee
Evaluating the summations, we obtain
\bes{
&H_\CC [ \eref{mirrtwocopiesD2a}] (t; x_1, \ldots, x_{2N+1}) =  \left[ H_\CH [D_2[SU(2N+1)]] (x_1, \ldots, x_{2N+1}) \right]^2~,
}
where the Higgs branch Hilbert series $H_\CH [D_2[SU(2N+1)]]$ of $D_2[SU(2N+1)]$ is given in Appendix \ref{sec:D2SUodd}.  We have tested this expression for $N=1,2,3$.   Similarly to the remark below \eref{twocopiesminorbSU3}, the full Coulomb branch symmetry is expected to be $SU(2N+1)^2$; however, it is possible to see only the diagonal subgroup $SU(2N+1)$, corresponding to the fugacities $x_1, \ldots, x_{2N+1}$ in the Hilbert series.

The Higgs branch Hilbert series, on the other hand, can be computed as follows:
\bes{
& H_\CH [ \eref{mirrtwocopiesD2a}] (t; x, y) = \int d \mu_{U(2N)} (\vec z)\, \int d \mu_{USp(2N)} (\vec v)  \times \\
&\quad H_\CH[T_{[1^{2N}]}(SU(2N))] (t; \vec z) \times \\
& \quad H_\CH [[1]_x-[2N]_{\vec z}]\, H_\CH [[2N]_{\vec z}-[USp(2N)]_{\vec v}] \times \\
& \quad  H_{\CH}[ (USp(2N))_{\vec v}-[SO(2)]_y] \times \\
& \quad \PE \left[ -\left( \sum_{i,j=1}^{2N} z_i z_j^{-1} \right) t^2 - \chi^{USp(2N)}_{[2,0,\ldots,0]} (\vec v) t^2\right]~,
}
where the Higgs branch Hilbert series of $T_{[1^{2N}]}(SU(2N))$ is given by \cite[(3.4)]{Hanany:2011db}
\be
H_\CH[T_{[1^{2N}]}(SU(2N))] (t; \vec z) = \PE\left[ t^2  \sum_{i,j=1}^{2N} z_i z_j^{-1} \right] \prod_{p=1}^{2N}  (1-t^{2p})~.
\ee
For example, in the case of $N=2$, we obtain
\bes{
H_\CH [ \eref{mirrtwocopiesD2a}_{N=2}] (t; x, y)  = H_\CH [ \eref{mirrD2wNfree}_{N=2} ] (t; x y ) H_\CH [ \eref{mirrD2wNfree}_{N=2} ] (t; x y^{-1} )~,
}
where $H_\CH [ \eref{mirrD2wNfree}_{N=2} ] (t; x )$, which is the Coulomb branch Hilbert series of the $S^1$ reduction of the $D_2[SU(2N+1)]$ theory, is given by \eref{HiggsHmirrD2wNfree}.

\subsection{A sphere with punctures $[N+1,N]$, $[N+1,N]$, $[2N]_t$, $[2N]_t$}
We study a $USp(2N)$ gauge theory coupled to one flavour of the fundamental hypermultiplets and two copies of the $D_2[SU(2N+1)]$ theory, where a $USp(2N)$ subgroup of the $SU(2N+1)$ global symmetry of each copy is gauged.   The class $\mathsf{S}$ description (without an irregular puncture) was proposed in \cite[(6.5)]{Beem:2020pry}.  It is a twisted $A_{2N}$ theory associated with a sphere with punctures
\bes{
[N+1,N]~, \qquad [N+1,N]~, \qquad [2N]_t~, \qquad [2N]_t~.
}
The case of $N=1$ was studied in section \ref{sec:T23h3h}.  The 3d mirror of the reduction of this 4d theory on $S^1$ can be described by the following quiver:
\be
\begin{tikzpicture}[baseline]
\tikzstyle{every node}=[font=\footnotesize]
\node[draw, circle] (node2) at (-2,1) {$N$};
\node[draw, circle, fill=lightblue] (node3) at (0,1) {$2N$};
\node[draw, circle, fill=lightred] (node4) at (2,1) {$1$};
\node[draw, circle, fill=lightred] (node5) at (0,-0.5) {$1$};
\node[draw, rectangle] (sqnode1) at (-4,1) {$1$};
\node[draw, rectangle] (sqnode2) at (0,4) {$1$};
\node[draw, circle] (node6) at (0,2.5) {$N$};
\draw[draw=black,solid]  (node2) to (node3) to (node4);
\draw[draw=black,solid]  (node2) to (sqnode1);
\draw[draw=black,solid]  (node3) to (node5);
\draw[draw=black,solid]  (node3) to (node6);
\draw[draw=black,solid]  (node6) to (sqnode2);
\end{tikzpicture}
\ee
Since the red circular node denotes $SO(1)$, this quiver can be rewritten as
\be \label{mirrD2D2USp}
\begin{tikzpicture}[baseline]
\tikzstyle{every node}=[font=\footnotesize]
\node[draw, circle] (node2) at (-2,0) {$N$};
\node[draw, circle, fill=lightblue] (node3) at (0,0) {$2N$};
\node[draw, rectangle, fill=lightred] (node4) at (2,0) {$2$};
\node[draw, rectangle] (sqnode1) at (-2,1.5) {$1$};
\node[draw, rectangle] (sqnode2) at (2,1.5) {$1$};
\node[draw, circle] (node6) at (0,1.5) {$N$};
\draw[draw=black,solid]  (node2) to (node3) to (node4);
\draw[draw=black,solid]  (node2) to (sqnode1);
\draw[draw=black,solid]  (node3) to (node6);
\draw[draw=black,solid]  (node6) to (sqnode2);
\end{tikzpicture}
\ee
The quaternionic dimension of the Higgs branch dimension of the 4d theory is $2N(N+1)+2N -\frac{1}{2}(2N)(2N+1)= 3N$, where $N(N+1)$ is the Higgs branch dimension of the $D_2[SU(2N+1)]$ theory.  The Coulomb branch of quiver \eref{mirrD2D2USp} is $N+N+N=3N$ quaternionic dimensional, in agreement with that of the Higgs branch of the 4d theory.  The Higgs branch of quiver \eref{mirrD2D2USp} is $N+2N^2+2N^2+N+2N-\left[ N^2+N^2+\frac{1}{2}(2N)(2N+1) \right]=3N$.  This is in agreement with the fact that each copy of the $D_2[SU(2N+1)]$ theory is of rank $N$ and the $USp(2N)$ gauge group has rank $N$, and so in total we have $3N$ dimensional Coulomb branch as expect.  Observe that the Coulomb and Higgs branches of \eref{mirrD2D2USp} have the same dimension.  Indeed, as we shall discuss below, theory \eref{mirrT323h3h} is {\bf self-mirror} for any $N$, where the case of $N=1$ was indeed self-mirror as shown in Section \ref{sec:T23h3h}.  


Let us first examine the Higgs branch. The Higgs branch Hilbert series of \eref{mirrD2D2USp} is given by
\bes{ \label{HiggsmirrD2D2USp}
&H_\CH [ \eref{mirrD2D2USp}] (t; x,y,q) = \\
&\int d\mu_{U(N)} (\vec u) \int d\mu_{U(N)} (\vec w)  \int d\mu_{USp(2N)} (\vec v) \times \\
& \qquad H_\CH [[N]_{\vec u}-[1]_x] (t; u, x) \, H_\CH [[N]_{\vec u}-[USp(2)]_v] (t; u, v) \times \\
& \qquad H_\CH [[N]_{\vec w}-[1]_y] (t; u, x) \, H_\CH [[N]_{\vec w}-[USp(2)]_v] (t; w, v) \times \\
& \qquad H_\CH [[USp(2)]_v-[SO(2)]_q] (t; v, q) \times \\
& \qquad \PE \left[ -\left( \sum_{i,j=1}^{N} u_i u_j^{-1} \right) t^2 -\left( \sum_{i,j=1}^{N} w_i w_j^{-1} \right) t^2 - \chi^{USp(2N)}_{[2,0,\ldots,0]} (\vec v) t^2\right]~,
}
Let us compute the integrals in the case of $N=2$, we obtain
\bes{
& H_\CH [ \eref{mirrD2D2USp}_{N=2} ] (t; x,y,q) =\\
&\PE \Big[ 3t^2 + t^3 \left(q x+\frac{q}{x}+\frac{1}{q x}+\frac{x}{q}+q y+\frac{q}{y}+\frac{1}{q y}+\frac{y}{q}\right) \\
& \qquad +t^4 \left(5+q^2+\frac{1}{q^2}+x y+\frac{y}{x}+\frac{x}{y}+\frac{1}{x y}\right) \\
& \qquad +2t^5 \left(q x+\frac{q}{x}+\frac{1}{q x}+\frac{x}{q}+q y+\frac{q}{y}+\frac{1}{q y}+\frac{ y}{q}\right) \\
& \qquad +t^6 \left(3+q^2+\frac{1}{q^2}+x y+\frac{y}{x}+\frac{x}{y}+\frac{1}{x y}\right) + \ldots \Big]~.
}
On the other hand, the Higgs branch Hilbert series of the 4d theory is given by
\bes{
& H_\CH [ \text{4d theory}](t; x,y, q) = \\ 
 &\int d\mu_{USp(2N)} (\vec z)  \,  \PE \left[ -t^2 \chi^{USp(2N)}_{[2,0,\ldots,0]} (\vec z) \right] \, \PE\left[\chi^{USp(2N)}_{[1,0,\ldots,0]} (\vec z)(q+q^{-1}) t \right]  \times \\
& ~H_\CH[ D_2[SU(2N+1)]] (t; x^{\frac{1}{3}} (z_1, \ldots, z_N) , x^{\frac{1}{3}} (z^{-1}_{1}, \ldots, z^{-1}_{N}), x^{\frac{2N}{3}}  )\times \\
& ~H_\CH[D_2[SU(2N+1)]]  (t; y^{\frac{1}{3}} (z_1, \ldots, z_N) , y^{\frac{1}{3}} (z^{-1}_{1}, \ldots, z^{-1}_{N}), y^{\frac{2N}{3}}  )~,
}
where each of $x, y, q$ is a $U(1)$ fugacity and $\vec z =(z_1, \ldots, z_{N})$ are the $USp(2N)$ gauge fugacity.  The expression for $H_\CH[ D_2[SU(2N+1)]]$ is given in Appendix \ref{sec:D2SUodd}. Here, under the decomposition $SU(2N+1) \supset SU(2N) \times U(1) \supset USp(2N) \times U(1)$, we have
\bes{
SU(2N+1) ~&\rightarrow~ USp(2N) \times U(1) \\
[1,0,\ldots,0]~ &\rightarrow~ [1,0,\ldots,0]_{\frac{1}{3}}+ [0,\ldots,0]_{\frac{2N}{3}}~.
}
If we write the character of the fundamental representation $[1,0,\ldots,0]$ of $SU(2N+1)$ as $\sum_{j=1}^{2N+1} u_{j}$ (with $\prod_{j=1}^{2N+1} u_j =1$) and that of the fundamental representation $[1,0,\ldots, 0]$ of $USp(2N)$ as $\sum_{j=1}^N (v_j +v^{-1}_j)$, then a fugacity map is
\be
u_k = \begin{cases} q^{\frac{1}{3}} v_k~,  &k = 1,2,\ldots, N \\
q^{\frac{1}{3}}v_{k-N}^{-1}~, & k=N+1, N+2\ldots, 2N \\
q^{\frac{2N}{3}}~, & k=2N+1~,
\end{cases}
\ee
where $q$ is the fugacity for the $U(1)$ symmetry.  Since theory \eref{mirrD2D2USp} is self-mirror, its Higgs branch Hilbert series can be equated to that of the 4d theory as follows:
\bes{
H_\CH [ \text{4d theory}](t; x,y, q) = H_\CH [ \eref{mirrD2D2USp} ] (t; x,y,q)~.
}

The Coulomb branch Hilbert series of \eref{mirrD2D2USp} can be computed using either the monopole formula or the the Hall-Littlewood formula.  The latter reads
\be \label{HLmirrD2D2USp}
\scalebox{0.9}{$
\begin{split}
&H_\CC [ \eref{mirrD2D2USp}] (t; x_1, x_2, y_1, y_2) =\\
&\sum_{n_1 \geq n_2 \geq n_N \geq 0} t^{-2 \left[ \sum_{j=1}^N |2n_j| + \sum_{1 \leq i < j \leq N} (|n_i-n_j|+|n_i+n_j|) \right]} P_{USp(2N)} (t; n_1,\ldots, n_N) \times \\
&~ H_\CC[T_{[N+1,N]} (SU(2N+1))](t; x_1, x_2; n_1, n_2, \ldots, n_N,0,-n_N, -n_{N-1}, \ldots, -n_1) \times \\
&~ H_\CC[T_{[N+1,N]} (SU(2N+1))](t; y_1, y_2; n_1, n_2, \ldots, n_N,0,-n_N, -n_{N-1}, \ldots, -n_1) \times \\
& ~  H_\CC[T_{[2N]}(USp'(2N))]  (t;  n_1, \ldots, n_N)  \, H_\CC[T_{[2N]}(USp'(2N))]  (t;  n_1, \ldots, n_N) ~,
\end{split}$}
\ee
where 
\be
x_1^{N+1} x_2^N = y_1^{N+1} y_2^N =1 ~.
\ee
Taking into account these constraints on the fugacities, we see that there are only two $U(1)$ fugacities that are manifest in the Hilbert series \eref{HLmirrD2D2USp}, whereas the full Coulomb branch symmetry is $U(1)^3$.  This phenomenon is similar to what we have encountered in Section \ref{sec:T23h3h}.  In order to match \eref{HiggsmirrD2D2USp} with \eref{HLmirrD2D2USp}, we need to unrefine one fugacity in the former:
\bes{
&H_\CC [ \eref{HLmirrD2D2USp}] (t; x^N, x^{-N-1}, y^N, y^{-N-1}) \\
&=H_\CH [ \eref{mirrD2D2USp}] (t; x^{2N+1},y^{2N+1} ,q = 1) \\
&=H_\CH [ \text{4d theory}] (t; x^{2N+1},y^{2N+1} ,q = 1)~,
}
where we have checked this relation for $N=1, 2$.

\section{Conclusion} \label{sec:conclusion}

In this paper we propose a description of the 3d mirror theories of the $S^1$ reduction of twisted $A_{2N}$ theories of class $\mathsf{S}$.  This is the `almost' star-shaped quiver with the central gauge node being $USp(2N)$ and with one flavour of the fundamental hypermultiplet under the unitary group located next to the central $USp(2N)$ node.  In general, the quiver contains unitary, symplectic and special orthogonal gauge groups.  We use the 4d theories recently proposed by Beem and Peelaers \cite{Beem:2020pry} as a testing ground of our proposal.  Using the Hilbert series, we show that the Higgs branch of the 4d SCFT matches with the Coulomb branch of the corresponding 3d mirror theory. Furthermore, we study the Higgs branch of such mirror theories; its quaternionic dimension matches with the rank of the 4d SCFT. In many cases, there are more than one description of the mirror theory, where one is constructed using the proposal of this paper and the other involves only unitary gauge groups.  We conjecture that these theories are dual to each other and this can be tested by matching the Hilbert series.  One of the important features of these dualities is that in many cases not all Coulomb branch symmetries of the unitary-orthosymplectic quiver is manifest in the quiver; in other word, one cannot turn on in the Hilbert series all of the fugacities associated with the full symmetry of the SCFT in the IR.  The full Coulomb branch symmetry can be seen as an enhanced topological symmetry in the quiver containing only unitary gauge groups.  Moreover, we study deformations of proposed mirror theories by mass and FI terms.  Several of them flow to expected theories in the IR, providing a highly non-trivial check of our proposal.  Making use of the mirror quiver description, we find a new supersymmetry enhancement RG flow from the $\tilde{T}_3$ theory to the $SO(4)$ super-Yang-Mills.

Let us discuss some of the open questions that arise from the findings of this paper.  First of all, some of the quiver descriptions that we proposed are `bad' theories in the sense of \cite{Gaiotto:2008ak}.  Even though we manage to use such a description to compute various quantities, such as the Coulomb branch dimension and the Higgs branch Hilbert series, and to study various deformations, it would be nice to come up with a `good' description for such theories.   Secondly, it would be nice to understand better the dualities between different descriptions of the mirror theory of the $S^1$ reduction of the same 4d SCFT, such that as how to `derive' one description from the others.  Finally, as pointed out in \cite[(3.8)]{Ohmori:2018ona}, the $R_{2, 2N}$ SCFT can be obtained from a compactification of the 6d $(D_{2N+3}, D_{2N+3})$ minimal conformal matter theory \cite{DelZotto:2014hpa} on $T^2$ with non-trivial Stiefel-Whitney classes.  One description of the mirror theory associated with $R_{2, 2N}$ is given by \eref{mirrR22N}.  However, there is another description of such a theory in terms of a quiver with a non-simply-laced edge  \cite{Bourget:2020asf}.  It would be interesting to find out whether the other mirror theories discussed in this paper also admit non-simply-laced quiver descriptions.

\acknowledgments
N.M. thanks Stefano Cremonesi, Amihay Hanany and Alberto Zaffaroni for the collaboration on the previous work, especially on the $T_{\vec \rho} (USp'(2N))$ theory, which constitutes the main results of this paper. The work of S.G. is supported by the ERC Consolidator Grant 682608 “Higgs bundles: Supersymmetric Gauge Theories and Geometry (HIGGSBNDL).”  M.S. is partially supported by the ERC-STG grant 637844-HBQFTNCER, by the University of Milano-Bicocca grant 2016-ATESP0586, by the MIUR-PRIN contract 2017CC72MK003, and by the INFN. 

\appendix
\section{A brief review of $T^{\vec \sigma}_{\vec \rho}(SU(N))$ and $T^{\vec \sigma}_{\vec \rho}(USp'(2N))$} \label{app:Ttheories}
In this appendix we briefly review the crucial results on the $T^{\vec \sigma}_{\vec \rho}(SU(N))$ and $T^{\vec \sigma}_{\vec \rho}(USp'(2N))$ theories that are important for the discussion in the main text.
\subsection{$T^{\vec \sigma}_{\vec \rho}(SU(N))$} \label{sec:TsigrhoSUN}
Let $\vec \rho= (\rho_1, \ldots, \rho_{\ell'})$ and $\vec \sigma= (\sigma_1, \ldots, \sigma_\ell)$ be two 
partitions of $N$: 
\bea\label{partitionsSU}
 \sigma_1 \geq \ldots \geq  \sigma_\ell > 0~, \qquad  \rho_1 \geq \ldots \geq  \rho_{\ell'} > 0~, \qquad \sum_{i=1}^\ell \sigma_i = \sum_{i=1}^{\ell'} \rho_i =N~.  
\eea
The quiver diagram for $T^{\vec \sigma}_{\vec \rho}(SU(N))$ is depicted in \eref{quivTsigrhoSUN} \cite{Gaiotto:2008ak}, where each circular node denotes a gauge group and each rectangular node denotes a flavour group. The label $m$ at each node denotes a $U(m)$ group and $\ell'$ is the length of the partition $\vec \rho$.  This type of quivers has first appeared in \cite{nakajimaquiver}. 
\bea 
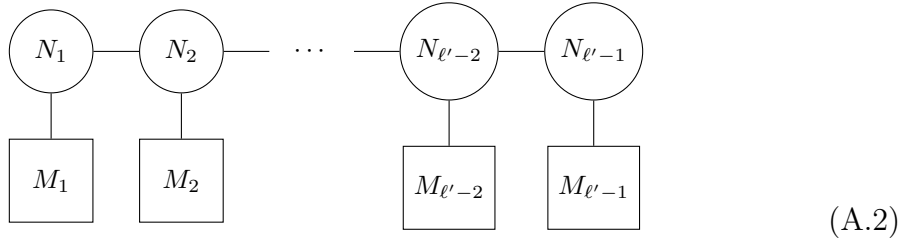
\label{quivTsigrhoSUN}
\begin{tikzpicture}[font=\footnotesize]
\begin{scope}[auto,%
  every node/.style={draw, minimum size=1.1cm}, node distance=0.6cm];
\node[circle] (UN1) at (0, 0) {$N_1$};
\node[circle, right=of UN1] (UN2) {$N_2$};
\node[draw=none, right=of UN2] (dots) {$\cdots$};
\node[circle, right=of dots] (UNlm1) {$N_{\ell'-2}$};
\node[circle, right=of UNlm1] (UNl) {$N_{\ell'-1}$};
\node[rectangle, below=of UN1] (UM1) {$M_1$};
\node[rectangle, below=of UN2] (UM2) {$M_2$};
\node[rectangle, below=of UNlm1] (UMlm1) {$M_{\ell'-2}$};
\node[rectangle, below=of UNl] (UMl) {$M_{\ell'-1}$};
\end{scope}
\draw (UN1) -- (UN2)
(UN2)--(dots)
(dots)--(UNlm1)
(UNlm1)--(UNl)
(UN1)--(UM1)
(UN2)--(UM2)
(UNlm1)--(UMlm1)
(UNl)--(UMl);
\end{tikzpicture}
\eea
The flavour symmetries $U(M_j)$, with $1\leq j \leq \ell'-1$, are determined from the transpose $\vec \sigma^T=(\hat \sigma_1, \ldots, \hat \sigma_{\hat \ell})$, with $\hat \sigma_1 \geq \ldots \geq  \hat \sigma_{\hat \ell} > 0$, of $\vec \sigma$ as follows:
\bea \label{flvgroups}
M_{j} &= 
\hat \sigma_{j} -\hat \sigma_{j+1} , \qquad \text{with}  \\
\hat \sigma_i &=0, \quad \text{for all}~ i \geq \hat \ell+1~.
\eea
Observe that $M_i=0$ for $i\geq \hat\ell +1$ so that  there are at most $\hat\ell$ flavour groups. 
The gauge symmetries $U(N_j)$, with $1\leq j \leq \ell'-1$, are given by
\bea
N_j &= \sum_{k=j+1}^{\ell'} \rho_k  - \sum_{i=j+1}^{\hat{\ell}} \hat \sigma_i~.
\eea
Note that the theories $T^{\vec \sigma}_{\vec \rho}(SU(N))$ are defined only for $\vec\sigma^T \, < \, \vec\rho$. 

In this paper, we focus on the case in which $\vec{\sigma} = [1^{N}]$ and the theory in question is denoted by $T_{\vec \rho}(SU(N))$.  The corresponding quiver reduces to
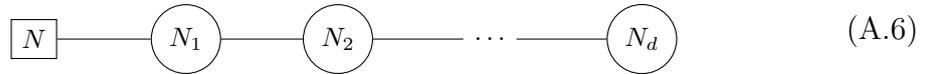
\be \label{quivTSUsigrho}
\begin{tikzpicture}[baseline]
\tikzstyle{every node}=[font=\footnotesize]
\node[draw, circle] (node1) at (-4,0) {$N_1$};
\node[draw, circle] (node2) at (-2,0) {$N_2$};
\node[draw=none] (node3) at (0,0) {$\ldots$};
\node[draw, circle] (node4) at (2,0) {$N_d$};
\node[draw, rectangle] (node6) at (-6,0) {$N$};
\draw[draw=black,solid]  (node6) to (node1) to (node2) to (node3) to (node4);
\end{tikzpicture}
\ee
where the number of gauge groups is $d= \ell'-1$.  In the main text, we mainly consider the following two examples of $T_{\vec \rho}(SU(2N+1))$.
\paragraph{The partition $\vec{\rho} = [1^{2N+1}]$.}  We denote the theory in this case by $T(SU(2N+1))$ and the corresponding quiver is
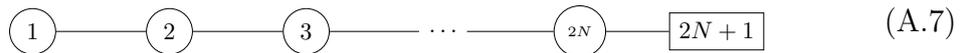
\be
\scalebox{0.9}{
\begin{tikzpicture}[baseline]
\tikzstyle{every node}=[font=\footnotesize]
\node[draw, circle] (node1) at (-4,0) {$1$};
\node[draw, circle] (node2) at (-2,0) {$2$};
\node[draw, circle] (node3) at (0,0) {$3$};
\node[draw=none] (node4) at (2,0) {$\ldots$};
\node[draw, circle] (node5) at (4,0) {\tiny $2N$};
\node[draw, rectangle] (node6) at (6,0) {$2N+1$};
\draw[draw=black,solid]  (node1) to (node2) to (node3) to (node4) to (node5) to (node6);
\end{tikzpicture}}
\ee
This is a self-mirror theory whose Higgs and Coulomb branches are isomorphic to the nilpotent cone of $SU(2N+1)$ \cite{Gaiotto:2008ak}, whose quaternionic dimension is 
\be
\frac{1}{2}\left[ (2N+1)^2- (2N+1) \right] = 2N(2N+1)~.
\ee

\paragraph{The partition $\vec{\rho} = [N+1,N]$.} The corresponding quiver is
\be
\scalebox{0.9}{ 
\begin{tikzpicture}[baseline]
\tikzstyle{every node}=[font=\footnotesize]
\node[draw, circle] (node6) at (6,0) {\tiny $N$};
\node[draw, rectangle] (node7) at (8,0) {$2N+1$};
\draw[draw=black,solid] (node6)  to (node7);
\end{tikzpicture}}
\ee

\subsubsection*{The Coulomb branch Hilbert series}
For a partition $\vec \rho$ of $N$, the Coulomb branch Hilbert series of the $T_{\vec \rho} (SU(N))$ theory can be computed from its quiver description using the monopole formula, described in \cite{Cremonesi:2013lqa}.  Alternatively, one can compute this quantity using the Hall-Littlewood formula \cite{Cremonesi:2014kwa}, without using the quiver description.  In the following, let us review the latter.

The Hall-Littlewood formula for the Coulomb branch Hilbert series of the $T_{\vec \rho} (SU(N))$ theory reads
\bes{ \label{HCTSUN}
&H[{T_{\vec \rho} (SU(N))}] (t; x_1, \ldots, x_{d+1} ; n_1, \ldots, n_N) \\
&= t^{\delta_{U(N)}(\vec n)}  (1-t^2)^N K^{U(N)}_{\vec \rho} (\vec x;t)\Psi_{U(N)}^{\vec n}(\vec x t^{\vec w_{\vec \rho}}; t) ~,
}
where the notations are as follows:
\ben
\item The integer $d+1$ is the length of the partition $\vec \rho$. It turns out that $d$ is also equal to the number of gauge group in quiver \eref{quivTSUsigrho}.
\item The Coulomb branch symmetry of the $T_{\vec \rho}(SU(N))$ theory is $\left[ \prod_k U(r_k) \right]/U(1)$, where $r_k$ is the number of times that part $k$ appears in the partition $\vec \rho$ and $d+1 =\sum_{k} r_k$.  The fugacities associated to this symmetry are $x_1, \, x_2, \,\ldots, \, x_{d+1}$ subject to the constraint:
\be \label{fugconstr}
\prod_{i=1}^{d+1} x_i^{\rho_i} =1~.
\ee
\item The power of $t$ in the prefactor is
\bes{
\delta_{U(N)}(\vec n) = \sum_{j=1}^N (N+1-2j)n_j~.
}
\item The Hall-Littlewood polynomial associated with the group $U(N)$ is given by
\bea
\Psi^{\vec n}_{U(N)} (x_1,\dots,x_N;t)=\sum_{\sigma \in S_N}
x_{\sigma(1)}^{n_1} \dots x_{\sigma(N)}^{n_N}
\prod_{1 \leq i<j \leq N}   \frac{  1-t x_{\sigma(i)}^{-1} x_{\sigma(j)} } {1-x_{\sigma(i)}^{-1} x_{\sigma(j)}}~.
\eea
\item $\vec w_{{r}}$ denotes the weights of the $SU(2)$ representation of dimension $r$:
\bea
\vec w_{{r}} = (r-1, r-3, \ldots, 3-r, 1-r)~.
\eea
Hence the notation $t^{\vec w_{r}}$ represents the vector
\bea
t^{\vec w_{r}} = (t^{(r-1)}, t^{(r-3)}, \ldots, t^{-(r-3)},t^{-(r-1)})~.
\eea
We abbreviate
\bea
\Psi_{U(N)}^{\vec n}(\vec x t^{\vec w_{\vec \rho}}; t) := \Psi_{U(N)}^{(n_1, \ldots, n_N)}(x_1 t^{\vec w_{\rho_1}}, x_2 t^{\vec w_{\rho_2}} , \ldots, x_{d+1} t^{\vec w_{\rho_{d+1}}};t)~.
\eea
\item The prefactor $K^{U(N)}_{\vec \rho} (\vec x;t)$  is given by \cite{Gadde:2011uv}
\bea \label{KUN}
K^{U(N)}_{\vec \rho} (\vec x;t) = \prod_{i=1}^{\text{length}({\vec \rho}^T)} \prod_{j,k=1}^{\rho^T_i} \frac{1}{1-a^i_j \bar{a}^i_k}~,
\eea
where ${\vec \rho}^T$ denotes the transpose of the partition $\vec{\rho}$ and
\be
\begin{split}
a^i_j &= x_j \;\; t^{ \rho_j-i+1}~, \qquad  i=1,\dots,\rho_j \\
{\bar a}^i_k &= x_k^{-1} t^{ \rho_k-i+1}~, \qquad  i=1,\dots,\rho_k \label{defaabar}
\end{split}
\ee
For example:
\ben
\item For $\vec \rho= [1^N]$, we have $\vec \rho^T = [N]$ and so
\bea
K^{U(N)}_{[1^N]} (\vec x; t)= \prod_{1\leq j, k\leq N} \frac{1}{1- x_j x_k^{-1} t^2} =  \PE[t^2 \chi^{U(N)}_{\bf Adj} (\vec x) ]~.
\eea
\item For the partition $\vec \rho=[N+1,N]$ of $2N+1$, we have $\vec \rho^T = [2^N,1]$ and so 
\be
\scalebox{0.9}{$
\begin{split}
K^{U(2N+1)}_{[N+1,N]} (\vec x; t)= \PE \left[t^{2N+2}+(x_1 x_2^{-1} + x_2 x_1^{-1}) \sum_{j=1}^N t^{2j+1}  + 2 \sum_{l=1}^N t^{2l} \right]~.
\end{split} $}
\ee
\een
\een

\subsection{$T^{\vec \sigma}_{\vec \rho}(USp'(2N))$} \label{sec:TsigrhoUSpP}
The $T^{\vec \sigma}_{\vec \rho}(USp'(2N))$ theory was studied in \cite{Cremonesi:2014uva} (see also \cite{Feng:2000eq}). From the string theory perspective, it can be realised on the worldvolume of $N$ D3 branes parallel to an orientifold $\widetilde{\mathrm{O3}}^+$ plane and ending on systems of half D5 branes and of half NS5 branes. Similar to the theories discussed in \cite{Gaiotto:2008ak}, the partitions ${\vec \sigma}$ and ${\vec \rho}$ determine how the D3 branes end on the half D5 branes and on the half NS5 branes respectively. In this case both $\vec \sigma$ and $\vec \rho$ are $C$-partitions of $USp(2N)$, of lengths $\ell$ and $\ell'$ respectively.

The quiver diagram for $T^{\vec \sigma}_{\vec \rho}(USp'(2N))$ consists of alternating $(S)O/USp$ groups depicted in \eref{quivTCsigrhoP}, where each red node with a label $N$ denotes an $O(N)$ or $SO(N)$ group and each blue node with an even label $2N$ denotes a $USp(2N)$ group.
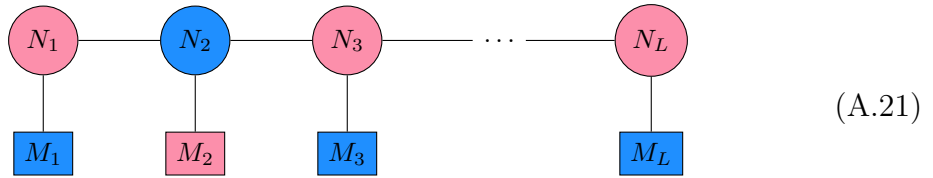
\be \label{quivTCsigrhoP}
\begin{tikzpicture}[baseline]
\tikzstyle{every node}=[font=\footnotesize]
\node[draw, circle, fill=lightred] (node1) at (-4,1) {$N_1$};
\node[draw, circle, fill=lightblue] (node2) at (-2,1) {$N_2$};
\node[draw, circle, fill=lightred] (node3) at (0,1) {$N_3$};
\node[draw=none] (node4) at (2,1) {$\ldots$};
\node[draw, circle, fill=lightred] (node5) at (4,1) {$N_L$};
\node[draw, rectangle,fill=lightblue] (sqnode1) at (-4,-0.5) {$M_1$};
\node[draw, rectangle, fill=lightred] (sqnode2) at (-2,-0.5) {$M_2$};
\node[draw, rectangle, fill=lightblue] (sqnode3) at (0,-0.5) {$M_3$};
\node[draw, rectangle, fill=lightblue] (sqnode5) at (4,-0.5) {$M_L$};
\draw[draw=black,solid]  (node1) to (node2) to (node3) to (node4) to (node5);
\draw[draw=black,solid]  (node1) to (sqnode1);
\draw[draw=black,solid]  (node2) to (sqnode2);
\draw[draw=black,solid]  (node3) to (sqnode3);
\draw[draw=black,solid]  (node5) to (sqnode5);
\end{tikzpicture}
\ee
We defined 
\bea
L= \begin{cases} \ell' -1  & \quad \text{$\ell'$ is even} \\
\ell'  & \quad \text{$\ell'$ is odd} \end{cases} 
\eea
and if both $N_L$ and $M_L$ are zero, the nodes are removed from the quiver and the length of quiver \eref{quivTCsigrhoP} is $L-1$.

The labels $M_j$, with $1\leq j \leq L$, for the flavour symmetries are determined from the transpose $\vec \sigma^T=(\hat \sigma_1, \ldots, \hat \sigma_{\hat \ell})$, with $\hat \sigma_1 \geq \ldots \geq  \hat \sigma_{\hat \ell} > 0$, of $\vec \sigma$ as follows:
\bea \label{flvgroups}
M_{j} &= 
\hat \sigma_{j} -\hat \sigma_{j+1} , \qquad \text{with}  \\
\hat \sigma_i &=0, \quad \text{for all}~ i \geq \hat \ell+1~.
\eea
On the other hand, the labels $N_j$, with $1\leq j \leq L$, for the gauge symmetries are given by
{\small
\bea
N_j = \begin{cases} 
\left[1+ \sum_{k=j+1}^{\ell'} \rho_k  \right]_{\tilde{+}} - \left( \sum_{i=j+1}^{\hat{\ell}} \hat \sigma_i \right) \quad \text{for the $O/SO$ node}~, & \text{if $\ell'$ is even}~, \\
 \left[\sum_{k=j+1}^{\ell'} \rho_k  \right]_{-} - \left( \sum_{i=j+1}^{\hat{\ell}} \hat \sigma_i \right) \quad \text{for the $USp$ node}~, & \text{if $\ell'$ is even}~, \\ 
\left[ \sum_{k=j+1}^{\ell'} \rho_k  \right]_{\tilde{+}} - \left( \sum_{i=j+1}^{\hat{\ell}} \hat \sigma_i \right) \quad \text{for the $O/SO$ node}~, & \text{if $\ell'$ is odd}~,\\ 
\left[ \sum_{k=j+1}^{\ell'} \rho_k  \right]_{+} - \left( \sum_{i=j+1}^{\hat{\ell}} \hat \sigma_i \right) \quad \text{for the $USp$ node}~, & \text{if $\ell'$ is odd}~.
\end{cases}
\eea}

In this paper, we focus on the case in which $\vec{\sigma} = [1^{2N}]$ and the theory in question is denoted by $T_{\vec \rho}(USp'(2N))$.  The corresponding quiver reduces to
\be \label{quivTCrhoP}
\scalebox{0.9}{
\begin{tikzpicture}[baseline]
\tikzstyle{every node}=[font=\footnotesize]
\node[draw, circle, fill=lightred] (node1) at (-4,0) {$N_1$};
\node[draw, circle, fill=lightblue] (node2) at (-2,0) {$N_2$};
\node[draw, circle, fill=lightred] (node3) at (0,0) {$N_3$};
\node[draw=none] (node4) at (2,0) {$\ldots$};
\node[draw, circle, fill=lightred] (node5) at (4,0) {$N_L$};
\node[draw, rectangle, fill=lightblue] (node6) at (-6,0) {$2N$};
\draw[draw=black,solid]  (node6) to (node1) to (node2) to (node3) to (node4) to (node5);
\end{tikzpicture}}
\ee
As we justify in the main text, the red circular node with a label $N$ in this quiver should be taken as the {\bf special orthogonal $SO(N)$ gauge group}.  We mainly consider the following two examples.
\paragraph{The partition $\vec{\rho} = [1^{2N}]$.}  We denote the theory in this case by $T(USp'(2N))$ and the corresponding quiver is
\be
\scalebox{0.9}{
\begin{tikzpicture}[baseline]
\tikzstyle{every node}=[font=\footnotesize]
\node[draw, circle, fill=lightred] (node1) at (-4,0) {$3$};
\node[draw, circle, fill=lightblue] (node2) at (-2,0) {$2$};
\node[draw, circle, fill=lightred] (node3) at (0,0) {$5$};
\node[draw, circle, fill=lightblue] (node4) at (2,0) {$4$};
\node[draw=none] (node5) at (4,0) {$\ldots$};
\node[draw, circle, fill=lightred] (node6) at (6,0) {\tiny $2N+1$};
\node[draw, rectangle, fill=lightblue] (node7) at (8,0) {$2N$};
\draw[draw=black,solid]  (node1) to (node2) to (node3) to (node4) to (node5) to (node6)  to (node7);
\end{tikzpicture}}
\ee
We remark this quiver is a `bad' theory in the sense of \cite{Gaiotto:2008ak}.  Nevertheless, one can use this description to compute many quantities, such as the Coulomb branch dimension and the Higgs branch Hilbert series.  Moreover, we can bypass the `badness' of the quiver and computing the Coulomb branch Hilbert series using the Hall-Littlewood formula as will be explained below.  Since $\vec \rho = \vec \sigma = [1^{2N}]$, the theory is indeed self-mirror in this case.  In fact, both Higgs and Coulomb branches of this theory is isomorphic to the nilpotent cone of $USp(2N)$, whose quaternionic dimension is
\be
\frac{1}{2}\left[\frac{1}{2}(2N)(2N+1) - N \right] = N^2~.
\ee

\paragraph{The partition $\vec{\rho} = [2N]$.} The corresponding quiver is
\be
\scalebox{0.9}{ \label{quivrho2N}
\begin{tikzpicture}[baseline]
\tikzstyle{every node}=[font=\footnotesize]
\node[draw, circle, fill=lightred] (node6) at (6,0) {$1$};
\node[draw, rectangle, fill=lightblue] (node7) at (8,0) {$2N$};
\draw[draw=black,solid] (node6)  to (node7);
\end{tikzpicture}}
\ee
where the red circular node denotes the $SO(1)$ group, and so the gauge symmetry is trivial in this case.  This is simply a theory of free $2N$ half-hypermultiplets.

\subsubsection*{The Coulomb branch Hilbert series}
It is possible to compute the Coulomb branch Hilbert series from the quiver description using the monopole formula \cite{Cremonesi:2013lqa}, provided that the quiver is not a `bad' theory in the sense of \cite{Gaiotto:2008ak}.  Alternatively, for a given $C$-partition $\vec \rho$ of $2N$, one can directly compute the Coulomb branch Hilbert series of $T_{\vec \rho}(USp'(2N))$ using the Hall-Littlewood formula \cite{Cremonesi:2014kwa}, without using the quiver and regardless whether the quiver is `bad' or not.  Although the Hall-Littlewood formula for the Coulomb branch Hilbert series $T_{\vec \rho}(USp'(2N))$ was not discussed in \cite{Cremonesi:2014kwa}, the following simple modification of (4.9) of \cite{Cremonesi:2014kwa} yields the required formula for the Coulomb branch Hilbert series in question:
\bes{ \label{HCTUSpp}
&H_\CC[T_{\vec \rho}(USp'(2N))](t; \vec x, n_1,  \ldots, n_N) \\
&= t^{\delta_{C_N}(\vec n)} (1-t^2)^N K^{C_N}_{\vec \rho} (\vec x,t)  \Psi^{\vec n}_{C_N} (\vec{a}(t,\vec x),t)~,
}
where the notations are as follows:
\ben
\item The power of $t$ in the prefactor is
\bes{
\delta_{C_N}(\vec n) = \sum_{j=1}^N (2N+2-2j)n_j~.
}
\item
The function $ \Psi^{\vec \lambda}_{C_N} (\vec x,t)$ is the Hall-Littlewood polynomial associated with the $C_N$ algebra and the partition $\vec \lambda$ is subject to $\lambda_1 \geq \lambda_2 \geq \ldots \geq \lambda_N \geq 0$, with all $\lambda_i$ integers.  It is given by
\bes{
 \Psi^{\vec \lambda}_{C_N} (x_1,\dots,x_N;t) &= \sum_{s_1,\ldots, s_N= \pm1} \; \sum_{\sigma \in S_N}
 \left( \prod_{i=1}^N x_{\sigma(i)}^{s_i \lambda_i} \frac{1-t^2 x_{\sigma(i)}^{-2s_i} } {1-x_{\sigma(i)}^{-2s_i} } \right) \times \nn\\
& \qquad \left( \prod_{1 \leq i<j \leq N}   \frac{1-t^2 x_{\sigma(i)}^{-s_i} x_{\sigma(j)}^{s_j} } {1-x_{\sigma(i)}^{-s_i} x_{\sigma(j)}^{s_j}} \cdot  \frac{1-t^2 x_{\sigma(i)}^{-s_i} x_{\sigma(j)}^{-s_j} } {1-x_{\sigma(i)}^{-s_i} x_{\sigma(j)}^{-s_j} } \right)~.
}
\item  The argument $\vec{a}(t,\vec x)$, which shall be abbreviated as $\vec a$, of the Hall-Littlewood polynomial can be determined by considering the decomposition
\be
x^{C_N}_{\text{\bf fund}} (\vec a) = \sum_{j=1}^N (a_j+a_j^{-1}) = \sum_{k} x^{G_{\rho_k}}_{\text{\bf fund}} ({\vec x}_k) \chi^{SU(2)}_{[\rho_k-1]} (t)~,
\ee
where the group $G_{\rho_k}$ depends on the part $k$ of the partition $\vec \rho$ that appears $r_k$ times and is defined as
\bes{
G_{\rho_k} = \begin{cases} USp(r_k) \qquad & \text{if $k$ is odd} \\
SO(r_k) \qquad & \text{if $k$ is even}~. \\
\end{cases}
}
For example, for $\vec \rho = [1^{2N}]$, we have $a_j = x_j$ for $j=1,\ldots, N$, and for $\vec \rho = [2N]$, we have $a_j =t^{2j-1}$ for $j=1,\ldots, N$.
\item The prefactor $K^{C_N}_{\vec \rho} (\vec x,t) $ can be determined in two steps.  First of all, we need to identify the representations $R_j$ of the group
\be \label{Grho}
G_{\vec \rho} = \prod_k G_{\rho_k} = \prod_{\text{$k$ odd}} USp(r_k) \times  \prod_{\text{$k$ even}} SO(r_k)
\ee 
from the following decomposition:
\bes{
\chi^{C_N}_{\bf Adj} (\vec a) 
&= \sum_{j =0, \frac{1}{2}, 1, \frac{3}{2}, \ldots}  \chi^{G_{\vec \rho}}_{R_j}(\vec x_j)  \chi^{SU(2)}_{[2j]}(t)~. \label{decompadj} 
}
Once $R_j$ are determined, the prefactor in question is then given by
\bes{
K^{C_N}_{\vec \rho} (\vec x,t) =\PE \left[\sum_{j =0, \frac{1}{2}, 1, \frac{3}{2}, \ldots}  t^{2j+2} \chi^{G_{\vec \rho}}_{R_j}({\vec x}_j )\right].  \label{K}
}
For example,  for $\vec \rho = [1^{2N}]$, we have 
\be
K^{C_N}_{[1^{2N}]} (\vec x,t) = \PE \left[\chi^{C_N}_{\bf Adj}(\vec x)  t^2 \right]~ 
\ee
and for $\vec \rho = [2N]$, we have
\be
K^{C_N}_{[1^{2N}]} (\vec x,t) = \PE \left[ t^4 + t^8 +\ldots + t^{4N} \right]~.
\ee
\een
For a given partition $\vec \rho$, the Coulomb branch symmetry of $T_{\vec \rho}(USp'(2N))$ is $G_{\vec \rho}$, determined by \eref{Grho}.  In the Coulomb branch Hilbert series \eref{HCTUSpp}, the fugacities $\vec x$ are those associated with the symmetry $G_{\vec \rho}$, and $n_1, \ldots, n_N$ are the background magnetic fluxes associated with the flavour symmetry $USp(2N)$ of the theory.  

Note that in the special case of $\vec \rho = [2N]$, the Hall-Littlewood polynomial is
\bes{
 \Psi^{\vec \lambda}_{C_N} (x_1,\dots,x_N;t)  = t^{-\sum_{j=1}^N (2N+1-2j)n_j} \PE[ N t^2 - t^4 - t^8-\ldots -t^{4N}]
}
and so the Hilbert series \eref{HCTUSpp} becomes
\be
H_\CC[T_{[2N]}(USp'(2N))](t; \vec x, n_1,  \ldots, n_N) = t^{n_1+ n_2+\ldots+ n_N}~.
\ee
This is indeed the Coulomb branch Hilbert series of the theory of free $2N$ half-hypermultiplets, as described in \eref{quivrho2N}.

\section{Reduction of the $D_2[SU(2N+1)]$ theory on $S^1$} \label{sec:D2SUodd}
The 4d $\CN=2$ $D_2[SU(2N+1)]$ theory was first studied in \cite{Cecotti:2012jx, Cecotti:2013lda}. For $N=1$, the $D_2[SU(3)]$ theory is simply the $(A_1, D_4)$ Argyres-Douglas theory. Upon reduction on $S^1$ to 3d, the mirror theory is described by the following quiver \cite[Figure 3]{Xie:2016uqq}:
\be \label{mirrD2SU2Np1}
\begin{tikzpicture}[baseline]
\tikzstyle{every node}=[font=\footnotesize]
\node[draw, circle] (node1) at (-6,0) {$1$};
\node[draw, circle] (node2) at (-4.5,0) {$2$};
\node[draw=none] (node3) at (-3,0) {\Large $\ldots$};
\node[draw, circle] (node4) at (-1.5,0) {\tiny $N-1$};
\node[draw, circle] (node5) at (0,0) {$N$};
\node[draw, circle] (node6) at (1.5,0) {$N$};
\node[draw, circle] (node7) at (3,0) {\tiny $N-1$};
\node[draw=none] (node8) at (4.5,0) {\Large $\ldots$};
\node[draw, circle] (node9) at (6,0) { $2$};
\node[draw, circle] (node10) at (7.5,0) {$1$};
\node[draw, rectangle] (sqnode1) at (0,1.2) {$1$};
\node[draw, rectangle] (sqnode2) at (1.5,1.2) {$1$};
\draw[draw=black,solid]  (node1) to(node2) to (node3) to (node4) to (node5) to (node6) to (node7) to (node8) to (node9) to (node10);
\draw[draw=black,solid]  (node5) to (sqnode1);
\draw[draw=black,solid]  (node6) to (sqnode2);
\end{tikzpicture}
\ee
The quaternionic dimension of the Coulomb branch of \eref{mirrD2SU2Np1} is $2\sum_{j=1}^N j = N(N+1)$, in agreement with the dimension of the Higgs branch of the 4d $D_2(SU(N+1))$ theory, which is given by
\bes{
\dim_\BH \, \CH[D_2(SU(2N+1))] &= 24(c-a) \\
&= 24 \left[ \frac{1}{3}N(N+1)-\frac{7}{24}N(N+1) \right] \\
&= N(N+1) ~,
}
where $a=\frac{7}{24}N(N+1) $ and $c=\frac{1}{3}N(N+1)$ are the conformal anomalies \cite{Cecotti:2013lda}.  The Higgs branch of \eref{mirrD2SU2Np1} is $2\sum_{j=1}^{N-1} j(j+1) +N^2 +2N -2 \sum_{j=1}^N j^2 = N$ quaternionic dimensional; this is in agreement with the fact that the $D_2[SU(2N+1)]$ theory is a rank $N$ theory.

The Coulomb branch and Higgs branch Hilbert series can be computed as described in the main text.  The Coulomb branch symmetry is $SU(2N+1)$, whereas the Higgs branch symmetry is $U(1)$.  In this paper we focus mainly on the case of $N=1, 2$. The case of $N=1$ was discussed in the main text in the context of the $(A_1, D_4)$ theory.  For $N=2$, the highest weight generating function of the Coulomb branch Hilbert series of theory \eref{mirrD2SU2Np1} admits the following simple closed form:
\bes{
\mathrm{HWG} \left[ H_\CC [ \eref{mirrD2SU2Np1}_{N=2}] \right] &=\mathrm{HWG} [H_\CH [D_2[SU(5)]]] \\
&= \PE \left[ t^2 \mu_1 \mu_4 + t^4 \mu_2 \mu_3\right]~.
}
If we set the fugacities in the $SU(5)$ characters to unity, we obtain the closed form for the following unrefined Coulomb branch Hilbert series for $N=2$:
\bes{
\frac{1 + 12 t^2 + 53 t^4 + 88 t^6 + 53 t^8 + 12 t^{10} + t^{12}}{(1 - t)^{12} (1 + t)^{12}}~.
}
Observe that the order of the pole at $t=1$ is $12$, equal to the complex dimension of the Coulomb branch.

\bibliographystyle{ytphys}
\bibliography{ref}
\end{document}